\documentclass[aps, pra, twocolumn, longbibliography, superscriptaddress]{revtex4-1}

\usepackage[table]{xcolor}
\usepackage{physics}
\usepackage{amsmath, amssymb}
\usepackage{graphicx}
\usepackage{dcolumn}		    
\usepackage{bm} 				
\usepackage{bbold}
\usepackage{floatrow}
\usepackage{xfrac}

\usepackage{mathtools}	        

\usepackage[caption=false]{subfig}
\usepackage{verbatim}
\usepackage{hyperref}

\usepackage{amsthm} 

\begin{document}

\title{Consistent Quantization of Nearly Singular Superconducting Circuits}

\author{Martin Rymarz}
\email{martin.rymarz@rwth-aachen.de}
\affiliation{JARA-Institute for Quantum Information, RWTH Aachen University, D-52056 Aachen, Germany}

\author{David P. DiVincenzo}
\affiliation{JARA-Institute for Quantum Information, RWTH Aachen University, D-52056 Aachen, Germany}
\affiliation{Peter Gr\"unberg Institute, Theoretical Nanoelectronics, Forschungszentrum J\"ulich,\\ D-52425 J\"ulich, Germany}

\begin{abstract}
The theory of circuit quantum electrodynamics has successfully analyzed superconducting circuits on the basis of the classical Lagrangian, and the corresponding quantized Hamiltonian, describing these circuits.
In many simplified versions of these networks, the modeling involves a Lagrangian that is \textit{singular}, describing an inherently constrained system.
In this work, we demonstrate the failure of the Dirac-Bergmann theory for the quantization of realistic, nearly singular superconducting circuits, both reciprocal and nonreciprocal.
The correct treatment of nearly singular systems involves a perturbative Born-Oppenheimer analysis. We rigorously prove the validity of the corresponding perturbation theory using Kato-Rellich theory.
We find that the singular limit of this regularized analysis is, in many cases, completely unlike the singular theory.
Dirac-Bergmann, which uses the Kirchhoff's (and Tellegen's) laws to deal with constraints, predicts dynamics that depend on the detailed parameters of nonlinear circuit elements, e.g., Josephson inductances.
By contrast, the limiting behavior of the low-energy dynamics obtained from the regularized Born-Oppenheimer approach exhibits a fixed point structure, flowing to one of a few universal fixed points as parasitic capacitance values go to zero. 
\end{abstract}

\maketitle

\section{Introduction}
Superconducting circuits \cite{Google, Zuchongzhi2_0, Zuchongzhi2_1} facilitate a highly promising architecture for the realization of a universal quantum computer \cite{Feynman, Nielsen}, whose potential to outperform a classical computer in special tasks is the driving force of an entire area of research. However, although superconducting qubits exist for more than two decades \cite{ShnirmanCPB, DevoretCPB, NakamuraCPB}, state-of-the-art quantum technology is still too noisy to allow for accurate calculations of arbitrary length \cite{PreskillNISQ}.
Many efforts are put into the improvement of currently existing superconducting qubits as well as into the invention and fabrication of entirely new designs. Purposeful design of new circuits has been successful; for example, the fluxonium \cite{Manucharyan} is observed to be the most coherent superconducting qubit to date \cite{Somoroff}.

Theoretical work has been a successful contributor to this effort. Ideally, superconducting circuits are described by a lossless dynamics of a discrete set of degrees of freedom. These circuits are described classically on the Hamiltonian level, where fluxes and charges are considered as pairs of conjugate variables. The quantization of these macroscopic Hamiltonians has been successful \cite{Devoret1, Devoret2, Girvin_cQED, cQED_Review}, with quantitatively accurate predictions of many observed phenomena. 

Especially in more advanced qubit designs, with increasing numbers of independent dynamical variables, one sees the emergence of a {\em hierarchy} of energy (or time) scales; it is the consequences of this hierarchy that will be the subject of this paper. 

The purposeful use of this hierarchy is one aspect of a set of four very simple design principles, which have enabled the large number of successful circuit designs that are in use today. 1) Use only the standard lossless circuit elements, the capacitor and the inductor, (obviously) avoiding resistors. (And not using, up until now, the standard gyrator, but see, e.g., Ref.~\cite{Rymarz}.) 2) Achieve long-distance coupling by transmission lines, but used in such a way that they again can be effectively represented by a small assembly of capacitors and inductors. 3) Use metallization to, on purpose, make some node-to-node capacitances very large, while keeping many node-to-node capacitances at their small, parasitic values, resulting in a range of capacitance values of perhaps seven orders of magnitude \cite{Roth_IEEE}. This is the hierarchy we will study here. 4) Use linear as well as nonlinear inductors. 

Of course, principle 4 is a centerpiece of qubit circuits, with the use of a particular nonlinearity, that given by the Josephson junction. In contrast to a linear inductance described by a linear current-versus-flux characteristic $I=\phi/L$, the Josephson nonlinear inductor has the two-terminal characteristic $I=I_c\sin(2\pi\phi/\Phi_0)$. The availability of this low-loss nonlinearity permits the quantum eigenspectrum of these circuits to be atomic-like, in that it can make the $\ket{0}\!\!-\!\!\ket{1}$ energy difference unique, making it possible to perform quantum logic gates by resonant Rabi driving.

We will show here that principles 3 and 4 interact in a novel and, potentially, dangerous way. A first, seemingly natural, step in the analysis of circuits is, given the capacitance hierarchy of real structures, to declare a certain capacitance threshold $C_{th}$, and to set all capacitances below this value equal to zero. This considerably simplifies analysis, and leads to a simple way of conceptualizing very useful composite effective inductance structures, including the superinductor \cite{Manucharyan, Brooks, ViolaCatelani}, and the SNAIL \cite{SNAIL1, SNAIL2, SNAIL3}, to be discussed below. 

But this simplification often leads to an important consequence for the mechanics of the circuit. Adopting the common procedure of describing this mechanics using a Lagrangian $\mathcal{L}$ with node fluxes $\phi_i$ as dynamical variables \cite{Devoret1, Devoret2, Girvin_cQED, cQED_Review}, one finds that some dynamical variables can become constrained. A constrained variable is one whose conjugate variable $\partial\mathcal{L}/\partial\dot\phi_i$ cannot be properly inverted to obtain $\dot\phi_i$, and whose classical dynamics is slaved to other independent variables, i.e., $\phi_1(t)=g(\phi_2(t),\phi_3(t),\ldots)$, at all times. Constrained (or ``frozen") variables are indeed considered a useful simplification in current treatments of superconducting circuits, singled out in currently available software \cite{Koch_Chitta}.

Singular mechanics and its quantization have received considerable attention in modern physics. First systematic treatments of singular Lagrangians and proposals for their quantization were proposed independently by Dirac and Bergmann in the early 1950's \cite{Bergmann1, Bergmann2, Dirac1, Bergmann3, Dirac2, Dirac3}. Since then, the proposed procedure of progressively identifying and classifying certain constraints of the system, known as the Dirac-Bergmann algorithm \cite{HenneauxBook, RotheBook, Brown}, has been frequently applied to various singular gauge theories. Often these theories are applied in cases where the singularity is viewed as fundamental, for example when a particle is expected to have exactly zero mass.
There also exist singular Lagrangians that approximate a limiting case of a non-singular system, e.g., when a particle has very small but nonzero mass. We consider the circuit problem to be in the latter category: capacitances play the role of masses, and, according to basic electrostatics, node-to-node capacitances are never exactly zero.

The Dirac-Bergmann algorithm can be worked out for general lossless electric networks, as we present in full generality (including nonreciprocity) in Appendix~\ref{sec_general_formalism}. For our circuits, the Dirac-Bergmann algorithm amounts to applying Kirchhoff's current law, eliminating variables by using basic series combination rules. For example, the algorithm says that a series combination of inductances $L_1$ and $L_2$ can be replaced by a single inductance $L_1+L_2$ (thus neglecting any capacitance to the joining node). This indeed turns out to be correct from all points of view.

But, the Dirac-Bergmann algorithm makes predictions also for nonlinear circuits, i.e., involving Josephson elements. Do its predictions also agree with a ``regularized" approach, in which one considers the limit as all small capacitances are taken to zero? The answer is, absolutely, \underline{\em \bf no}. A sign of trouble already appears when we consider a series combination of a linear inductance and a Josephson inductor. While the resulting effective inductor depends in detail on the parameters of the two elements, for certain parameters the effective inductive energy is predicted to be {\em multivalued}. 

Suggestions exist in the literature for how such multivaluedness should be interpreted; see the theory of ``branched Hamiltonians" \cite{Henneaux, Wilczek_branched}. However, we find no existing approach that matches the result of regularizing the singularity by taking small capacitances $C_{s} < C_{th}$ into account. Our result is in complete contrast to Dirac-Bergmann, where the effective Hamiltonian depends in detail on the parameters of the nonlinear element; in the regularized treatment, the result is more akin to a renormalization flow, in the sense that the limit $C_{s}\rightarrow 0$ gives a universal result with only a few possible fixed points.

To understand this fixed point structure more comprehensively, we find it valuable to adopt the point of view, perhaps due to Heaviside, stated routinely in many textbooks on electrical theory \cite{Peikari, DesoerKuh1, DesoerKuh2}: an inductor is a two terminal element exhibiting an instantaneous relationship between current and flux, $I(t)=f(\phi(t))$, with arbitrary function $f(\cdot)$. (A similar formulation of nonlinear capacitor is also given, which we will use much less in this paper.) While distinctions are made between bijective, current-controlled, and flux-controlled inductors (the Josephson characteristic is flux-controlled), in all cases this generalized inductor is a proper energy storage device. Thus, such an element can be incorporated into a circuit Lagrangian for arbitrary $f(\cdot)$ \footnote{When a flux-controlled inductor has a characteristic $f(\cdot)$ that is periodic in $\phi$, it may be appropriate (and important for quantization) to consider $\phi$ to be a compact rather than extended variable \cite{PhysRevLett.103.217004}. This subtle issue does not affect any of the conclusions of this paper, and we will not consider the compactness question any further here}.

This generalization is highly valuable in that it reveals that there are generically three fixed points as $C_{s}\rightarrow 0$. They are exemplified by our simple series combination scenario, in which a linear inductance $L$ is in series with a nonlinear inductor with anti-symmetric characteristic $I\sim \rm{sign}(\phi)|\phi|^\beta$. The renormalization flow is determined by $\beta$. For all ``sublinear" cases ($0<\beta<1$), the flow erases the two elements from the circuit, i.e., they are replaced by an {\em open circuit}. For the ``superlinear" cases ($\beta>1$), the flow results in the nonlinear element being replaced by a {\em short circuit}. The linear $\beta=1$ case is marginal, and is the one case where the combination procedure given by Dirac-Bergmann is correct.

The Josephson case is in the sublinear universality class and flows to the open-circuit fixed point. To show this, and to determine the universality class of a large set of $f(\cdot)$, we calculate as follows: 1) For sufficiently small $C_{s}$, the variable to be eliminated becomes ``fast", and can be accurately dealt with using the Born-Oppenheimer approximation. 2) With suitable rescaling, the fast-variable Schr{\"o}dinger equation is one in which one term can be treated perturbatively. 

To prove that the flow goes to our fixed points, we must prove the convergence of the resulting perturbation problem. Particularly for the sublinear case, we successfully treat a large class of functions $f(\cdot)$, dealing with the perturbation theory rigorously, using primarily the Kato-Rellich theorem \cite{Kato} as provided by Reed and Simon \cite{ReedSimon}. We cannot prove that all $f(\cdot)$ flow to one of the fixed points; we find that the flow has additional complexities when non-symmetric characteristics are studied. We also show an amusing example of a self-similar $f(\cdot)$ for which the flow is successively attracted by two different fixed points as $C_{s}\rightarrow 0$, but never reaches either of them. We find that the flows have additional complexity in nonreciprocal circuits, and we work out several examples of singular circuits involving gyrators.

One can finally say that the physics of our results has to do with the diverging quantum fluctuations of the variables to be eliminated as $C_{s}\rightarrow 0$. This is the complete opposite of the Dirac-Bergmann treatment, in which these variables have no independent quantum fluctuations, being simply slaved to other variables in the circuit. But while these zero-point fluctuations diverge, the character of these divergences shows three different varieties, giving rise to the three fixed points that we have identified. 

The remainder of this paper is organized as follows.
In Sec.~\ref{sec_singular_circuits}, we review both the concepts of singular Lagrangians and the application of the Dirac-Bergmann algorithm. Based on two concrete examples, we demonstrate that the results of the systematically applied Dirac-Bergmann algorithm have to be handled with care if the system is supposed to be quantized.
In Sec.~\ref{sec_failure_Dirac}, we analyse the series combination of a linear inductance and a generic nonlinear inductor, and we provide an expression for the effective replacement of this series combination. In particular, we compare the results obtained from the Dirac-Bergmann algorithm with the limiting case of the low-energy dynamics derived from the Born-Oppenheimer approximation after the inclusion of small parasitic capacitances that lift the singularity of the system.
In Sec.~\ref{sec_SPA}, we revise the frequently used single-phase approximation for the simplified analysis of arrays of Josephson junctions. We show that such a single-phase approximation is akin to the application of the Dirac-Bergmann algorithm although an opposite limit of capacitances is considered. In particular, we provide a leading order correction term to the single-phase approximation due to the finite intrinsic capacitances of the Josephson junctions.
In Sec.~\ref{sec_Nonreciprocal_Circuits}, we extend our analysis to nonreciprocal circuits,
and we illustrate that conventionally used replacement laws for nonreciprocal circuits are not applicable for a quantized description of the system.
Finally, we summarize our results and provide a perspective for possible future work in Sec.~\ref{sec_conclusion}.

\section{Singular Superconducting Circuits and the Dirac-Bergmann Algorithm}\label{sec_singular_circuits}
The theory of circuit quantum electrodynamics \cite{Devoret1, Devoret2, Girvin_cQED, cQED_Review} provides a very powerful tool for the description of superconducting circuits. Generally, it starts with a circuit modeling the electrical network under consideration. With a particular choice of variables, each circuit element usually can be associated with a contribution to the total Lagrangian describing the system \footnote{There are exceptions to this rule. For example, as we see in Sec.~\ref{sec_Tellegen_transformer}, ideal transformers do not enter the Lagrangian of a system as an additional term but enforce a constraint on the variables.}. After the assembly of the total Lagrangian, a Legendre transformation converts the Lagrangian formalism to the Hamiltonian formalism, which, in turn, is the starting point for a quantized theory.
 
However, depending on both the physical precision and the details of the model that describes the system, the Legendre transformation is not always applicable, viz. invertible. Given a Lagrangian $\mathcal{L}(\{ x_i \}, \{ \dot{x}_i \}, t)$, which depends on generalized positions $x_i$, generalized velocities $\dot{x}_i = dx_i/dt$ and time $t$, the canonical momenta are defined as $p_i = \partial \mathcal{L} / \partial \dot{x}_i$ and the corresponding Hamiltonian
\begin{equation}\label{eq_Legendre_transformation}
	H \left( \{ x_i \}, \{ p_i \}, t \right) = 
	\sum_i p_i \dot{x}_i - \mathcal{L}(\{ x_i \}, \{ \dot{x}_i \}, t)
\end{equation}
must be expressed as a function of $x_i$, $p_i$ and $t$. In this process, the correct application of the Legendre transformation requires that every generalized velocity can be expressed as function of the generalized positions, the conjugate momenta and the time, i.e.,
\begin{equation}\label{eq_generalized_velocities}
	\dot{x}_i \equiv \dot{x}_i \left( \{ x_j \}, \{ p_j \}, t \right).
\end{equation}
If it is not possible to obtain such a functional dependence for each generalized velocity, the Lagrangian is said to be \textit{singular}, and a Legendre transformation is not well-defined and thus not applicable. The terminology arises from the observation that for many physical systems, the Lagrangian contains a kinetic part that is quadratic in the generalized velocities, and solving for the velocities as in Eq.~\eqref{eq_generalized_velocities} corresponds to the inversion of a quadratic coupling matrix, which is not possible if this matrix is singular \footnote{Typically, the derivation of a Hamiltonian describing a superconducting circuit involves the inversion of a capacitance matrix, which conventional textbooks usually assume to be accomplishable.}.

Accordingly, we refer to a superconducting circuit as singular if it is described by a singular Lagrangian. A singular Lagrangian implies that the physical system that is described has some underlying constraints \cite{Dirac1, Dirac2, Dirac3, Bergmann1, Bergmann2, Bergmann3, HenneauxBook, RotheBook, Brown} and that the canonical variables $x_i$ and $p_j$ within the Hamiltonian description are not independent as assumed for the application of the variational principle.
In particular, the classical phase-space variables are no longer necessarily canonical as the constraints restrict the dynamics to a subspace of the entire phase space. We stress that it might depend on the level of details of the system's description and on the choice of variables whether the corresponding Lagrangian is singular or not, as will be seen in examples below.

As elaborated in Refs.~\cite{Dirac1, Dirac2, Dirac3, Bergmann1, Bergmann2, Bergmann3, HenneauxBook, RotheBook, Brown}, a possible strategy to derive a quantized theory on a Hamiltonian level, starting from a singular Lagrangian, is accomplished by determining and classifying the system's underlying constraints and involves a subsequent reduction of the number of variables, remaining with independent variables only. This, however, is accompanied by a redefinition of the conventional Poisson brackets -- defining the Dirac brackets -- and hence of the commutator in quantum mechanics as well. In general, this approach, which is known as the Dirac-Bergmann algorithm, can be rather involved, even for seemingly simple systems \footnote{We note that an alternatively proposed procedure -- the Faddeev-Jackiw method \cite{RotheBook, Faddeev_Jackiw}, which is equivalent to the Dirac-Bergmann algorithm \cite{Garcia, Non_Equivalence} -- promises a simplified approach as no classification of constraints is required. In this paper, however, we work with the Dirac-Bergmann algorithm as its application turns out to be straightforward for our systems.}. 

However, as pointed out by Dirac \cite{Dirac3}, the arguably simplest class of singular Lagrangians is the one in which one generalized momentum vanishes, say $p_1=0$, while the corresponding generalized position can be expressed as a function of all the other canonical variables, i.e., $x_1 \equiv x_1 ( \{ x_i \}_{i \neq 1}, \{ q_i \}_{i \neq 1}, t)$. In this case, $x_1$ can be substituted in the Hamiltonian such that this degree of freedom can be discarded.

In the remainder of the paper, we focus on this simple class of singular Lagrangians in the setting of circuit quantization. In this context, the generalized positions are usually taken to be the magnetic fluxes associated to the nodes of the circuit,
\begin{equation}
    \phi_i = \int_{t_0}^{t} dt' V_i(t'),
\end{equation}
where $V_i(t')$ is the voltage of the $i^{\rm{th}}$ node with respect to ground \footnote{Note that different approaches of circuit quantization might also involve loop charges or mixtures of both sets of variables; see Ref.~\cite{Ulrich}. Again, we stress that the singularity of a system might depend on the chosen set of variables describing the system.}. For singular superconducting circuits, which are described by a Lagrangian that gives rise to vanishing generalized momenta, a full algebraic application of the Dirac-Bergmann algorithm, leading to a circuit Hamiltonian, is provided in Appendix~\ref{sec_general_formalism}. But note that, as detailed later in this paper, we find this Dirac-Bergmann Hamiltonian to be an incorrect description of the circuit dynamics in many cases.

For completeness, the presented general formalism includes the systematic description of \textit{nonreciprocal} superconducting circuits, i.e., circuits with broken time-reversal symmetry. As we discuss in Sec.~\ref{sec_Nonreciprocal_Circuits}, these circuits naturally constitute a large class of singular circuits, and, with only a few exceptions (e.g., Refs.~ \cite{Rymarz_Master, Rymarz, Parra1, Parra2, Parrax, Parra4}), nonreciprocal circuits are not covered in the conventional literature on circuit quantization. 

In order to get familiar with the Dirac-Bergmann algorithm, and to indicate its limitations when applied to superconducting circuits, we analyze two exemplary electrical networks that give rise to singular Lagrangians.

\begin{figure}[b!]  
	\centering
	\subfloat{\parbox{0.3\linewidth}{a) \hspace{30pt} \hfill \null \\[-3ex] \includegraphics[]{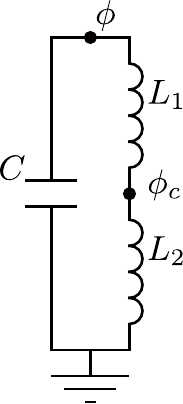}\label{fig_example_singular_loop_a}}}
	\hfill
	\subfloat{\parbox{0.3\linewidth}{b) \hspace{30pt} \hfill \null \\[-3ex] \includegraphics[]{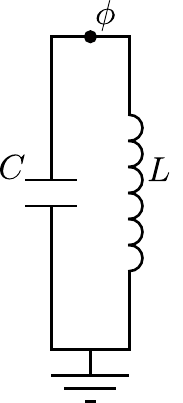}\label{fig_example_singular_loop_b}}}
	\hfill
	\subfloat{\parbox{0.3\linewidth}{c) \hspace{30pt} \hfill \null \\[-3ex] \includegraphics[]{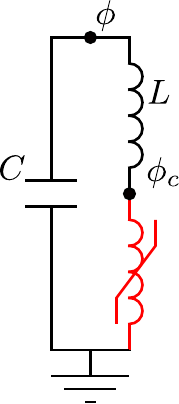}}\label{fig_example_singular_loop_c}}
	\caption{
	a) Series combination of two linear inductances $L_1$ and $L_2$ with a shunting capacitance $C$, and b) the effective equivalent LC circuit obtained by eliminating the constrained variable and adding the inductances, $L = L_1 + L_2$. c) Series combination of a linear inductance $L$ and a nonlinear inductor (red) with a shunting capacitance $C$.}
\end{figure}

\subsection{Addition of Linear Inductances in Series}\label{subsec_addition_lin_L}
First, we consider an apparently ``trivial" example, a series combination of two linear inductances $L_1$ and $L_2$ that is shunted by a capacitance $C$; see Fig.~\ref{fig_example_singular_loop_a}. The two-dimensional Lagrangian of this electrical network, 
\begin{equation}\label{eq_example_singular_loop_Lagrangian}
	\mathcal{L} = \frac{C\dot{\phi}^2}{2} - \frac{(\phi-\phi_c)^2}{2L_1} - \frac{\phi_c^2}{2L_2},
\end{equation}
is singular because one cannot solve for the generalized velocity $\dot{\phi}_c$ as function of the generalized positions and momenta. However, since $\dot{\phi}_c$ does not appear in the Lagrangian, we find that the corresponding generalized conjugate momentum vanishes, i.e., $Q_c = \partial \mathcal{L} / \partial \dot{\phi}_c = 0$. Exploiting the classical Euler-Lagrange equation of motion for the $\phi_c$-degree of freedom,
\begin{equation}
	0 = \frac{d}{dt} \left( \frac{\partial \mathcal{L}}{\partial \dot{\phi}_c} \right) 
	-\frac{\partial \mathcal{L}}{\partial \phi_c},
\end{equation}
we find the holonomic constraint
\begin{equation}\label{eq_holonomic_constraint}
	\phi_c = \frac{L_2}{L_1 + L_2} \phi,
\end{equation}
which essentially is Kirchhoff's current conservation law at the node $\phi_c$. Inserting this expression in the Lagrangian in Eq.~\eqref{eq_example_singular_loop_Lagrangian} renders it one-dimensional and regular,
\begin{equation}
	\mathcal{L} = \frac{C\dot{\phi}^2}{2} - \frac{\phi^2}{2L},
\end{equation}
with the total inductance $L = L_1 + L_2$. Thus, the elimination of the constrained variable $\phi_c$ reproduces what one would expect, the addition of two inductances in a series connection. This shows that the circuit in Fig.~\ref{fig_example_singular_loop_a} is effectively equivalent to an ordinary LC resonator; see Fig.~\ref{fig_example_singular_loop_b}. Finally, defining the conjugate charge $Q=\partial \mathcal{L} / \partial \dot{\phi}=C\dot{\phi}$, the Legendre transformation is applicable and results in the harmonic-oscillator Hamiltonian,
\begin{equation}
	H = \frac{Q^2}{2C} + \frac{\phi^2}{2L},
\end{equation}
which is quantized by imposing the canonical commutation relation $[\phi,Q]=i\hbar$.

This analysis demonstrates the application of the Dirac-Bergmann algorithm for a simple linear system, and the resulting total inductance $L$ agrees with the well-known series-combination formula. If, however, the system is not linear, the Dirac-Bergmann algorithm will possibly result in a bizarre description of the dynamics. In the following, we highlight emerging inconsistencies in the Dirac-Bergmann algorithm by replacing one of the linear inductances with a nonlinear inductor -- specifically, a Josephson junction. 

\subsection{Addition of a Linear and a Nonlinear Inductor in Series}\label{sec_branched_Hamiltonian}
In the previous subsection, we considered a system with a constraint in the form of a one-to-one functional dependence between variables; see Eq.~\eqref{eq_holonomic_constraint}. However, the effective description of singular electrical networks might involve constraints of a different type as well. In this subsection, we demonstrate the possible emergence of multi-valued constraints. In particular, we consider a series combination of a linear inductance $L$ and a generic nonlinear inductor that is shunted by a capacitance $C$; see Fig.~\ref{fig_example_singular_loop_c}.

\begin{figure*}[t!]  
	\centering
	\subfloat{\parbox[t]{0.2\linewidth}{a) \hspace{20pt} \hfill \null \\[-3ex] \includegraphics[]{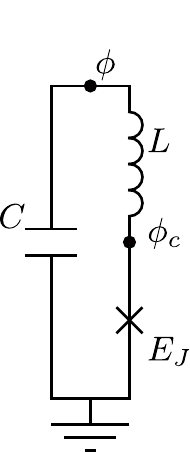}\label{fig_multivalued_a}}}
	\hfill
	\subfloat{\parbox[t]{0.39\linewidth}{b) \hspace{10pt} \hfill \null \\[-3ex] \includegraphics[width=0.39\textwidth]{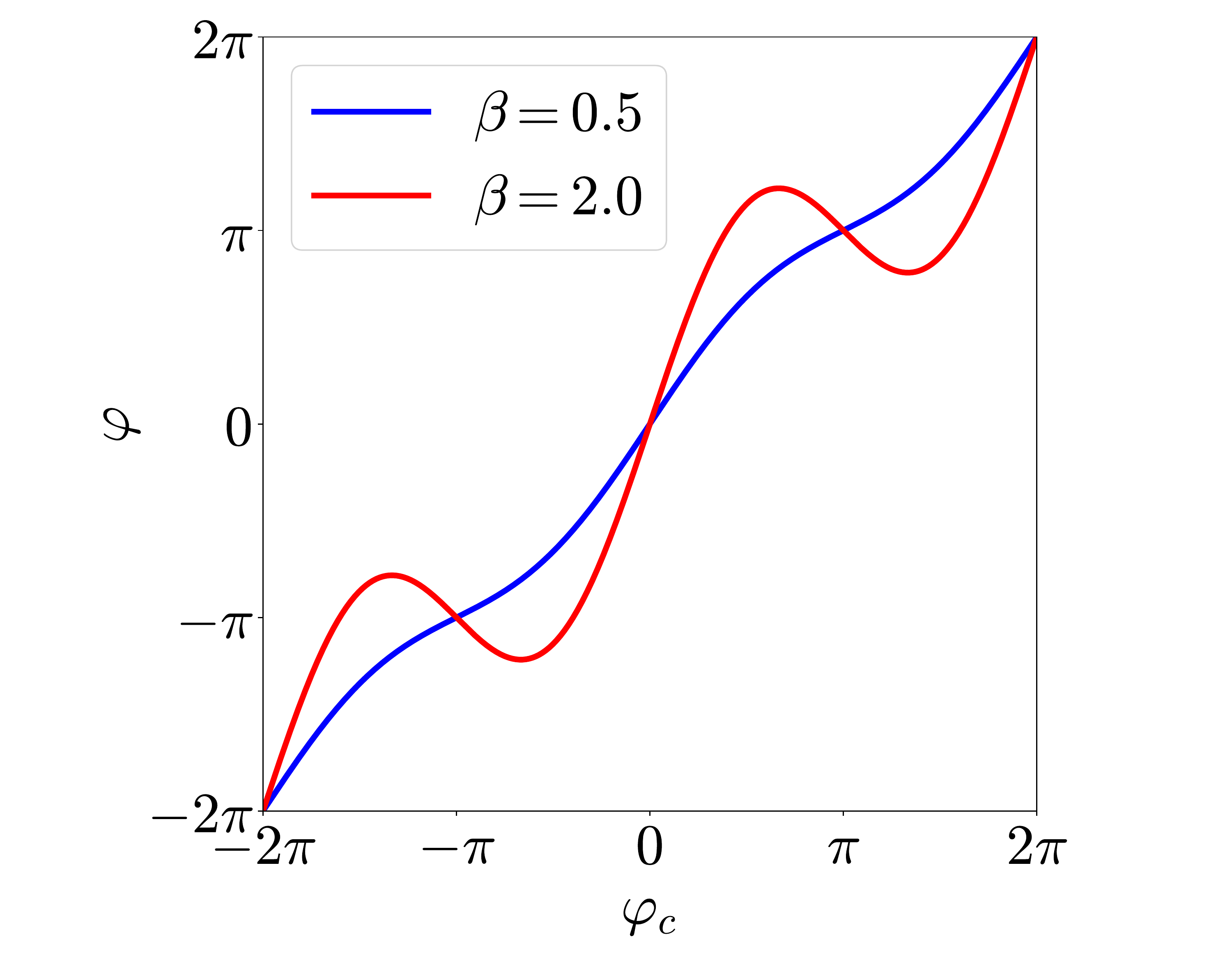}\label{fig_multivalued_b}}}
	\hfill
	\subfloat{\parbox[t]{0.39\linewidth}{c)  \hspace{50pt} \hfill \null \\[-3ex] \includegraphics[width=0.39\textwidth]{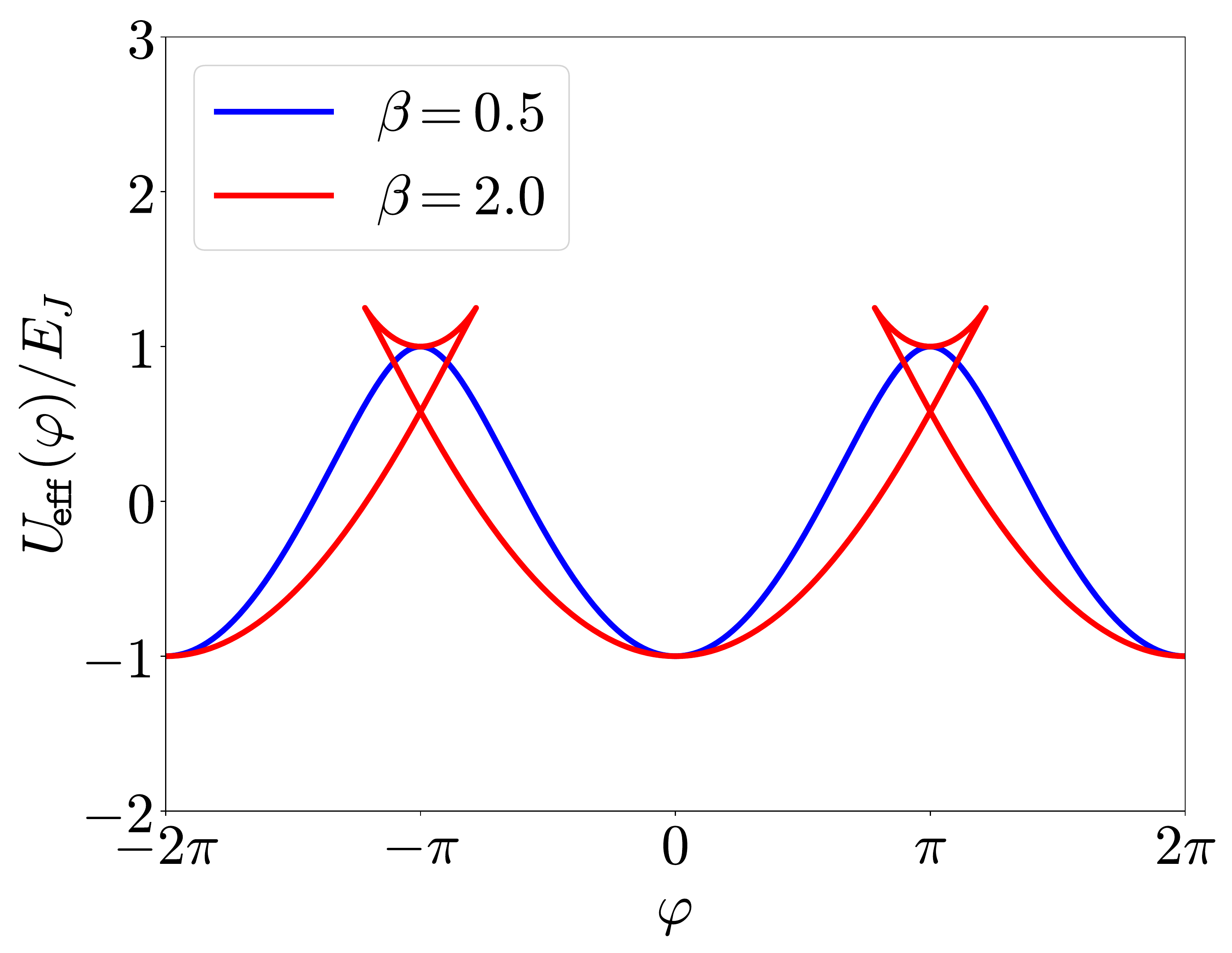}}\label{fig_multivalued_c}}
	\caption{
	Series combination of a linear inductance $L$ and a Josephson junction $E_J$ with a shunting capacitance $C$.
	a) Circuit model.
	b) Constraint relating the rescaled flux variables $\varphi$ and $\varphi_c$ [see Eq.~\eqref{eq_multivalued_constraint_1}] for the different cases $\beta \leq 1$ (blue) and $\beta>1$ (red), in which the constrained variable $\varphi_c(\varphi)$ is either a well-defined function of $\varphi$ or multi-valued, respectively.
	c) Effective one-dimensional potential [cf. Eq.~\eqref{eq_effective_1d_potential}] obtained by substituting the constrained variable for the different cases $\beta \leq 1$ (blue) and $\beta>1$ (red), in which $U_\text{eff}(\varphi)$ is either a well-defined function of $\varphi$ or multi-valued, respectively.}
	\label{fig_multivalued}
\end{figure*}

The Lagrangian of the electrical network, 
\begin{equation}\label{eq_Lagrangian_multivalued_potential}
	\mathcal{L} = \frac{C\dot{\phi}^2}{2} - \frac{(\phi-\phi_c)^2}{2L} - U_{nl}(\phi_c),
\end{equation}
in which $U_{nl}(\phi_c)$ describes the nonlinear inductor, is singular because it does not contain the generalized velocity $\dot{\phi}_c$, and, as a consequence, the associated generalized momentum vanishes. Following the scheme that we present in Appendix~\ref{sec_general_formalism}, the Dirac-Bergmann algorithm effectively reduces to an evaluation of the classical Euler-Lagrange equation of motion, i.e., Kirchhoff's law of current conservation, for the $\phi_c$-degree of freedom in order to eliminate it. Thus, setting the current through the linear inductor equal to that through the nonlinear one, we find the constraint
\begin{equation}\label{eq_lin_nonlin_constraint}
    \phi = \phi_c + L U'_{nl}(\phi_c),
\end{equation}
which must be inverted in order to obtain the functional dependence $\phi_c(\phi)$. Then, after eliminating the constrained variable $\phi_c$ in the Lagrangian, the series combination of both the inductors can be replaced by an effective inductor that is described by the effective potential
\begin{equation}\label{eq_effective_1d_potential}
	U_\text{eff}(\phi) = 
	\frac{\big[ \phi-\phi_c(\phi) \big]^2 }{2L} 
	+ U_{nl}[\phi_c(\phi)].
\end{equation}
By construction, the resulting Hamiltonian of the initially singular system,
\begin{equation}\label{eq_Hamiltonian_s}
	H_s = \frac{Q^2}{2C} + U_\text{eff}(\phi),
\end{equation}
depends on one pair of conjugate variables only. However, as we show in the following, both the classical Hamiltonian description of the system as well as its quantization is not always straightforward.

To this end, we specify the nonlinear inductor as a Josephson junction with Josephson energy $E_J$, i.e., we set $U_{nl}(\phi_c) = - E_J \cos(2\pi \phi_c/\Phi_0)$. The corresponding total circuit is shown in Fig.~\ref{fig_multivalued_a}. After introducing the rescaled phase variables $\varphi=2\pi\phi/\Phi_0$, $\varphi_c=2\pi\phi_c/\Phi_0$ and the screening parameter $\beta = L E_J \left( 2\pi/\Phi_0 \right)^2$ \cite{Squid_handbook}, Eq.~\eqref{eq_lin_nonlin_constraint} reduces to Kepler's transcendental equation \cite{Minev}
\begin{equation}\label{eq_multivalued_constraint_1}
	\varphi = \varphi_c + \beta \sin \left( \varphi_c \right),
\end{equation}
which can be inverted numerically in order to solve for the constrained variable $\varphi_c \equiv \varphi_c(\varphi)$. Note that for $\beta \leq 1$, the right-hand side of Eq.~\eqref{eq_multivalued_constraint_1} is strictly monotonically increasing as $\varphi_c$ increases. However, for $\beta > 1$, it can be separated into infinitely many regimes in which it is either monotonically increasing or decreasing, respectively; see Fig.~\ref{fig_multivalued_b}.

As a result, for $\beta \leq 1$, the constrained variable $\varphi_c$ is a well-defined single-valued function of $\varphi$, whereas it is multi-valued for $\beta > 1$. In the latter case, for a given value of $\varphi$, there might exist several values of $\varphi_c$ satisfying the constraint in Eq.~\eqref{eq_multivalued_constraint_1}. Consequently, while the effective potential $U_\text{eff}(\varphi)$ [cf. Eq.~\eqref{eq_effective_1d_potential}] can be single-valued, it can also be multi-valued, depending on the value of $\beta$ \citep{Richer2, Rymarz_Master}; see Fig.~\ref{fig_multivalued_c}.

In the single-valued case ($\beta \leq 1$), the Hamiltonian $H_s$ in Eq.~\eqref{eq_Hamiltonian_s} is a mathematically well-defined function of a pair of two conjugate variables, and it can be used in the usual way to describe the dynamics of the system \cite{Richer2, Rymarz_Master}. In particular, a quantized description is obtained by promoting the canonical variables to operators and imposing the canonical commutation relation $[\phi, Q] = i\hbar$.

In contrast, in the multi-valued case ($\beta > 1$), the alternative might be to describe the system by a so-called \textit{branched} Hamiltonian \cite{Henneaux, Wilczek_branched}; but both the classical as well as the quantum description become subtle. Branched Hamiltonians emerge in various other contexts outside of electrical network theory, e.g., in extensions of Einstein's theory of gravitation \cite{Teitelboim} or in effective models of systems with finite response times \cite{Wilczek_classical}. All branched Hamiltonians have in common that the system is not uniquely described by its phase-space coordinates; one requires further information to determine the state of the system. As a consequence, the classical motion of the system might not be predictable for a given set of initial variables \cite{Henneaux}.

Although most branched Hamiltonians in the literature are multi-valued in the generalized momentum \footnote{The multi-valuedness of a branched Hamiltonian in the momentum might originate from a non-convex velocity dependence of the corresponding Lagrangian.}, the multi-valuedness can be transferred to the generalized position by a canonical (Fourier) transformation \cite{Wilczek_branched}.
Classically, such systems can give raise to a non-continuous velocity as a function of time \cite{Henneaux, Wilczek_classical}, or can even exhibit a non-vanishing velocity in the ground state, and therefore breaking time reversal symmetry as well as time translation symmetry \cite{Wilczek_classical, Zhao}. The analysis of such systems comprises the closely related concept of time crystals \citep{Wilczek_classical, Wilczek_quantum, Bruno, Watanabe}.

Generally, it is advisable to avoid the appearance of branched Hamiltonians, if possible. For example, a proposal for the classical treatment of such systems is to manually introduce further coordinates, and to start with a different, higher-dimensional but singular Lagrangian such that the enlarged, constrained phase space contains the phase space of the original system as an unconstrained subspace \cite{Zhao}. This approach comes along with the drawback of replacing the conventional Poisson brackets by Dirac brackets, which reveal that the set of chosen phase-space variables is not necessarily canonical. In particular, such variables are not Darboux coordinates, and the phase space does not necessarily provide a symplectic manifold anymore \cite{Zhao}. Similarly, one can avoid the description of the circuit in Fig.~\ref{fig_multivalued_a} in terms of a branched Hamiltonian by choosing  non-canonical phase-space variables; see Appendix~\ref{app_multivalued}.

A possible quantization of a general branched Hamiltonian involves the definition of an effective Hamiltonian as a topological combination of the various branch Hamiltonians \cite{Henneaux}. This approach is exclusively applicable for the quantum description as the effective Hamiltonian is derived via the path integral formalism and classically no unique history of the system is necessarily singled out \cite{Henneaux}. Alternatively, another possibility to quantize a branched Hamiltonian involves the evaluation of wave functions on the individual branches of the Hamiltonian, which are linked via appropriate boundary conditions \cite{Wilczek_branched}. This procedure effectively defines a total wave function over an expanded space. Note that such an expansion is required because neither the position, nor the momentum, provides a complete set of commuting observables \cite{Wilczek_classical, Wilczek_branched}.

The aim of the present work, however, is not to provide the general description of systems that potentially involve branched Hamiltonians. Instead, focusing on the quantized description of electrical networks, we note that from the point of view of electrostatics, nonzero (``parasitic") capacitances occur between every node of a physical network \footnote{Working with flux variables as generalized positions, the parasitic capacitance of an inductive element adds a term to the Lagrangian that is quadratic in the generalized velocity. This is basically equivalent to adding a mass to an otherwise massless particle, which, of course, is not a correct step in all physical theories.}, e.g., those of Josephson junctions, which, in practical realizations, always exist \cite{BKD}. Thus, a more physical description of the system renders the Lagrangian regular, and, within this approach, the physical origin and interpretation of the multi-valuedness becomes clear as the individual branches of the Hamiltonian correspond to classical (meta-)stable points. {\em But the limit of small but finite capacitances throughout the network reveals a qualitative mismatch between the effective dynamics of the system and that obtained from the Dirac-Bergmann algorithm applied to the singular counterpart} \cite{Rymarz_Master}.

\section{Failure of the Dirac-Bergmann Algorithm}\label{sec_failure_Dirac}
In the previous section, we applied the Dirac-Bergmann algorithm, given in detail in Appendix~\ref{sec_general_formalism}, to derive the Hamiltonian description of two simple superconducting circuits. For singular circuits with nonlinearities, however, the system's quantum dynamics resulting from this approach differs from a more appropriate treatment in which the singularities are lifted. In electrical networks, the singularity of the capacitance matrix is lifted by taking into account the small but finite intrinsic (or parasitic) capacitance of one or several network elements in the corresponding branch of the circuit. In this section, we determine in detail the discrepancy mentioned above between the singular and the regular approach, and we classify different types of nonlinearities. Our results justify the conclusion that one should {\em not} use Kirchhoff's current law to eliminate variables in the Lagrangian.

\begin{figure}[b!]  
	\centering
	\includegraphics[]{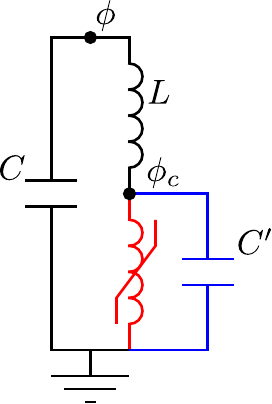}
	\caption{Series combination of a linear inductance $L$ and a nonlinear inductor (red) in parallel to a shunting capacitance $C$. The blue branch highlights the intrinsic capacitance $C'$ of the nonlinear inductor, which we consider to be either vanishingly small (regular case) or absent (singular case).}
	\label{fig_Non_Linear_Inductance}
\end{figure}

To provide a simple example that demonstrates the failure of the Dirac-Bergmann algorithm when applied to electrical networks, we consider the series combination of a linear inductance $L$ and a generic nonlinear inductor with intrinsic capacitance $C'$, all in parallel with a total shunting capacitance $C$; see Fig.~\ref{fig_Non_Linear_Inductance}.

In the following, we analyze and compare the two cases: 1) absent intrinsic capacitance, $C'=0$, indicating the application of the Dirac-Bergmann algorithm; 2) extremely small but nonzero intrinsic capacitance, $C'>0$. In the first case the Lagrangian of the circuit,
\begin{equation}\label{eq_Lagrangian_non_linear_inductance}
	\mathcal{L} = \frac{C\dot{\phi}^2}{2} + \frac{C'\dot{\phi}_c^2}{2} - \frac{(\phi-\phi_c)^2}{2L} - U_{nl}(\phi_c),
\end{equation}
is singular, while in the second case it is regular. In particular, we allow the nonlinear inductor to be a generic flux-controlled inductor \cite{Peikari, DesoerKuh1, DesoerKuh2} that can be modeled via the potential $U_{nl}(\phi_c)$, which we do not further specify at this point.

The treatment of the singular case is already presented in Sec.~\ref{sec_branched_Hamiltonian}. There, we also discussed the potential ambiguities in the construction of the effective potential $U_\text{eff}(\phi)$ in Eq.~\eqref{eq_effective_1d_potential}. In the remainder of this work, the effective potential of the singular system will serve for a comparison with the limiting behavior of the regular case, which we analyze next.

\subsection{Approaching the Singular Limit -- Born-Oppenheimer Analysis}\label{sec_BO_approx_introduction}
The consideration of nonzero but finite values of $C'$ (see Fig.~\ref{fig_Non_Linear_Inductance}) is motivated by the observation that any physical realization of a network element contains some residual intrinsic or stray capacitance. For $C'>0$, the Lagrangian in Eq.~\eqref{eq_Lagrangian_non_linear_inductance} describing the circuit shown in Fig.~\ref{fig_Non_Linear_Inductance} is regular, and the Hamiltonian is straightforwardly obtained via an ordinary Legendre transformation, resulting in
\begin{equation}\label{eq_Hamiltonian_r}
	H_r = \frac{Q^2}{2C} + \frac{Q_c^2}{2C'} + \frac{(\phi-\phi_c)^2}{2L} + U_{nl}(\phi_c).
\end{equation}
Here, $\phi, Q$ and $\phi_c, Q_c$ denote two independent pairs of conjugate variables, and $H_r$ is quantized by imposing the canonical commutation relations $[\phi_{(c)}, Q_{(c)}] = i \hbar$.

In what follows, we compare the Hamiltonian of the regular circuit ($H_r$ for $C'>0$) with that of the singular one ($H_s$ for $C'=0$), and thus, we consider the limit of vanishingly small but finite $C'$ in the regular system. We immediately note that $H_r$ is two-dimensional, whereas $H_s$ describes the dynamics of one effective degree of freedom only. The fact that models with different numbers of dynamical variables could describe the same system can be understood by the observation that for $C'/C \ll 1$ the time scales on which the dynamics of $\phi$ and $\phi_c$ change, as mediated by $H_r$, are vastly different.

In light of this, the Born-Oppenheimer approximation \cite{Born_Oppenheimer,PhysRevB.74.014514} will allow us to derive an effective low-energy Hamiltonian as a function of $\phi$ and $Q$ only. To this end, we first solve the stationary Schrödinger equation associated with the fast degree of freedom, $\phi_c$, for fixed values of $\phi$ and $Q$. Thus, we identify the fast part \footnote{In the molecular physics literature, $H_\text{fast}$ is called the electronic Hamiltonian as it describes the motion of the fast (light) electrons for fixed positions of the slow (heavy) nuclei.} of $H_r$ as
\begin{equation}\label{eq_def_H_fast}
    H_\text{fast} = \frac{Q_c^2}{2C'} + \frac{(\phi-\phi_c)^2}{2L} + U_{nl}(\phi_c),
\end{equation}
and we solve
\begin{equation}\label{eq_fast_SE_BO}
	H_\text{fast} \psi_{\phi,n}(\phi_c) = E_{\phi,n} \psi_{\phi,n}(\phi_c)
\end{equation}
for the eigenstates $\psi_{\phi,n}(\phi_c)$ and the associated eigenenergies $E_{\phi,n}$, which both are labeled by $n \in \mathbb{N}_0$ and parametrized by $\phi$. The ground state energy ($n=0$) is then considered as an effective low-energy potential for the slow variable, $\phi$, whose dynamics is captured by the effective Hamiltonian
\begin{equation}\label{eq_H_r_eff}
	H_{r,\text{eff}} = \frac{Q^2}{2C} + U_\text{BO}(\phi),
\end{equation}
with the Born-Oppenheimer potential that we define as
\begin{equation}\label{eq_def_BO_potential}
    U_\text{BO}(\phi) = E_{\phi,0} - E_{0,0}.
\end{equation}
Here, we have chosen the energy offset of $U_\text{BO}(\phi)$ such that $U_\text{BO}(0) = 0$ in order to avoid divergent additive constants.

In summary, the Born-Oppenheimer approximation provides an effective Hamiltonian $H_{r,\text{eff}}$ for the regular case ($C'>0$), which is suitable for a comparison with $H_s$ that is obtained in the singular case ($C'=0$). Note that the Born-Oppenheimer approximation becomes more accurate the smaller the ratio $C'/C$, which is exactly the regime of interest for the aforementioned comparison.

\subsection{Types of Network Branches Leading to the Failure of the Dirac-Bergmann Algorithm}\label{sec_types_of_inductors}
In the following, we evaluate the Born-Oppenheimer potential for a generic nonlinear inductor. Unless stated otherwise, we generally restrict our considerations to potentials $U_{nl}(\phi_c)$ that are symmetric in $\phi_c$, i.e., $U_{nl}(\phi_c)=U_{nl}(-\phi_c)$, and that do not diverge for $|\phi_c| < \infty$. Furthermore, we assume that the nonlinear inductor can be categorized into one of the following three types, depending on the behavior of its potential for large values of $\phi_c$ \footnote{These definitions are structured to exclude from consideration potential functions that deviate very slowly from a linear inductance, e.g., $U_{nl}(\phi_c) \sim \phi_c^2/\log{(\phi_c^2+4)}$. We suspect that this potential behaves like a type-1 case, but the proofs of our theorems do not apply. Note that such functions are also excluded by the second part of the definition of type-L inductors.}:
\newcounter{fnnumber}   
\begin{itemize}
	\item type~1 (sublinear \footnote{Sublinear and superlinear refer to the form of the classical two-terminal current-versus-flux characteristic. When this relation is linear, the coefficient of proportionality is an inverse inductance, referred to as a reluctance in the electrical literature.}\setcounter{fnnumber}{\thefootnote}): \\
	$\exists\,\gamma\in(0,2): \lim_{\phi_c \to \pm \infty} U_{nl}(\phi_c)/\phi_c^{\gamma} = 0 $
    \begin{enumerate}
		\item[] (a) $\gamma \in (0, 2)$, $U_{nl}(\phi_c) = U_{nl}(-\phi_c)$
		\item[] (b) $\gamma \in (0, 1)$, $U_{nl}(\phi_c) \neq U_{nl}(-\phi_c)$
	\end{enumerate}
	
	\item type~2 (superlinear \footnotemark[\thefnnumber]): \\
	$\lim_{\phi_c \to \pm \infty} \phi_c^2/U_{nl}(\phi_c) = 0$
	
	\item type~L (linear): \\
	$\exists\,\mathfrak{L}>0: \lim_{\phi_c \to \pm \infty} U_{nl}(\phi_c)/\phi_c^2=1/{2 \mathfrak{L}}$ 
	\quad and  \\
	$\exists\,\gamma\in(0,2): \lim_{\phi_c \to \pm \infty} [U_{nl}(\phi_c) - \phi_c^2/2\mathfrak{L}] /\phi_c^{\gamma} = 0$
\end{itemize}

Note that not all possible nonlinear inductors can be classified into one of the three types we provide. In Sec.~\ref{subsec_pathological_potential}, we discuss a nonlinear inductor whose potential does not have a well-defined leading term for large values of $\phi_c$. In Sec.~\ref{subsec_asymmetric_potential}, we provide an example for a nonlinear inductor with an asymmetric potential that is not of type~1(b). \\

As we show in the following, in the limit $C'/C \rightarrow 0$, the dynamics of the regular circuit shown in Fig.~\ref{fig_Non_Linear_Inductance} strongly depends on which type of nonlinear inductor is considered. In particular, we prove the validity of a perturbative treatment in which, depending on the type of the nonlinear inductor in the circuit in Fig.~\ref{fig_Non_Linear_Inductance}, either the potential of the linear inductor or that of the nonlinear one can be identified as the perturbation to the rest of the Hamiltonian. Finally, for each type of nonlinear inductor, we provide expressions for $U_\text{BO}(\phi)$ in the limit of $C'/C \rightarrow 0$, and we associate an effective inductor with the Born-Oppenheimer potential in that limit. 

\subsubsection{Effective Potential for Sublinear Inductors (Type~1)}\label{subsec_type_1}
We start with the analysis of nonlinear inductors of type~1. Before analyzing the behaviour of the Born-Oppenheimer potential in the limit of a small intrinsic capacitance $C'$, we provide two helpful lemmas. \\

\textbf{Lemma~1.} \quad
\textit{Suppose that a nonlinear inductor of type~1 is described by the potential $U_{nl}(\phi_c)$. Then, for some $\gamma \in (0,2)$ and all $\phi_c$}
\begin{equation*}
    \forall \, \, \beta > 0 \, \, \exists \, \, M > 0 : \quad
    |U_{nl}(\phi_c)| \leq \beta |\phi_c|^\gamma +M.
\end{equation*}

\begin{proof}
Fix $\beta > 0$ for the remainder of the proof. From the definition of a nonlinear inductor of type~1, it follows that there exists some $\gamma \in (0,2)$ such that $\lim_{|\phi_c| \to \infty} |U_{nl}(\phi_c)| / \beta |\phi_c|^{\gamma} = 0$. Thus, there exists $\widetilde{\phi}_c \in \mathbb{R}^+$ such that $|U_{nl}(\phi_c)| \leq \beta |\phi_c|^{\gamma}$ for all $|\phi_c| \geq \widetilde{\phi}_c$. With $M = \max_{\phi_c \in [-\widetilde{\phi}_c, \widetilde{\phi}_c ] } |U_{nl}(\phi_c)|$, the inequality $|U_{nl}(\phi_c)| \leq \beta |\phi_c|^\gamma +M$ holds for all $\phi_c \in \mathbb{R}$.
\end{proof}

In the next lemma, we introduce a new dynamical variable $y$, in anticipation of the rescaling that will be done in the proof of the upcoming theorem: \\

\textbf{Lemma~2.} \quad
\textit{Suppose that two Hamiltonians $H_1$ and $H_2$ satisfy $H_2 = H_1 + \delta V(y)$ with $\delta V(y) \geq 0$ for all $y$. Then, the ground state energies of $H_1$ and $H_2$ satisfy $E_0(H_2) \geq E_0(H_1)$.}

\begin{proof}
Let $\ket{\psi}$ be the normalized ground state of $H_2$. Then, with the variational method applied to $H_1$, one obtains for the ground state energy of $H_2$:
\begin{equation}
\begin{split}
    E_0 (H_2)
    &= \bra{\psi} H_2 \ket{\psi}
    = \bra{\psi} H_1 \ket{\psi} + \bra{\psi} \delta V(y) \ket{\psi}     \\
    &\geq E_0 (H_1).
\end{split}
\end{equation}
Thus, the ground state energy of $H_2$ is lower-bounded by the ground state energy of $H_1$.
\end{proof}

It follows that the ground state energy of a particle in a potential $V_2(y)$, with $y$ being the position variable, is always larger than or equal to the ground state energy of the same particle in a potential $V_1(y)$ if $V_2(y) \geq V_1(y)$ for all values of $y$. Furthermore, suppose that a third Hamiltonian $H_3$ satisfies $H_3 = H_2 + \Delta V (y)$ with $\Delta V (y) \geq 0$. Then, using same reasoning, we obtain the useful ``sandwich" $E_0 (H_3) \geq E_0 (H_2) \geq E_0(H_1)$.

In the following Theorem, we show that for a nonlinear inductor of type~1, the Born-Oppenheimer potential {\em vanishes} as $C'/C \rightarrow 0$; the nonlinear branch is replaced by an {\em open circuit}. \\

\textbf{Theorem~1.} \quad
\textit{Consider $H_\text{fast}$ as defined in Eq.~\eqref{eq_def_H_fast} with $U_{nl}(\phi_c)$ describing a nonlinear inductor of type~1(a). Then, $U_\text{BO}(\phi)$ as defined in Eq.~\eqref{eq_def_BO_potential} satisfies}
\begin{equation*}
	\forall \, \, \phi \in \mathbb{R} : \quad
	\lim_{C' \rightarrow 0} U_\text{BO}(\phi) = 0 .
\end{equation*}

The general strategy of the proof is as follows: while all the eigenvalues of $ H_\text{fast}$ in Eq.~\eqref{eq_def_H_fast} diverge like $1/\sqrt{C'}$ as $C'\rightarrow 0$, we note that if this diverging factor is scaled out, the Hamiltonian can be brought into the form of a standard harmonic oscillator plus an additional term that can be considered a perturbation for all potentials $U_{nl}(\phi_c)$ of type~(a). With the use of several auxiliary bounding Hamiltonians, we show that results from analytic perturbation theory can be used (despite the fact that $U_{nl}(\phi_c)$ may not be analytic in $\phi_c$) to show that the resulting Rayleigh-Schrödinger series is well behaved and absolutely convergent. Evaluation of the appropriate terms in this series gives the result of the theorem.

\begin{proof}
We introduce the $LC'$-resonator frequency and the flux zero point fluctuation, defined as
\begin{equation}\label{eq_defs_ZPF_and_omega_LC}
    \omega'_r = \frac{1}{\sqrt{LC'}},
    \qquad
    \Phi_\textit{ZPF}=\sqrt{\hbar}\sqrt[4]{\frac{L}{C'}},
\end{equation}
respectively, and we express $H_\text{fast}$ as
\begin{equation}\label{eq_H_fast_rewritten}
    H_\text{fast} =
    \hbar \omega'_r
    \left[
    \frac{p^2+(y-\phi/\Phi_\textit{ZPF})^2}{2} + \frac{U_{nl}(y \Phi_\textit{ZPF})}{\hbar \omega'_r}
    \right].
\end{equation}
The dimensionless conjugate variables $y$ and $p$ are defined as $y = \phi_c / \Phi_\textit{ZPF}$ and $p = Q_c \Phi_\textit{ZPF} / \hbar$, and they satisfy the canonical commutation relation $[y,p]=i$.

We define the parameter $\epsilon = 1 / \sqrt{\hbar\omega'_r} = \sqrt{L}/\Phi_\textit{ZPF}$, and we divide out the prefactor in Eq.~\eqref{eq_H_fast_rewritten}, obtaining
\begin{equation}\label{eq_eps2_H_fast_type_1}
    \epsilon^2 H_\text{fast} = H_0 + \frac{\epsilon^2\phi^2}{2L} - \frac{\epsilon \phi y}{\sqrt{L}} + \epsilon^2 U_{nl} \left( \sqrt{L}y / \epsilon \right),
\end{equation}
with the dimensionless harmonic-oscillator Hamiltonian $H_0 = (p^2+y^2)/2$. By definition of a type-1 inductor, there exists a parameter $\gamma \in (0, 2)$ such that
\begin{equation}\label{eq_type_1_condition}
    \lim_{\epsilon \rightarrow 0} \epsilon^\gamma U_{nl} \left( \sqrt{L}y / \epsilon \right) = 0
\end{equation}
for all values of $y$. With this property of $U_{nl}(\phi_c)$ in mind, we introduce the auxiliary Hamiltonian
\begin{equation}
    \epsilon^2 H_\text{aux} =
    H_0 
    + \frac{\epsilon^2\phi^2}{2L} 
    - \frac{\epsilon \phi y}{\sqrt{L}} 
    + \epsilon^{2-\gamma} \alpha^\gamma U_{nl} \left( \sqrt{L}y / \alpha \right),
\end{equation}
which generalizes $H_\text{fast}$ as it involves a new independent parameter $\alpha$. The fast Hamiltonian is recovered from the auxiliary one by setting $\alpha = \epsilon$ as $H_\text{aux} = H_\text{fast}$ in that case.

Without loss of generality, we choose $\gamma \in \mathbb{Q}$, and therefore we set $\gamma = p/q$ with $p,q \in \mathbb{N}$ satisfying $2q-p \in \mathbb{N}$. We substitute $\epsilon = \lambda^q$ in the auxiliary Hamiltonian to obtain
\begin{equation}\label{eq_for_Kato}
    \lambda^{2q} H_\text{aux} = 
    H_0 
    + \frac{\lambda^{2q} \phi^2}{2L} 
    - \frac{\lambda^q \phi y}{\sqrt{L}} 
    + \lambda^{2q-p} \alpha^\gamma U_{nl} \left( \sqrt{L}y / \alpha \right).
\end{equation}
The operator domain $D$ of $\lambda^{2q} H_\text{aux}$ is independent of $\lambda$, and for each $\ket{\psi} \in D$, $\lambda^{2q} H_\text{aux} \ket{\psi}$ is a vector-valued analytic function of $\lambda$. Thus, $\lambda^{2q} H_\text{aux}$ is an analytic family in the sense of Kato (in particular, an analytic family of type (A); see p.~16 in Ref.~\cite{ReedSimon}) and the Kato-Rellich theorem applies; see p.~15, Theorem~XII.8 in Ref.~\cite{ReedSimon}. It follows that for any values of $\phi$ and $\alpha>0$ the $n^{\rm{th}}$ eigenenergy $\lambda^{2q} E_{\phi, n}(\lambda, \alpha)$ of $\lambda^{2q} H_\text{aux}$ is an analytic function in $\lambda$ with a non-vanishing radius of convergence $\lambda_n(\phi, \alpha)$, i.e., for all $\lambda < \lambda_n (\phi, \alpha)$ we can write the eigenenergy as Rayleigh-Schrödinger series (p.~1 in Ref.~\cite{ReedSimon}):
\begin{equation}\label{eq_RS_series_type_1}
    \lambda^{2q} E_{\phi, n}(\lambda, \alpha) 
    = \sum_{k=0}^\infty E_{\phi, n}^{(k)}(\alpha) \lambda^k.
\end{equation}

By construction, and since the auxiliary Hamiltonian in Eq.~\eqref{eq_for_Kato} is a polynomial in $\phi$, the Rayleigh-Schrödinger coefficients $E_{\phi, n}^{(k)}(\alpha)$ are polynomials in $\phi$ of the order $j \leq k/q$. 

We have not yet established that the Rayleigh-Schrödinger coefficients are well behaved as $\alpha\rightarrow 0$. We now show this for the ground state:
consider the pair of new auxiliary Hamiltonians
\begin{equation}\label{eq_bounds_H_aux}
    \lambda^{2q} H^\pm                             
    = 
    H_0 
    + \frac{\lambda^{2q} \phi^2}{2L} 
    - \frac{\lambda^q \phi y}{\sqrt{L}}                 
    \pm \lambda^{2q-p}
    \left[
    \beta |\sqrt{L} y|^{\gamma} + \alpha^\gamma M
    \right],
\end{equation}
with $\beta, M > 0$. Since $\alpha^\gamma$ enters Eq.~\eqref{eq_bounds_H_aux} as the prefactor of the identity operator at the right-hand side, the eigenenergies $\lambda^{2q} E^\pm_{\phi, n}(\lambda, \alpha)$ of $\lambda^{2q} H^\pm$ depend linearly on $\alpha^\gamma$. According to Lemma~1, the parameters $\beta$ and $M$ can be chosen such that $|U_{nl}(\phi_c)| \leq \beta |\phi_c|^\gamma+M$, and therefore $\lambda^{2q} (H^+ - H_\text{aux}) \geq 0$, $\lambda^{2q} (H_\text{aux} - H^-) \geq 0$. Applying Lemma~2 twice gives
\begin{equation}
    \lambda^{2q} E^-_{\phi, 0}(\lambda, \alpha)     \leq 
    \lambda^{2q} E_{\phi, 0}(\lambda, \alpha)       \leq
    \lambda^{2q} E^+_{\phi, 0}(\lambda, \alpha).    
\end{equation}
Furthermore, it follows from Sturm-Liouville theory and its extensions (p.~719ff in Ref.~\cite{1953MF}) that $|\lambda^{2q} E^\pm_{\phi, 0}(\lambda, \alpha)| < \infty$. Thus, the ground state energy $\lambda^{2q} E_{\phi, 0}(\lambda, \alpha)$ remains finite for any $\alpha < \infty$, and, within the radius of convergence, the Rayleigh-Schrödinger coefficients satisfy $\lim_{\alpha \rightarrow 0} |E_{\phi, 0}^{(k)}(\alpha)| < \infty$; thus, these coefficients are guaranteed to be ``well behaved", including when $\alpha\rightarrow 0$.

Since $\lambda^{2q} H_\text{aux}$ is an analytic family of type (A), the radius of convergence of Eq.~\eqref{eq_RS_series_type_1}, $\lambda_n(\phi, \alpha)$, can be lower bounded (see p.~379, Remark 2.9 in Ref.~\cite{Kato}) by a function $r_n(\phi, \alpha) > 0$ that remains finite as $\alpha\rightarrow 0$ (in fact, it increases monotonically as $\alpha$ decreases; see Appendix~\ref{app_lower_bound_radius} for more details). Thus, there exists $\widetilde{\alpha}_0(\phi) > 0$ such that
\begin{equation}
    r_0\left(\phi, \widetilde{\alpha}_0(\phi) \right) 
    = \widetilde{\alpha}_0(\phi)^{1/q} \equiv \widetilde{\lambda}_0(\phi).
\end{equation}
It follows that the Rayleigh-Schrödinger series in Eq.~\eqref{eq_RS_series_type_1} converges at least if $\lambda < \widetilde{\lambda}_0(\phi)$ and $\alpha < \widetilde{\alpha}_0(\phi)$. Thus, for $\lambda < \widetilde{\lambda}_0(\phi)$, the ground state energy of $\epsilon^2 H_\text{fast}$ is given by Eq.~\eqref{eq_RS_series_type_1} with the substitution $\alpha = \lambda^q$.

Due to the symmetry of $U_{nl}(\phi_c)$ for nonlinear inductors of type~1(a), the eigenenergies $\lambda^{2q} E_{\phi, n}(\lambda, \alpha)$ are symmetric in $\phi$, i.e.,
\begin{equation}\label{even}
    E_{\phi, n}(\lambda, \alpha) = E_{-\phi, n}(\lambda, \alpha)
\end{equation}
for any values of $\lambda$ and $\alpha>0$. Thus, the Rayleigh-Schrödinger coefficients $E_{\phi, n}^{(k)}(\alpha)$ must contain only even powers of $\phi$. It follows that $E_{\phi, n}^{(k)}(\alpha) - E_{0, n}^{(k)}(\alpha) = 0$ for $k<2q$. Thus, the Born-Oppenheimer potential as defined in Eq.~\eqref{eq_def_BO_potential} can be expressed as
\begin{equation}\label{eq_U_BO_type_1_before_lim}
\begin{split}
    U_\text{BO}(\phi) 
    &= E_{\phi, 0}(\lambda, \alpha) - E_{0, 0}(\lambda, \alpha) \Big\rvert_{\alpha = \lambda^q} \\
    &= \sum_{k=0}^\infty 
    \left[E_{\phi, 0}^{(2q+k)}(\lambda^q) - E_{0, 0}^{(2q+k)}(\lambda^q) \right] \lambda^{k}.
\end{split}
\end{equation}
The limit of vanishingly small intrinsic capacitance $C'$ corresponds to the limit $\lambda \rightarrow 0$. To analyze $U_\text{BO}(\phi)$ in this limit, the addition and multiplication rules for limits as well as a straightforward evaluation of $E_{\phi, 0}^{(2q)}(\lambda^q)$ yield
\begin{equation}\label{eq_lim_U_BO_type_1}
    \lim_{\lambda \rightarrow 0} 
    U_\text{BO}(\phi)
    = 
    \lim_{\lambda \rightarrow 0}
    \left[
    E_{\phi, 0}^{(2q)}(\lambda^q) - E_{0 ,0}^{(2q)}(\lambda^q)
    \right] = 0.
\end{equation}
Thus, for any value of $\phi$, the Born-Oppenheimer potential vanishes in the limit $C' \rightarrow 0$.
\end{proof}

One can lift the restriction that $U_{nl}(\phi_c)$ is symmetric and also obtain a Born-Oppenheimer potential that vanishes in the limit $C' \rightarrow 0$ if Eq.~\eqref{eq_type_1_condition} is satisfied with $\gamma \in (0,1)$, as the following theorem shows. \\

\textbf{Theorem~2.} \quad
\textit{Consider $H_\text{fast}$ as defined in Eq.~\eqref{eq_def_H_fast} with $U_{nl}(\phi_c)$ describing a nonlinear inductor of type~1(b). Then, $U_\text{BO}(\phi)$ as defined in Eq.~\eqref{eq_def_BO_potential} satisfies}
\begin{equation*}
	\forall \, \, \phi \in \mathbb{R} : \quad
	\lim_{C' \rightarrow 0} U_\text{BO}(\phi) = 0.
\end{equation*}

Note that in this theorem the potential $U_{nl}(\phi_c)$ need not be symmetric, i.e., $U_{nl}(\phi_c) \neq U_{nl}(-\phi_c)$.

\begin{proof}
For nonlinear inductors of type~1(b), all arguments in the proof of Theorem~1 remain applicable until Eq.~\eqref{even} -- but the further argument cannot rely on the even parity of the eigenvalues with respect to $\phi$. For $\gamma \in (0,1)$, an explicit and straightforward evaluation of the Rayleigh-Schrödinger coefficients shows that $E_{\phi, n}^{(k)}(\alpha) - E_{0, n}^{(k)}(\alpha) = 0$ for $k\leq2q$ \footnote{Assume $H = H_0 + \lambda^{2q}A + \lambda^{q}B + \lambda^{2q-p}C$ to be an analytic family in the sense of Kato, where the operators $H_0, A, B,C$ do not depend on $\lambda$. Then, within some finite radius of convergence, the Rayleigh-Schrödinger series is analytic in $\lambda$, and the only powers of $\lambda$ are given by $k=i(2q)+j(q)+l(2q-p)$ with $i,j,l \in \mathbb{N}_0$. Thus, with $q<2q-p<2q$, there are only four possible values of $k$ such that $k \leq 2q$.}. Thus, the Born-Oppenheimer potential can be expressed as in Eq.~\eqref{eq_U_BO_type_1_before_lim}, and Eq.~\eqref{eq_lim_U_BO_type_1} remains valid.
\end{proof}

We note that nonlinear inductors of type~1(b) include the large class of nonlinear inductors that are described by a bounded potential, i.e., $|U_{nl}(\phi_c)| \leq M$ for all $\phi_c$ and some $M>0$. The Josephson junction, the SQUID and the SNAIL are probably the most important representatives of this class of inductors. For example, besides the SNAIL, Josephson junctions with broken time-reversal symmetry, are described by an asymmetric potential $U_{nl}(\phi_c)$; see p.~414 in Ref.~\cite{RMP_Golubov_Kupriyanov_Ilichev} and references therein.

To recap, within the framework of the Born-Oppenheimer approximation, which becomes more accurate the smaller the capacitance ratio $C'/C$ is, we have shown that the inductive branch in the circuit in Fig.~\ref{fig_Non_Linear_Inductance}, i.e., the series combination of the linear inductance and the generic nonlinear inductor of type~1 (including its intrinsic capacitance), effectively becomes an open circuit as $C'/C \rightarrow 0$.

More information on analytic perturbation theory can be found in Refs.~\cite{Simon_article_long, Simon_article_short}, supplementing the results we have used directly in our proofs \cite{ReedSimon, Kato}. 

\subsubsection{Effective Potential for Superlinear Inductors (Type~2)}\label{subsec_type_2}
Next, we consider the Born-Oppenheimer potential for nonlinear inductors of type~2. Here, we focus on nonlinear inductors of type~2 that are described by the following infinite set of potentials:
\begin{equation}\label{eq_family_type_2_inductors}
	U_{nl}(\phi_c) = \beta |\phi_c|^\gamma,
	\qquad	\beta > 0, 
	\quad 	\gamma \in \mathbb{Q}^{>2}.
\end{equation}
In the following, we show that for a nonlinear inductor of type~2 with a potential of the form of Eq.~\eqref{eq_family_type_2_inductors}, the Born-Oppenheimer potential approaches the potential of the linear inductance $L$ as $C'/C \rightarrow 0$. Thus, in this limit, the nonlinear branch is replaced by a {\em short circuit}, meaning that one sets $\phi_c=0$. To show this, our strategy is to identify the contribution of the linear inductance $L$ in the fast Schrödinger equation in Eq.~\eqref{eq_fast_SE_BO} as a perturbation, the opposite of the type-1 scenario. To this end, we provide the following theorem. \\

\textbf{Theorem~3.} \quad
\textit{Consider $H_\text{fast}$ as defined in Eq.~\eqref{eq_def_H_fast} with $U_{nl}(\phi_c)$ as defined in Eq.~\eqref{eq_family_type_2_inductors} describing a subset of nonlinear inductors of type~2. Then, $U_\text{BO}(\phi)$ as defined in Eq.~\eqref{eq_def_BO_potential} satisfies}
\begin{equation*}
	\forall \, \, \phi \in \mathbb{R} : \quad
	\lim_{C' \rightarrow 0} U_\text{BO}(\phi) = \frac{\phi^2}{2L}.
\end{equation*}

\begin{proof}
We introduce the pair of rescaled conjugate variables $y = C'^{\frac{1}{\gamma+2}}\phi_c$ and $p = Q_c / C'^{\frac{1}{\gamma+2}}$, satisfying the canonical commutation relation $[y,p]=i \hbar$. The Hamiltonian $H_\text{fast}$ expressed in these variables reads
\begin{equation}\label{eq_fast_Hamiltonian_type_2}
    H_\text{fast} =
    C'^{-\frac{\gamma}{\gamma+2}}
    \left( \frac{p^2}{2} + \beta |y|^\gamma \right) 
    + \frac{y^2}{2 L C'^{\frac{2}{\gamma+2}}}
    - \frac{\phi y}{L C'^{\frac{1}{\gamma+2}}}
    + \frac{\phi^2}{2L}.
\end{equation}
We define the Hamiltonian $H_0 = p^2/2 + \beta |y|^\gamma $ and the parameter $\epsilon = C'^{\frac{1}{\gamma+2}}$. Multiplying Eq.~\eqref{eq_fast_Hamiltonian_type_2} with $\epsilon^\gamma$ results in
\begin{equation}
    \epsilon^\gamma H_\text{fast} 
    = H_0
    + \epsilon^{\gamma-2} \frac{y^2}{2 L}
    - \epsilon^{\gamma-1} \frac{\phi y}{L}
    + \epsilon^{\gamma} \frac{\phi^2}{2L}.
\end{equation}

Since $\gamma \in \mathbb{Q}^{>2}$, we set $\gamma=p/q$ with $p, q \in \mathbb{N}$ satisfying $p-2q \in \mathbb{N}$. We substitute $\epsilon = \lambda^q$ in the fast Hamiltonian and obtain
\begin{equation}
    \lambda^p H_\text{fast} 
    = H_0
    + \lambda^{p-2q} \frac{y^2}{2 L}
    - \lambda^{p-q} \frac{\phi y}{L}
    + \lambda^{p} \frac{\phi^2}{2L},
\end{equation}
which is an analytic family of type (A) in the sense of Kato. Since the spectrum of $H_0$ is non-degenerate (p.~719ff in Ref.~\cite{1953MF}), the Kato-Rellich theorem applies, and the $n^{\rm{th}}$ eigenenergy of $\lambda^p H_\text{fast}$ is an analytic function in $\lambda$ with a non-vanishing radius of convergence $\lambda_n(\phi) > 0$, i.e., for all $\lambda < \lambda_n (\phi)$ we can write the eigenenergy as the Rayleigh-Schrödinger series
\begin{equation}\label{eq_RS_series_type_2}
    \lambda^{p} E_{\phi, n}(\lambda) 
    = \sum_{k=0}^\infty E_{\phi, n}^{(k)} \lambda^k.
\end{equation}
For $k < p$, the Rayleigh-Schrödinger coefficients $E_{\phi, n}^{(k)}$ do not depend on $\phi$, and it follows that $E_{\phi, n}^{(k)} - E_{0, n}^{(k)}=0$ for $k < p$.

Within the radius of convergence, i.e., for $\lambda < \lambda_0(\phi)$, the Born-Oppenheimer potential as defined in Eq.~\eqref{eq_def_BO_potential} can be expressed as
\begin{equation}\label{eq_U_BO_type_2_before_lim}
\begin{split}
    U_\text{BO}(\phi) 
    = E_{\phi, 0}(\lambda) - E_{0, 0}(\lambda)
    = \sum_{k=0}^\infty 
    \left[E_{\phi, 0}^{(p+k)} - E_{0, 0}^{(p+k)} \right] \lambda^{k}.
\end{split}
\end{equation}
The limit of vanishingly small intrinsic capacitance $C'$ corresponds to the limit $\lambda \rightarrow 0$. To analyze $U_\text{BO}(\phi)$ in this limit, the addition and multiplication rules for limits as well as a straightforward evaluation of $E_{\phi, 0}^{(p)}$ yield
\begin{equation}\label{eq_lim_U_BO_type_2}
    \lim_{\lambda \rightarrow 0} 
    U_\text{BO}(\phi)
    = 
    E_{\phi, 0}^{(p)} - E_{0, 0}^{(p)} 
    = \frac{\phi^2}{2L}.
\end{equation}
Thus, for any value of $\phi$, the Born-Oppenheimer potential approaches the potential of the linear inductance $L$ in the limit $C' \rightarrow 0$.
\end{proof}

To recap, within the framework of the Born-Oppenheimer approximation, which becomes more accurate the smaller the capacitance ratio $C'/C$ is, we have shown that the inductive branch in the circuit in Fig.~\ref{fig_Non_Linear_Inductance}, i.e., the series combination of the linear inductance and the type-2 nonlinear inductor described by the potential in Eq.~\eqref{eq_family_type_2_inductors} (including its intrinsic capacitance) is effectively replaced by a linear inductance $L$ connecting the $\phi$-node to ground as $C'/C \rightarrow 0$; i.e., the nonlinear inductor and its intrinsic capacitance are effectively replaced by a short circuit between the nodes $\phi_c$ and ground.

At this point, we do not attempt to provide a general proof, but we conjecture that Theorem~3 holds for any generic nonlinear inductor of type~2 and is not restricted to potentials of the form of Eq.~\eqref{eq_family_type_2_inductors}.

\subsubsection{Effective Potential for Type-L Inductors}\label{subsec_type_L}
Next, we analyze nonlinear inductors of type~L. Note that an inductor of type~L is \textit{not} in general linear, but it is clear from the definition that its potential is the sum of that of a linear inductance $\mathfrak{L}$ with that of a nonlinear inductor, with a potential $U_{nl}(\phi_c)-\phi_c^2/{2 \mathfrak{L}}$, that is of type~1. In other words, the type-L inductor can always be represented as the parallel combination of a linear inductance and a nonlinear inductor of type~1. This equivalence will be useful later.

In the following, we show that for a nonlinear inductor of type~L, the Born-Oppenheimer potential approaches the potential of a total linear inductance $L + \mathfrak{L}$ as $C'/C \rightarrow 0$.
Thus, in this limit, the nonlinear branch is replaced by a {\em linear inductance} $\mathfrak{L}$, and the node $\phi_c$ is removed by adding the linear inductances $L$ and $\mathfrak{L}$ in a series connection, resulting in a total inductance $L + \mathfrak{L}$ between the nodes $\phi$ and ground.

To show this, we combine ideas of the proofs of Theorem~1 and Theorem~3. In particular, with regard to the fast Hamiltonian in Eq.~\eqref{eq_def_H_fast}, we split the potential of the type-L inductor into two parts, with one part rescaling the underlying harmonic-oscillator Hamiltonian, while the other part is identified as contribution to the perturbation of that system. This analysis results in the following theorem. \\

\textbf{Theorem~4.} \quad
\textit{Consider $H_\text{fast}$ as defined in Eq.~\eqref{eq_def_H_fast} with $U_{nl}(\phi_c)$ describing a nonlinear inductor of type~L. Then, $U_\text{BO}(\phi)$ as defined in Eq.~\eqref{eq_def_BO_potential} satisfies}
\begin{equation*}
	\forall \, \, \phi \in \mathbb{R} : \quad
	\lim_{C' \rightarrow 0} U_\text{BO}(\phi) = \frac{\phi^2}{2(L + \mathfrak{L})}
\end{equation*}
\textit{with $\mathfrak{L} = \lim_{\phi_c \to \infty} \phi_c^2 / 2U_{nl}(\phi_c) > 0$.}

\begin{proof}
It follows from the definition of a type-L nonlinear inductor that there exists a constant $\mathfrak{L} > 0$ such that its potential can be written as
\begin{equation}
    U_{nl}(\phi_c) = \frac{\phi_c^2}{2 \mathfrak{L}} + U_{t1}(\phi_c),
\end{equation}
where $U_{t1}(\phi_c)$ describes a nonlinear inductor of type~1. We define the effective parallel combination inductance $l$ and the characteristic flux scale $\Phi$ as
\begin{equation}
    l = \frac{L \mathfrak{L}}{L + \mathfrak{L}},   \qquad
    \Phi = \sqrt{\hbar} \sqrt[4]{\frac{l}{C'}},
\end{equation}
respectively. We further introduce the dimensionless conjugate variables $y = \phi_c / \Phi$ and $p = Q_c \Phi / \hbar$ satisfying the canonical commutation relation $[y, p] = i$.

With $\epsilon = \sqrt{l}/\Phi$ and $H_0 = (p^2+y^2)/2$, the fast Hamiltonian in Eq.~\eqref{eq_def_H_fast} can be expressed as [cf. Eq.~\eqref{eq_eps2_H_fast_type_1}]
\begin{equation}\label{eq_eps2_H_fast_type_L}
    \epsilon^2 H_\text{fast} 
    = H_0 
    + \frac{\epsilon^2\phi^2}{2L} 
    - \frac{\epsilon \sqrt{l} \phi y}{L} 
    + \epsilon^2 U_{t1} \left( \sqrt{l} y / \epsilon \right).
\end{equation}
For the remainder of this proof, all arguments in the proof of Theorem~1 for nonlinear inductors of type~1 remain applicable until Eq.~\eqref{eq_lim_U_BO_type_1}. In particular, the Born-Oppenheimer potential can be expressed as in Eq.~\eqref{eq_U_BO_type_1_before_lim}. However, as opposed to type-1 nonlinear inductors, here, a straightforward evaluation of $E_{\phi, 0}^{(2q)}(\lambda^q)$ yields
\begin{equation}\label{eq_lim_U_BO_type_L}
    \lim_{\lambda \rightarrow 0} 
    U_\text{BO}(\phi)
    = 
    \lim_{\lambda \rightarrow 0}
    \left[
    E_{\phi, 0}^{(2q)}(\lambda^q) - E_{0 ,0}^{(2q)}(\lambda^q)
    \right]
    = \frac{\phi^2}{2(L + \mathfrak{L})}.
\end{equation}
Thus, for any value of $\phi$, the Born-Oppenheimer potential approaches the potential of a linear inductance $L + \mathfrak{L}$ in the limit $C' \rightarrow 0$.
\end{proof}

To recap, within the framework of the Born-Oppenheimer approximation, which becomes more accurate the smaller the capacitance ratio $C'/C$ is, we have shown that the inductive branch in the circuit in Fig.~\ref{fig_Non_Linear_Inductance}, i.e., the series combination of the linear inductance and the generic nonlinear inductor of type~L (including its intrinsic capacitance), is effectively replaced by a linear inductance $L + \mathfrak{L}$ connecting the $\phi$-node to ground as $C'/C \rightarrow 0$.

\subsection{A Pathological Potential}\label{subsec_pathological_potential}
Not all series combinations of a linear inductance and a nonlinear inductor (cf. Fig.~\ref{fig_Non_Linear_Inductance}) necessarily have a well-defined effective limiting behavior as the internal capacitance vanishes. To illustrate the potentially ambiguous limit, we analyze a pathological example of a nonlinear inductor with a potential energy that cannot be classified as falling into one of our categories. 

\begin{figure}[t!]  
	\centering
	\includegraphics[width=1.0\textwidth]{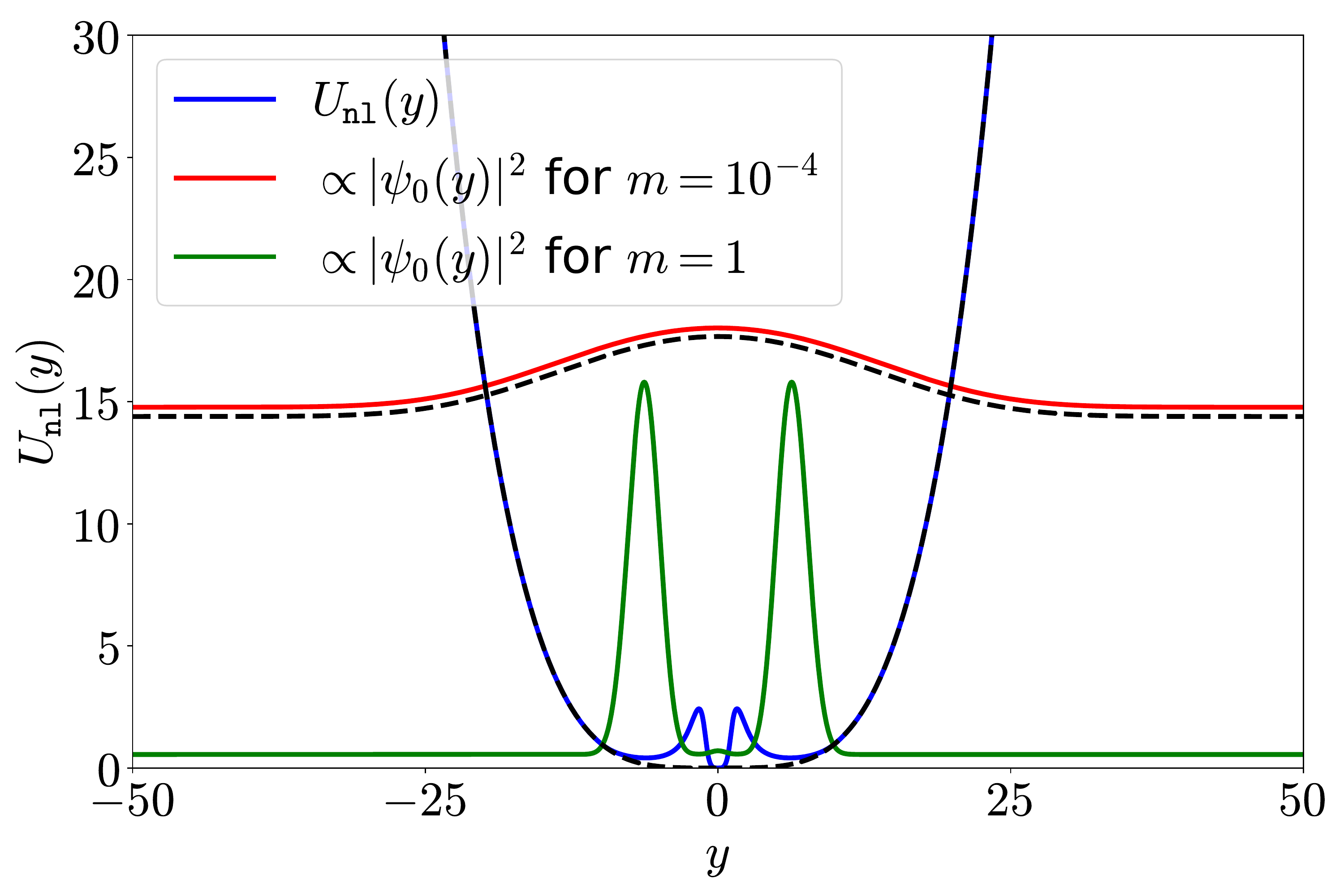}	
	\caption{Pathological potential $U_{nl}(y)$ in Eq.~\eqref{eq_pathological_potential} (blue) together with the ground state wave function of the Hamiltonian in Eq.~\eqref{eq_H_pathological} for $m=10^{-4}$ (red) and $m=1$ (green). Different choices of $m$ correspond to different values of the internal shunting capacitance. For $m=10^{-4}$, the ground state wave function of the potential $y^4/10^4$ is shown for comparison  (black dashed lines for both wave function and potential). For better visibility, all wave functions are scaled by a factor 10 and shifted by their corresponding ground state energies.}
	\label{fig_Self_Sim_PotentialWF}
\end{figure}

First, we focus on an isolated nonlinear inductor accompanied by its internal shunting capacitance. Working with dimensionless variables, the Hamiltonian of this system can be written as
\begin{equation}\label{eq_H_pathological}
    H = \frac{p_y^2}{2m} + U_{nl}(y),
\end{equation}
in which $m$ denotes the rescaled shunting capacitance. The rescaled canonical variables satisfy the dimensionless commutation relation $[y,p_y]=i$. In the following, we consider a nonlinear inductor that is described by the following symmetric, differentiable potential ($n \in \mathbb{Z}$):
\begin{equation}\label{eq_pathological_potential}
    U_{nl}(y) =
    \begin{cases}
    10^{(3-4n+2\log_{10}|y|)^3} \\
    \times 10^{8n-7}y^{-2} 
    & \text{\!\!\!\!\!\! for } 10^{2n-2} \leq |y| \leq 10^{2n-1}             \\
    10^{-4n} y^4       
    & \text{\!\!\!\!\!\! for } 10^{2n-1} \leq |y| \leq 10^{2n}.
    \end{cases}
\end{equation}
The potential $U_{nl}(y)$ and the corresponding ground state wave function of the Hamiltonian in Eq.~\eqref{eq_H_pathological} for different values of $m$ are shown in Fig.~\ref{fig_Self_Sim_PotentialWF}. Because of the self-similarity of the potential, 
\begin{equation}
    U_{nl}(10^2 y) = 10^4 U_{nl}(y),
\end{equation}
the eigensystem of $H$ associated with the mass $m$ relates to that with a rescaled mass $m' = 10^{-8}m$. In that case, the eigenenergies and the eigenstates satisfy $E'_\nu = 10^{4} E_\nu$ and $\psi'_\nu(y) \propto \psi_\nu(y/100)$, respectively.

\begin{figure}[t!]  
	\centering
	\includegraphics[width=1.0\textwidth]{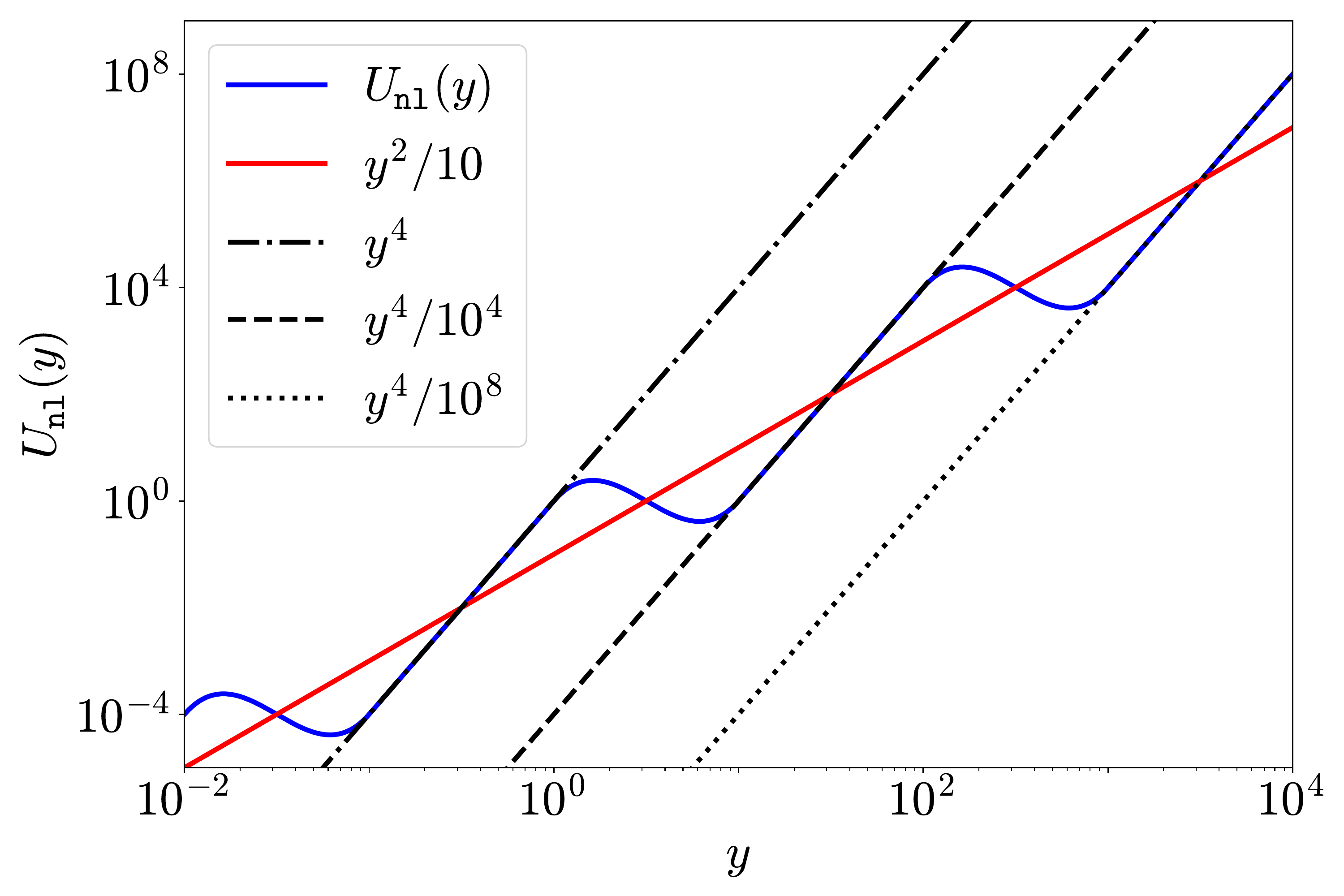}
	\caption{Pathological potential $U_{nl}(y)$ in Eq.~\eqref{eq_pathological_potential}  (blue) together with the 'global trend' $y^2/10$ (red). For $|y| \in [10^{2n-1}, 10^{2n}]$ with $n \in \mathbb{Z}$, $U_{nl}(y)$ grows faster than a second order polynomial; in particular, $U_{nl}(y) \propto y^4$ (indicated with black lines). However, in the limit $y \rightarrow \infty$, the ratio $U_{nl}(y)/y^2$ remains ill-defined.}
	\label{fig_Self_Sim_PotentialLogLog}
\end{figure}

We partition the range of $m$ into three distinct regions in which the eigensystem of $H$ behaves qualitatively differently; see also Fig.~\ref{fig_Self_Sim_PotentialLogLog}. First, there is a region of $m$ in which both the ground state energy and the ground state wave function are well approximated by that of a purely quartic potential $\propto y^4$ (red wave function in Fig.~\ref{fig_Self_Sim_PotentialWF}). Second, there is a disjoint region of $m$ in which the ground state wave function resembles that of a double well potential (green wave function in Fig.~\ref{fig_Self_Sim_PotentialWF}). Within these two regions, the scaling of, e.g., the eigenenergies with respect to $m$ is fundamentally different. Last, there are intermediate values of $m$ in which the system transitions between both the previously mentioned regions. Thus, by construction of $U_{nl}(y)$, there is no well-defined asymptotic behavior of the eigensystem as $m \rightarrow 0$.

\begin{figure}[b!]  
	\centering
	\includegraphics[width=1.0\textwidth]{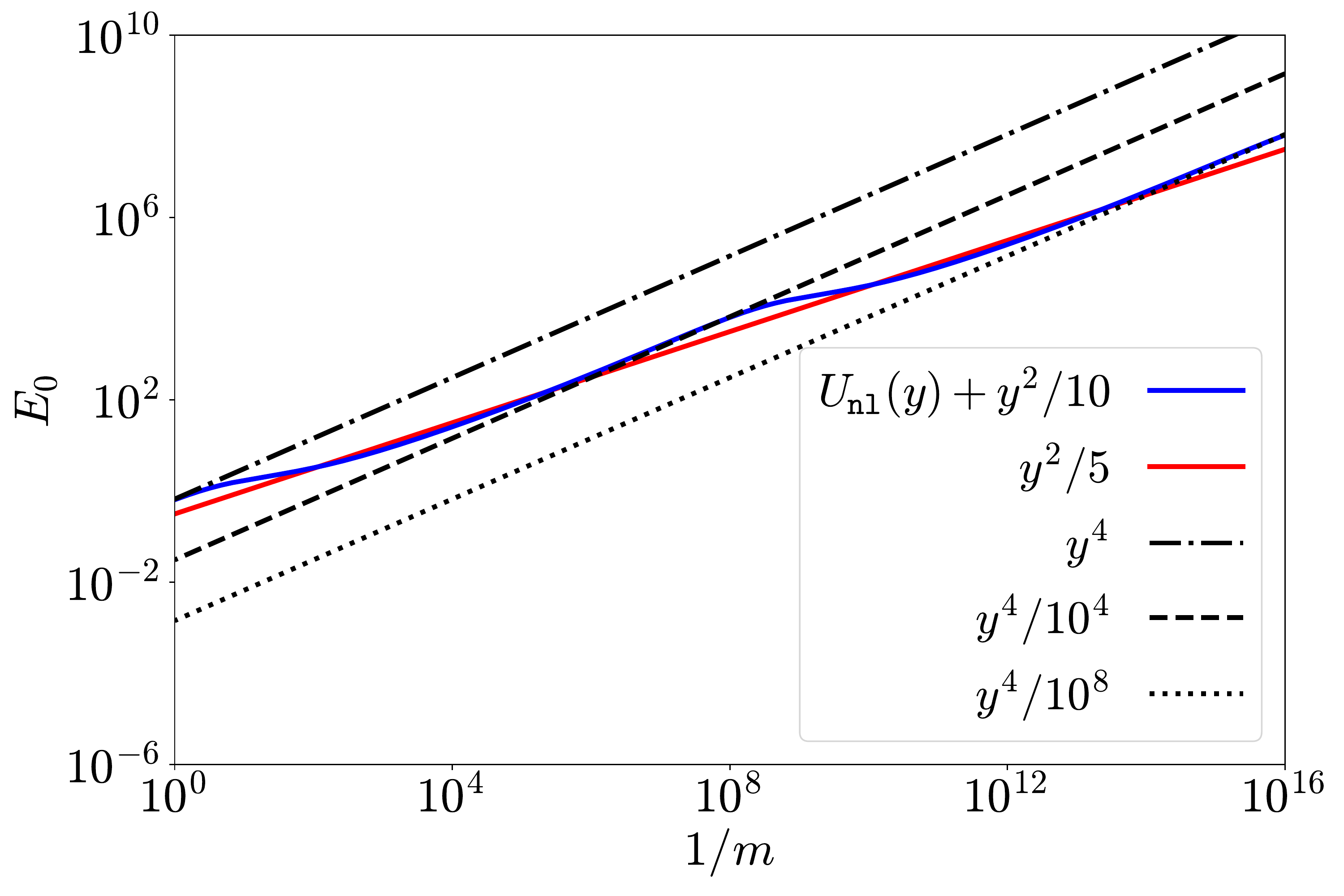}
	\caption{
	Ground state energy $E_0$ of $H_\text{fast}$ in Eq.~\eqref{eq_fast_Hamiltonian_pathological} for a fixed value $x=0$ as a function of $1/m$ (blue).
    For certain regimes of $m$, $E_0$ is well approximated by the ground state energy corresponding to a purely quartic potential (indicated with black lines).
    On large scales of $m$, however, $E_0$ follows the 'global trend' of the ground state energy corresponding to a purely quadratic potential (red).}
	\label{fig_Embedded_Self_Sim_Energies}
\end{figure}

Next, we embed such a nonlinear inductor in the circuit shown in Fig.~\ref{fig_Non_Linear_Inductance}. We choose the linear inductance such that the total system is described by the Hamiltonian [cf. Eq.~\eqref{eq_Hamiltonian_r}]
\begin{equation}\label{eq_Hamiltonian_embedded_self_sim_potential}
    H_r = \frac{p_x^2}{2M} + \frac{p_y^2}{2m} + U_{nl}(y) + \frac{(y-x)^2}{10} ,
\end{equation}
in which $M$ is the rescaled outer capacitance, and $x$ and $y$ can be interpreted as slow and fast variables, respectively. The ground state energy $E_0$ of the fast part of $H_r$ [cf. Eq.~\eqref{eq_def_H_fast}],
\begin{equation}\label{eq_fast_Hamiltonian_pathological}
     H_\text{fast} = \frac{p_y^2}{2m} + U_{nl}(y) + \frac{(y-x)^2}{10},
\end{equation}
is shown in Fig.~\ref{fig_Embedded_Self_Sim_Energies} for $x=0$. For the special choice $x=0$, the total potential entering $H_\text{fast}$ remains self-similar in $y$. As a consequence, $E_0$ scales linearly by $10^{4}$ as $1/m$ is scaled by a factor $10^8$. For certain regimes of $m$, the ground state energy is well approximated by that of a bare quartic potential; see black lines in Fig.~\ref{fig_Embedded_Self_Sim_Energies}. However, if the mass $m$ is considered over several orders of magnitude, $E_0$ follows the 'global trend' given by the ground state energy corresponding to that of a quadratic potential; see red line in Fig.~\ref{fig_Embedded_Self_Sim_Energies}.

\begin{figure}[b!]  
	\centering
	\includegraphics[width=1.0\textwidth]{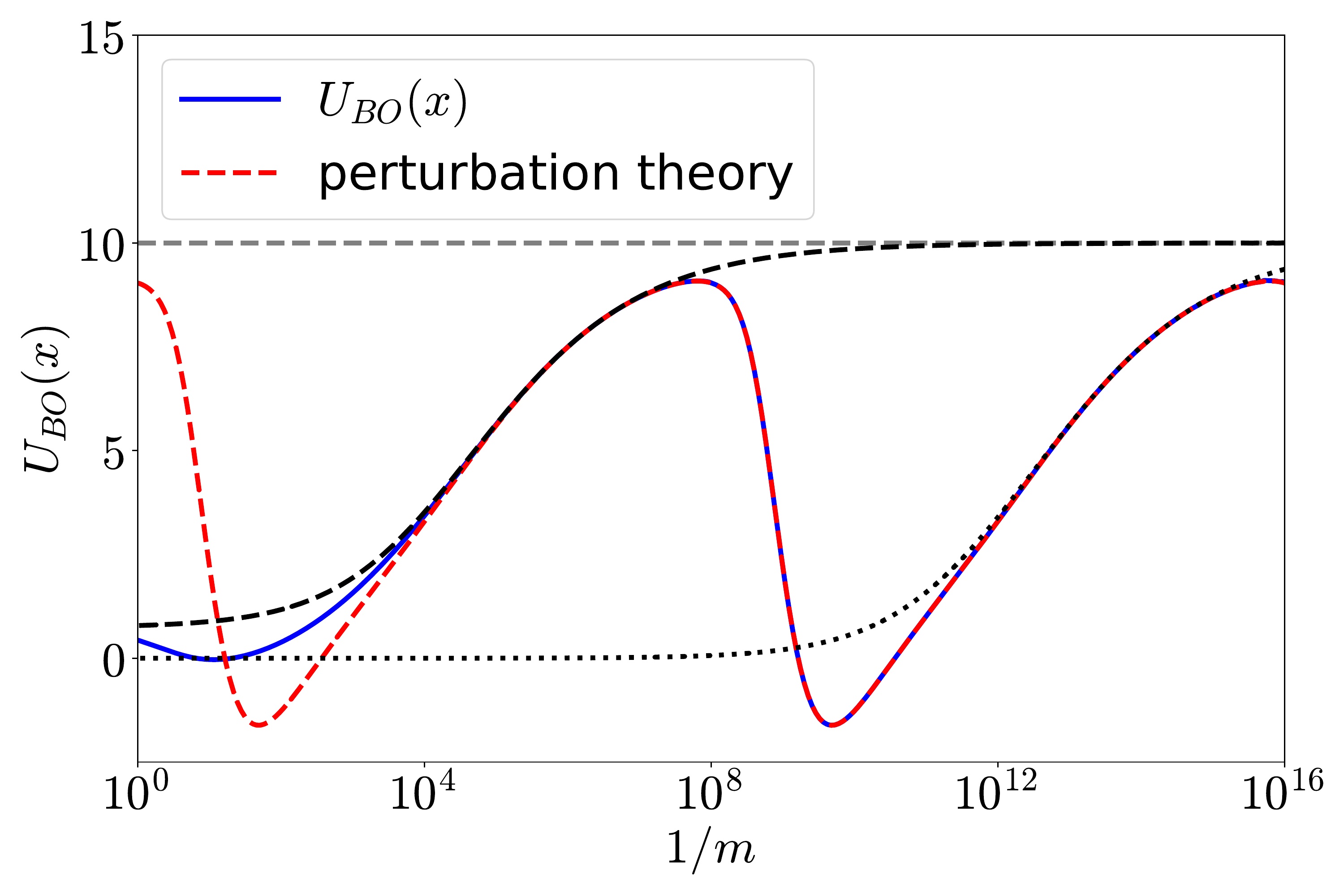}
	\caption{Born-Oppenheimer potential $U_\text{BO}(x)$ evaluated at $x=10$ as a function of $1/m$ (blue). 
	The red dashed line, showing the result of the second order perturbation theory in $x$, is periodic in $\log(1/m)$; see main text.
	For nonlinear inductors of type~2, $U_\text{BO}(10)$ approaches the value of $10$ in the limit of $m \rightarrow 0$ (gray dashed line). For comparison, black lines show the Born-Oppenheimer potential at $x=10$ with a nonlinear inductor that is described by a purely quartic potential (dashed: $y^4/10^4$, dotted: $y^4/10^8$).}
	\label{fig_Self_Sim_Example}
\end{figure}

For fixed values of $m$, we use the ground state energy of $H_\text{fast}$ at $x=0$ to shift the Born-Oppenheimer potential such that $U_\text{BO}(0)=0$. The Born-Oppenheimer potential $U_\text{BO}(x)$ for the specific choice $x=10$ is shown in Fig.~\ref{fig_Self_Sim_Example} as a function of $1/m$. As before, we note that for certain regimes of $m$, the Born-Oppenheimer potential associated with $U_{nl}(y)$ in Eq.~\eqref{eq_pathological_potential} is well approximated by that corresponding to a nonlinear inductor that is described by a purely quartic potential $\propto y^4$; see black curves in Fig.~\ref{fig_Self_Sim_Example}. In fact, recall that for nonlinear inductors of type~2 (to which quartic potentials belong) the Born-Oppenheimer potential approaches the value of $x^2/10$ [cf.~Theorem 3] in the limit of $m \rightarrow 0$. However, such a convergence is not observed for $U_\text{BO}(x)$ involving $U_{nl}(y)$ in Eq.~\eqref{eq_pathological_potential} as this potential does not describe a nonlinear inductor of type~2. In order to analyze the behavior of $U_\text{BO}(x)$ for small values of $m$, we note that $H_\text{fast}$ in Eq.~\eqref{eq_fast_Hamiltonian_pathological} corresponding to $m$ and $x$ relates to that corresponding to $m'=10^{-8}m$ and $x'=10^2x$. In particular, we find that $H'_\text{fast}=10^4 H_\text{fast}$. As a consequence, for small enough values of $m$, nonzero values of $x$ can be incorporated with second order perturbation theory 
(red dashed line in Fig.~\ref{fig_Self_Sim_Example}), which in fact becomes more precise as $m$ becomes smaller. As a result, we find that (up to small corrections) $U_\text{BO}(x)$ is periodic in $\log(1/m)$ for small values of $m$.

In total, the nonlinear inductor described by the pathological potential in Eq.~\eqref{eq_pathological_potential} exemplifies that not every series combination of inductances gives rise to a well defined single effective inductance as the internal capacitance vanishes. In particular, we conclude that if one cannot classify a nonlinear inductor at hand by means of the results presented in Secs.~\ref{subsec_type_1}~to~\ref{subsec_type_L}, one is required to know the particular value of its internal capacitance in order to derive a reliable one-mode replacement.

\subsection{An Asymmetric Potential}\label{subsec_asymmetric_potential}
The following shows that the result of Theorem~2 is tight: the Born-Oppenheimer potential does {\em not} vanish for a linear, asymmetric inductive potential. We provide the Born-Oppenheimer analysis for this simple asymmetric example: we consider the piecewise linear potential
\begin{equation}\label{eq_asym_potential_type_1}
    U_{nl}(\phi_c) = b \left[ 1 + a \Theta(\phi_c) \right] |\phi_c|,
    \qquad b > 0,
\end{equation}
in which $a>0$ tunes the asymmetry. Here, $b$ is some arbitrary positive prefactor, and $\Theta(\phi_c)$ denotes the Heaviside step function. Despite the asymmetry of $U_{nl}(\phi_c)$, the fact that $\lim_{\phi_c \to \pm \infty} U_{nl}(\phi_c)/\phi_c^2 = 0$ allows the Born-Oppenheimer potential following the steps in the proof of Theorem~1 (cf. Sec.~\ref{subsec_type_1}). In particular, recall that the zero-point fluctuation $\Phi_\textit{ZPF}$ as defined in Eq.~\eqref{eq_defs_ZPF_and_omega_LC} diverges as the intrinsic capacitance $C'$ vanishes. Thus, we proceed by treating $U_{nl}(\phi_c)$ in Eq.~\eqref{eq_asym_potential_type_1} as part of a perturbation around the harmonic-oscillator Hamiltonian; expanding the Born-Oppenheimer potential in powers of $1/\Phi_\textit{ZPF}$ yields
\begin{equation}\label{eq_U_BO_expanded}
    U_\text{BO}(\phi) 
    = \frac{ab\phi}{2}
    + \frac{(2+a)b\phi^2
        -a (2+a) b^2 L \phi}{2\sqrt{\pi}\Phi_\textit{ZPF}},
\end{equation}
where we omit terms of the order $\mathcal{O}(1/\Phi_\textit{ZPF}^2)$. We observe that $U_\text{BO}(\phi)$ does not vanish as $1/\Phi_\textit{ZPF} \rightarrow 0$ if the asymmetric case $a \neq 0$ is considered. To interpret this result, we rewrite $U_\text{BO}(\phi)$ in Eq.~\eqref{eq_U_BO_expanded} in normal form,
\begin{equation}
    U_\text{BO}(\phi)
    =
    \frac{(2+a)b}{2\sqrt{\pi}\Phi_\textit{ZPF}}
    \left[
    \phi 
    + \frac{a}{2}
    \left(
    \frac{\sqrt{\pi}\Phi_\textit{ZPF}}{2+a} - b L    
    \right)
    \right]^2,
\end{equation}
in which we dropped an additive constant that does not depend on $\phi$. Thus, if no further network element besides the shunting capacitance $C$ is attached to the node flux $\phi$ (see Fig.~\ref{fig_Non_Linear_Inductance}), the slow degree of freedom of the system is well approximated by $H_{r,\text{eff}}$ in Eq.~\eqref{eq_H_r_eff}, and its ground state wave function is a Gaussian whose center position and standard deviation are given by
\begin{equation}
    \phi_m 
    = - \frac{a}{2} \left( \frac{\sqrt{\pi}\Phi_\textit{ZPF}}{2+a} - b L \right),    \quad
    \Delta \phi 
    = \sqrt{\hbar} \sqrt[4]{\frac{\sqrt{\pi}\Phi_\textit{ZPF}}{(2+a)bC}},
\end{equation}
respectively. In the limit of large zero-point fluctuations, we find that $\Delta \phi/\phi_m \rightarrow 0$, while $\Delta \phi \rightarrow \infty$. Thus, when capacitively shunted, the series combination of a linear inductance and our non-symmetric, nonlinear inductor is effectively replaced by an open circuit, as in the case of a nonlinear inductor of type~1.

However, if the node $\phi$ is embedded into a larger circuit, the displacement $\phi_m$ has the effect of an effective magnetic flux through a closed loop formed by the inductive branch and further inductive elements. This effective magnetic flux does not affect the dynamics of the total system as long as the larger circuit involves linear inductances only. If, however, the system contains further nonlinear inductors, the actual value of $\phi_m$ and thus that of the small intrinsic capacitance $C'$ becomes of central importance. In that case, an effective replacement of the inductive branch in the circuit in Fig.~\ref{fig_Non_Linear_Inductance} is not well defined as the internal capacitance $C'$ vanishes.

\begin{figure}[b!]  
	\centering
	\includegraphics[]{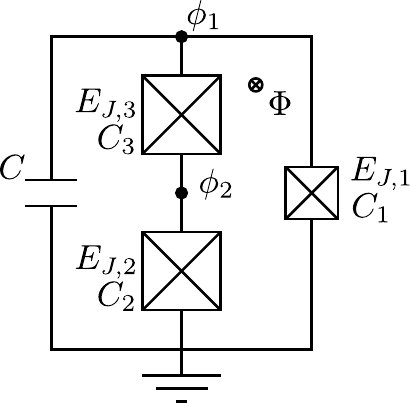}
	\caption{Capacitively shunted SNAIL with $N=2$ large Josephson junctions. The external magnetic flux $\Phi$ pierces the loop formed by the Josephson junctions.}
	\label{fig_SNAIL}
\end{figure}

\section{Josephson Junction Arrays -- Single-Phase Approximation Revised}\label{sec_SPA}
So far, we analyzed a series connection of a linear inductance and a nonlinear inductor. In this section, we generalize this analysis, and we consider the series connection of multiple nonlinear inductors. This is important because in practice, arrays of Josephson junctions are commonly fabricated to realize effective devices such as superinductances \cite{Manucharyan, Brooks, ViolaCatelani} or SNAILs (Superconducting Nonlinear Asymmetric Inductive eLements) \cite{SNAIL1,SNAIL2,SNAIL3}. Therefore, in the following, we revise the single-phase approximation \cite{NakamuraSNAIL} that is commonly used to simplify the description of these multi-node Josephson junction chains. Here, we focus our analysis to the case of a capacitively shunted SNAIL with $N=2$ large Josephson junctions \footnote{Although SNAILs are mostly operated with $N=3$ large Josephson junctions, we focus on the case $N=2$ for the sake of simplicity.}; see Fig.~\ref{fig_SNAIL}.

Typically, SNAILs are embedded into larger electrical networks in order to provide nonlinearity in the form of effective nonlinear inductors, or for the amplification of signals. In any case, the internal degrees of freedom of the SNAIL (here $\phi_2$) are commonly discarded such that it can be considered to be an element of one degree of freedom only (here $\phi_1$). This simplification is known as the single-phase approximation. 

We introduce the dimensionless parameters $k_i=C_i/C$ that relate the intrinsic capacitances $C_i$ of the Josephson junctions to the large capacitance $C$ of the shunt. In the following model, we consider all the intrinsic capacitances of the Josephson junctions to be finite, i.e., $k_i>0$. As a consequence, every branch of a SNAIL network such as shown in Fig.~\ref{fig_SNAIL} contains at least one capacitor, and the capacitance matrix
\begin{equation}
    \bm{C} = 
    \begin{pmatrix}
        1+k_1+k_3 & -k_3    \\
        -k_3    &   k_2+k_3
    \end{pmatrix}.
\end{equation}
is invertible. Thus, the circuit is regular and the corresponding Hamiltonian is straightforwardly obtained as
\begin{equation}\label{eq_Hamiltonian_SNAIL}
    H = 4 E_C \bm{n}^T \bm{C}^{-1} \bm{n} + U(\bm{\phi})
\end{equation}
with the charging energy $E_C=e^2/2C$, and the vector notation is $\bm{\phi}=(\phi_1,\phi_2)^T, \bm{n}=(n_1,n_2)^T$. The Josephson junctions constitute the total potential energy
\begin{equation}
\begin{split}
    U(\bm{\phi}) =
    - E_{J,1} \cos(\phi_1)
    - E_{J,2} \cos(\phi_2)                  \\
    - E_{J,3} \cos(\phi_1-\phi_2 + \Phi),
\end{split}
\end{equation}
in which $\Phi=\Phi^\text{ext}/\Phi_0$ is the rescaled external magnetic flux through the loop formed by the Josephson junctions. Similarly, the chosen variables $\phi_i$ and $n_j$ are dimensionless, and the system is quantized by imposing the usual commutation relations $[\phi_i, n_j] = i \delta_{ij}$.

In order to proceed, we diagonalize the kinetic term of the Hamiltonian in Eq.~\eqref{eq_Hamiltonian_SNAIL}. As we will see, in the limit of small intrinsic capacitances ($k_i \ll 1$), this diagonalization clearly separates the dynamics of the system into a fast variable and a slow one. In principle, such a decoupling can be achieved by means of several different variable transformations. Here, we define the canonical (but non-orthogonal) variable transformation obtained from a Cholesky decomposition:
\begin{equation}\label{eq_SNAIL_variable_transform}
    \bm{p} = \bm{A}^{-1} \bm{n}, \qquad
    \bm{x} = \bm{A}^T \bm{\phi}, \qquad
    \bm{A}
    =
    \begin{pmatrix}
        1 & \frac{-k_3}{k_2+k_3} \\
        0 & 1
    \end{pmatrix},   
\end{equation}
which transforms the Hamiltonian of the system to
\begin{equation}\label{eq_Hamiltonian_SNAIL_new}
    H = 4 E_C (d_1 p_1^2 + d_2 p_2^2) + U(x_1, x_2)
\end{equation}
with the diagonal kinetic matrix elements
\begin{equation}
    d_1 = \frac{k_2 + k_3}{k_2+k_3+k_1k_2+k_2k_3+k_3k_1},   \qquad
    d_2 = \frac{1}{k_2+k_3},
\end{equation}
and the total potential energy $U(x_1, x_2)$ in terms of the new position-like variables $x_1$ and $x_2$,
\begin{equation}\label{eq_potential_SNAIL}
\begin{split}
    U(x_1, x_2) =
    - E_{J,1} \cos(x_1)
    - E_{J,2} \cos(\frac{k_3}{k_2+k_3} x_1 + x_2)                                \\
    - E_{J,3} \cos(\frac{k_2}{k_2+k_3} x_1 - x_2 + \Phi).
\end{split}
\end{equation}

Note that, per construction, the variable transformation in Eq.~\eqref{eq_SNAIL_variable_transform} ensures that $[x_i, p_j] = i \delta_{ij}$ and $x_1=\phi_1$. The latter property will be crucial for the single-phase approximation of the SNAIL if it is coupled inductively to some further circuitry, as the relevant coupling variable will be $\phi_1$ in that case. The effect of the non-orthogonal transformation matrix $\bm{A}$ on the boundary conditions of the wave function will be discussed in Sec.~\ref{subsec_SNAIL_limit}.

In the following, we analyze the limit of small intrinsic capacitances, i.e., $k_i \ll 1$. In that limit, we find that $d_2 \gg d_1$ such that the dynamics in the $x_2$-direction becomes much faster than that in the $x_1$-direction. As elaborated in Sec.~\ref{sec_failure_Dirac}, such different time scales for the dynamics in the two directions make the Born-Oppenheimer approximation applicable. Thus, we first solve the Schrödinger equation for the fast variable $x_2$, keeping the slow variable $x_1$ as a fixed parameter. To further simplify the analysis, the following calculations will be carried out for a symmetric SNAIL, i.e., for the remainder of this section, we focus on the special case of $k_2 = k_3$ and $E_{J,2}=E_{J,3}$.

\subsection{Classical Approach}\label{subesec_BO_classical}
Instead of solving the fast Schrödinger equation for its quantum mechanical ground state energy, it has been typical to focus on the classical minimal energy of $U(x_1, x_2)$ in the $x_2$-direction for fixed values of $x_1$ \cite{SNAIL2,NakamuraSNAIL}. As illustrated in Sec.~\ref{sec_singular_circuits}, this simplification corresponds to the singular case ($k_2=0$), in which quantum fluctuations in $\phi_2$ are assumed to be absent, and the Dirac-Bergmann algorithm has to be applied in order to obtain the Hamiltonian. However, for the regular case ($k_2 > 0$), this classical approach is a good approximation only if the eigenfunction is well localized in $x_2$, which is true if all the individual Josephson junctions in the array are deeply in the transmon regime, $E_C / k_2 E_{J,2} \ll 1$ \cite{Koch}. Note that these two conditions on $k_2$ are not automatically compatible and need to be examined individually for each given set of circuit parameters; see also Appendix~A in Ref.~\cite{SNAIL2}. 

For the symmetric SNAIL and a fixed value of $x_1$, the condition for a minimal potential [Eq.~\eqref{eq_potential_SNAIL}] in the fast direction evaluates to $x_2=\Phi/2$. Inserting this value for $x_2$ in $U(x_1, x_2)$ results in an effective one-dimensional potential for the slow $x_1$-variable, namely
\begin{equation}\label{eq_V_BO_cl}
    U_\text{BO}^\text{cl}(x_1) = -E_{J,1} \cos(x_1) - 2E_{J,2} \cos(\frac{x_1+\Phi}{2}).
\end{equation}
This classically obtained Born-Oppenheimer potential is known as the single-phase approximation and simplifies the circuit in Fig.~\ref{fig_SNAIL} as it discards the dynamics of the slow internal degree of freedom.

However, note that $U(x_1, x_2)$ is minimal in the fast direction at $x_2=\Phi/2$ only if $|x_1 + \Phi| < \pi$ (discarding the periodicity). In particular, for $x_1=\pi-\Phi$, we find that $U(\pi-\Phi, x_2)$ does not depend on $x_2$, and thus $x_2$ is not unambiguously a fast variable compared to $x_1$, which might have consequences for the validity of the Born-Oppenheimer approximation. For this reason, one must also require the wave function in $x_1$ to be localized ``far enough away" from these critical points. Keeping this potential breakdown of the Born-Oppenheimer approximation in mind, we proceed with the analysis of finite but small quantum fluctuations in the fast variable $x_2$.

\subsection{Harmonic Oscillator Approach}\label{subesec_BO_HO}
In order to obtain a more accurate approximation for the Born-Oppenheimer potential than that in Eq.~\eqref{eq_V_BO_cl}, we expand the potential $U(x_1, x_2)$ up to second order in $x_2$ around its minimum in the fast direction,
\begin{equation}\label{eq_V_expanded_parabola}
    U(x_1, x_2) = U_\text{BO}^\text{cl}(x_1) + E_{J,2} \cos(\frac{x_1+\Phi}{2}) 
    \left( x_2 - \frac{\Phi}{2} \right)^2,
\end{equation}
and we omit terms of the order $\mathcal{O}[(x_2 - \Phi/2)^3]$. Then, the ground state energy of the resulting harmonic oscillator in the $x_2$-direction is taken to define the effective Born-Oppenheimer potential for $x_1$,
\begin{equation}\label{eq_V_BO_ho}
    U_\text{BO}^\text{h.o.}(x_1) =
    U_\text{BO}^\text{cl}(x_1) + 
    \sqrt{\frac{2}{k_2} E_C E_{J,2}\cos(\frac{x_1+\Phi}{2})}.
\end{equation}

As in the purely classical approach, the harmonic approximation is valid only if the wave function is well-localized in $x_2$, i.e., if $E_C/k_2 E_{J,2} \ll 1$ (the transmon limit mentioned above). Also, we again require $|x_1+\phi|<\pi$ (discarding the periodicity) in order to expand around an actual minimum and not around a maximum. Equation \eqref{eq_V_BO_ho} already improves the classical Born-Oppenheimer potential as the additional correction term takes account of the zero point energy due to finite quantum fluctuations.

The result for $U_\text{BO}^\text{h.o.}(x_1)$ can be used to improve the classical Born-Oppenheimer potential by renormalizing the Josephson energy $E_{J,2}$ instead of adding a correction term \footnote{A similar renormalization of the Josephson energy also occurs in the analysis of nonreciprocal circuits; see Sec.~\ref{SubSubSec_Gyrator_GKP}}. To this end, we note that the second order Maclaurin polynomial of both the functions $2\sqrt{\cos(\epsilon)}$ and $\cos(\epsilon)+1$ coincide. Therefore, after dropping a constant shift in energy, we approximate $U_\text{BO}^\text{h.o.}(x_1)$ as
\begin{equation}
    U_\text{BO}^\text{h.o.}(x_1) \approx
    U_\text{BO}^\text{cl}(x_1) + \sqrt{\frac{1}{2k_2} E_C E_{J,2}} \cos(\frac{x_1+\Phi}{2}).
\end{equation}
Finally, this approximation is used to identify the renormalized Josephson energy
\begin{equation}\label{eq_SNAIL_renormalization}
    \widetilde{E}_{J,2} = 
    E_{J,2} \left( 1 - \frac{1}{2} \sqrt{\frac{1}{2k_2} \frac{E_C}{E_{J,2}}} \right)
\end{equation}
such that $U_\text{BO}^\text{h.o.}(x_1) \approx \widetilde{U}^\text{cl}(x_1)$ with
\begin{equation}
    \widetilde{U}_\text{BO}^\text{cl}(x_1) = -E_{J,1} \cos(x_1) - 2\widetilde{E}_{J,2} \cos(\frac{x_1+\Phi}{2}).
\end{equation}

We conclude the analysis of small finite quantum fluctuations of the internal degrees of freedom in the SNAIL with the remark that a similar renormalization of the Josephson energy was reported in Ref.~\cite{DiPaoloBlais}. There, an effective single-mode theory for the fluxonium qubit is derived that incorporates possible capacitances to ground as well as disorder in the circuit elements. Specializing to a symmetric SNAIL with $N=2$ Josephson junctions in the array, the reported renormalization coincides with Eq.~\eqref{eq_SNAIL_renormalization} up to leading order in $\sqrt{E_C/k_2E_{J,2}}\ll1$.

\subsection{Limit of Small Internal Capacitances}\label{subsec_SNAIL_limit}
Both the Born-Oppenheimer potential based on the classical minimal energy (Sec.~\ref{subesec_BO_classical}) and that based on the harmonic-oscillator approximation (Sec.~\ref{subesec_BO_HO}) require the wave function to be localized at or close to the minimum in the fast $x_2$-direction, respectively. As discussed, this requirement is fulfilled if $E_C/k_2E_{J,2} \ll 1$. For intermediate values, $E_C/k_2E_{J,2} \simeq 1$, quantum fluctuations in $x_2$ are too large to allow for a quadratic expansion of the potential. In that case, one must solve the fast part of the Schrödinger equation numerically or attempt to find an (approximate) analytic solution.

However, for vanishingly small but finite internal capacitances, $E_C/k_2E_{J,2} \gg 1$, quantum fluctuations in $x_2$ dominate the eigenenergies of the fast Schrödinger equation. In the following, we compare this limit with the singular case ($k_2 = 0$) according to the Dirac-Bergmann algorithm. We expect the wave function to be widely extended, and therefore we first analyze its boundary conditions. Given the initial flux variables $\phi_1$ and $\phi_2$, the boundary conditions on the full two-dimensional wave function $\Psi(\phi_1,  \phi_2)$ read \footnote{Note that the boundary conditions in Eq.~\eqref{eq_SNAIL_BC_old} imply a preceding unitary gauge transformation of the initial wave function as the offset charges $\nu_1, \nu_2$ do not enter the Hamiltonian in Eq.~\eqref{eq_Hamiltonian_SNAIL}.}
\begin{subequations}\label{eq_SNAIL_BC_old}
\begin{align}
    \Psi(\phi_1 + 2\pi,  \phi_2) &= e^{i2 \pi \nu_1} \Psi(\phi_1,  \phi_2),   \\
    \Psi(\phi_1,  \phi_2 + 2\pi) &= e^{i2 \pi \nu_2} \Psi(\phi_1,  \phi_2).
\end{align}
\end{subequations}
Here, $\nu_1$ and $\nu_2$ take account of possible offset charges on the superconducting islands of the network in Fig.~\ref{fig_SNAIL}. However, we want to evaluate the wave function in the $x_1$-$x_2$-representation, and the non-orthogonal variable transformation in Eq.~\eqref{eq_SNAIL_variable_transform} imposes ``spiral" boundary conditions on $\Psi(x_1, x_2)$, namely
\begin{subequations}\label{eq_SNAIL_BC_new}
\begin{align}
    \Psi(x_1 + 2\pi, x_2 - \pi) &= e^{i2 \pi \nu_1} \Psi(x_1, x_2),      \\
    \Psi(x_1, x_2 + 2 \pi) &= e^{i2 \pi \nu_2} \Psi(x_1, x_2).
\end{align}
\end{subequations}

The Born-Oppenheimer approximation assumes that the total wave function factorizes,
\begin{equation}\label{eq_BO_wf_ansatz_SNAIL}
    \Psi(x_1, x_2) = \chi (x_1) \psi_{x_1}(x_2),
\end{equation}
where the individual factors describe the fast and the slow degree of freedom, respectively. In particular, $\psi_{x_1}(x_2)$ solves the fast part of the Schrödinger equation in which $x_1$ is treated as a fixed parameter. The resulting eigenenergy -- the Born-Oppenheimer potential -- is then used as the potential for the effective slow part of the Schrödinger equation, which is solved by $\chi(x_1)$. 

In the limit of vanishingly small internal capacitances, a convenient basis for solving the fast Schrödinger equation is set up by the plane waves
\begin{equation}
    u_n(x_2) = \frac{1}{\sqrt{2\pi}} e^{i (\nu_2+n) x_2},
    \qquad n \in \mathbb{Z},
\end{equation}
as they comply with the boundary conditions in Eq.~\eqref{eq_SNAIL_BC_new} and already diagonalize the kinetic term of the fast Hamiltonian,
\begin{equation}
    \bra{u_m} 4 E_C d_2 p_2^2 \ket{u_n} 
    = 4 E_C d_2 (\nu_2+n)^2 \delta_{m,n}.
\end{equation}
Furthermore, the potential energy $U(x_1, x_2)$ is tridiagonal in that basis. In particular, using the notation $U_{m,n} (x_1) = \bra{u_m} U(x_1, x_2) \ket{u_n}$, we find that
\begin{subequations}
\begin{align}
    U_{n,n} (x_1)
    &= - E_{J,1} \cos(x_1),      \\
    U_{n, n \pm 1} (x_1) &= - \frac{E_{J,2}}{2} 
    \left( e^{ \mp i x_1/2} + e^{\pm i x_1/2} e^{\pm i \Phi} \right),
\end{align}
\end{subequations}
while all the other matrix elements vanish. 

For simplicity, in the following analysis we focus on the case $\nu_2=0$. Then, in the limit of vanishingly small internal capacitances, $u_0(x_2)$ is a good approximation of the ground state of the fast Hamiltonian. In particular, in first order perturbation theory, the resulting Born-Oppenheimer potential for the slow $x_1$-variable,
\begin{equation}
    U_\text{BO} (x_1) = E_{J,1} \big[ 1 - \cos(x_1) \big],
\end{equation}
is essentially that of the Josephson junction shunting the Josephson junction array; see Fig.~\ref{fig_SNAIL}. Thus, we conclude that in the limit of vanishingly small intrinsic capacitances in the Josephson junction array, the entire branch comprising it can be efficiently modeled as an open circuit. We see that this is similar to the case of a type-1 inductor in series with a linear inductance; see Sec.~\ref{subsec_type_1}.

Also, the kinetic term corresponding to the $x_1$-variable, i.e., the $\phi_1$-node shunting capacitance, is in agreement with this result. In particular, in the limit of small $k_2, k_3$, we find
\begin{equation}
    d_1 E_C  \approx (1/E_C+1/E_{C,1})^{-1},
\end{equation}
which is the effective charging energy of the parallel connection of $C$ and $C_1$. This again coincides with the interpretation of the central branch in the circuit in Fig.~\ref{fig_SNAIL} being absent.

\section{Singular Nonreciprocal Circuits}\label{sec_Nonreciprocal_Circuits}
We now turn our attention to nonreciprocal networks. These networks provide many instances of singular systems. An electrical network is called nonreciprocal if the current at port $n$ due to the voltage at port $m$ is \textit{not} equal to the current at port $m$ if the same voltage is applied to port $n$. The gyrator, a linear two-port network element proposed by Tellegen in 1948 \cite{Tellegen}, is considered the most elementary nonreciprocal network element. We analyze various electrical networks containing gyrators.

In this section, we revise Tellegen's rule for the effective description of a gyrator that is shunted by a general (linear) network. First, we review Tellegen's original approach for the treatment of classical linear network elements that provide well-defined impedances. Afterwards, we present a different procedure that generalizes Tellegen's rule and enables the quantized description of general nonlinear network elements. In particular, contrary to Tellegen's approach, our analysis does not require the existence of a linear impedance, but builds on the Hamiltonian description of the system. 

\subsection{Tellegen's Replacement Rule for Linear Network Elements}\label{sec_Tellegen}

\begin{figure}[b!]  
	\centering
	\includegraphics[]{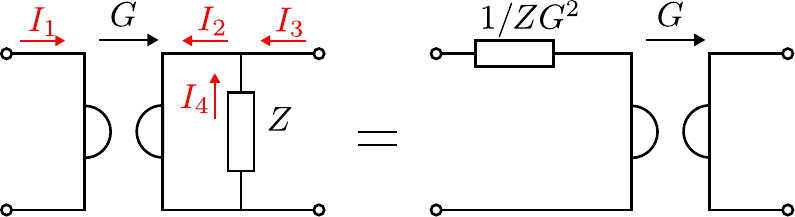}
	\caption{Equivalence of linear networks involving a gyrator.}
	\label{fig_Tellegen_linear}
\end{figure}

In Tellegen's original work \cite{Tellegen}, only linear nonreciprocal networks were considered. A very important conclusion of Tellegen is that a gyrator with gyration conductance $G$, shunted by an impedance $Z$ at its secondary port, is equivalent to a gyrator with the same gyration conductance and an impedance $1/ZG^2$ connected to a terminal of the primary port; see Fig.~\ref{fig_Tellegen_linear}. A proof of this statement is as follows: first, note the defining property of a gyrator: currents and voltages on opposite ports are related to each other via
\begin{equation}\label{eq_2nd_gyrator_relation}
    I_1 = - G V_2   , \qquad
    I_2 = G V_1     .
\end{equation}
Furthermore, the impedance at the secondary port of the gyrator provides a linear relation between voltage and current, $V_2 = Z I_4$. Finally, the proof crucially relies on Kirchhoff's law of current conservation, $I_2=I_3+I_4$; see Fig.~\ref{fig_Tellegen_linear}. Solving for $I_3$, we obtain
\begin{equation}
    I_3 
    = G V_1 - \frac{V_2}{Z} 
    = G \left( V_1 + \frac{1}{ZG^2} I_1 \right),
\end{equation}
which completes the proof. 

As a direct consequence of this replacement rule, Tellegen concluded that a gyrator terminated by a capacitance $C$ (or inductance $L$) is equivalent to an effective parallel shunting inductance $L_\text{eff}=C/G^2$ (or effective capacitance $C_\text{eff}=LG^2$). In light of our analysis above, it is clear that we need to critically examine this replacement rule for the quantum case.

\subsection{Hamiltonian Formalism for the Terminated Gyrator}
As Tellegen's replacement rule was derived based on manipulating Kirchhoff's current conservation law, we examine this rule critically in the following. In particular, we analyze the dynamics of various electrical networks involving a gyrator by means of the Hamiltonian description.

\subsubsection{Capacitive Shunt}
\begin{figure}[b!]  
	\centering
	\includegraphics[]{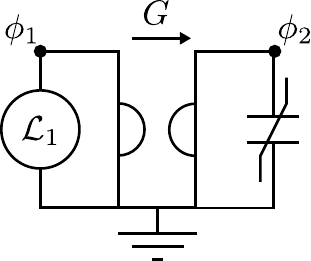}	
	\caption{Gyrator coupling a general network and a nonlinear capacitor.}
	\label{fig_Tellegen_nonlin_C}
\end{figure}

We start with the analysis of a gyrator that is terminated by a general network with Lagrangian $\mathcal{L}_1(\phi_1, \dot{\phi}_1)$ (discarding unimportant internal degrees of freedom) and a generic (potentially nonlinear) capacitor; see Fig.~\ref{fig_Tellegen_nonlin_C}.

Focusing on voltage-controlled capacitors \cite{Peikari}, the total Lagrangian of the system can be written as \cite{Rymarz_Master, Duinker, FestschriftChua, Parra4}
\begin{equation}\label{eq_Lag_nonlin_C}
    \mathcal{L} = 
    \mathcal{L}_1(\phi_1, \dot{\phi}_1)
    + G \phi_1 \dot{\phi}_2 
    + g ( \dot{\phi}_2 ),
\end{equation}
where the function $g(\dot{\phi}_2)$ accounts for the energy stored in the capacitor on the secondary port. Insisting on a monotone capacitor \cite{Peikari} ensures the invertibility of $g'(\dot{\phi}_2)$. In the following, we require the Lagrangian $\mathcal{L}_1(\phi_1, \dot{\phi}_1)$ to be regular. Note that such a requirement was not necessary in Tellegen's original consideration of linear networks; see Sec.~\ref{sec_Tellegen}.

In order to proceed, we move to the Hamiltonian formalism. We define the canonical momenta
\begin{equation}
    Q_1 = \frac{\partial \mathcal{L}_1}{\partial \dot{\phi}_1},
    \qquad
    Q_2 = G \phi_1 + g'(\dot{\phi}_2),
\end{equation}
which can be solved for the generalized velocities $\dot{\phi}_i$; in particular, we find
\begin{equation}\label{eq_phi2dot_nonlin_C}
    \dot{\phi}_2 = {g'}^{-1} (Q_2 - G \phi_1).
\end{equation}
Thus, we can derive the Hamiltonian via an ordinary Legendre transformation, and we obtain
\begin{equation}\label{eq_Hamiltonian_comparison_2}
\begin{split}
    H
    &= Q_1 \dot{\phi}_1 + Q_2 \dot{\phi}_2 - \mathcal{L}    \\
    &= H_1 (\phi_1, Q_1)
    + ( Q_2 - G \phi_1 ) {g'}^{-1} ( Q_2 - G \phi_1 ) \\
    &\qquad - g \left( {g'}^{-1} ( Q_2 - G \phi_1 ) \right),
\end{split}
\end{equation}
which does not depend on $\phi_2$. As a consequence, since $[H, Q_2] = 0$ (or $\partial H / \partial \phi_2 = 0$), we conclude that we can treat $Q_2$ as a constant parameter. Therefore, the total electrical network is equivalent to the network $\mathcal{L}_1$ that is shunted by an additional effective inductor.

To compare with Tellegen's results, we consider a linear shunting capacitance, $g ( \dot{\phi}_2 ) = C \dot{\phi}_2^2/2$, which results in the Hamiltonian
\begin{equation}\label{eq_Tellegen_capacitance_offset}
    H = H_1 (\phi_1, Q_1) + \frac{(\phi_1 - Q_2/G)^2}{2C/G^2}.
\end{equation}
For $Q_2=0$, this result agrees with Tellegen's conclusion that the capacitively terminated gyrator is equivalent to an effective inductance $L_\text{eff} = C/G^2$. However, for $Q_2 \neq 0$, there is a finite shift in the last term of Eq.~\eqref{eq_Tellegen_capacitance_offset} that can be interpreted as external magnetic flux through a loop formed by the electrical network at the primary port of the gyrator and the effective inductance. If the general network described by $\mathcal{L}_1$ (or $H_1$) is linear, the value of $Q_2$ has no effect. If, however, the system is nonlinear, the result based on the Hamiltonian formalism can differ from Tellegen's conclusion. 

\subsubsection{Inductive Shunt}\label{sec_inductive_shunt}
\begin{figure}[b!]  
	\centering
	\includegraphics[]{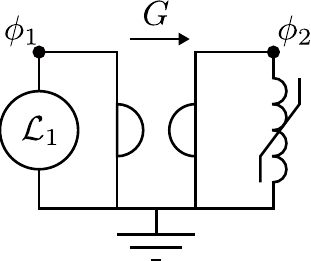}	
	\caption{Gyrator coupling a general network and a nonlinear inductor.}
	\label{fig_Tellegen_nonlin_L}
\end{figure}

Next, we consider a gyrator that is terminated by a general network with Lagrangian $\mathcal{L}_1(\phi_1, \dot{\phi}_1)$ and a generic (potentially nonlinear) inductor, respectively; see Fig.~\ref{fig_Tellegen_nonlin_L}.

We restrict our considerations to the analysis of general flux-controlled inductors \cite{Peikari} that store energy $f(\phi_2)$. Thus, the system is described by the total Lagrangian
\begin{equation}\label{eq_Lagrangian_nonlinear_L}
    \mathcal{L} = 
    \mathcal{L}_1(\phi_1, \dot{\phi}_1)
    - G \dot{\phi}_1 \phi_2
    - f ( \phi_2 ).
\end{equation}
Even though we assume the Lagrangian $\mathcal{L}_1(\phi_1, \dot{\phi}_1)$ to be regular, the total Lagrangian is still singular due to the absence of a shunting capacitance at the secondary port of the gyrator. Because of the absence of $\dot{\phi}_2$ in the Lagrangian due to the chosen gauge of the gyrator \footnote{We have confirmed that the Dirac-Bergmann result here is gauge invariant.}, the corresponding conjugate charge vanishes ($Q_2=0$), and, according to the Dirac-Bergmann algorithm (Appendix~\ref{sec_general_formalism}), we use the classical Euler-Lagrange equation of motion (recovering Kirchhoff's law of current conservation),
\begin{equation}\label{eq_gyrator_nonlin_inductance_Kirchhoff}
    0 = G \dot{\phi}_1 + f'(\phi_2),
\end{equation}
to find an expression for $\phi_2$ as a function of $\dot{\phi}_1$. Assuming the general inductor to be monotone \cite{Peikari}, we find:
\begin{equation}
    \phi_2 = {f'}^{-1} (-G\dot{\phi}_1).
\end{equation}
Inserting this expression in the Lagrangian in Eq.~\eqref{eq_Lagrangian_nonlinear_L}, we obtain the one-dimensional Lagrangian
\begin{equation}\label{eq_1d_Lagrangian_nl_L}
    \mathcal{L} = 
    \mathcal{L}_1(\phi_1, \dot{\phi}_1)
    - G \dot{\phi}_1 {f'}^{-1} (-G\dot{\phi}_1)
    - f ( {f'}^{-1} (-G\dot{\phi}_1) ),
\end{equation}
which is equivalent to the Lagrangian of the network $\mathcal{L}_1$ with an additional effective nonlinear shunting capacitor.

In order to derive the corresponding Hamiltonian, one must evaluate the canonical charge $Q_1 = \partial \mathcal{L} / \partial \dot{\phi}_1$, which depends not only on $\mathcal{L}_1$ but also on the function $f(\phi_2)$. As an example, consider a nonlinear inductor that is described by $f(\phi_2) = \beta |\phi_2|^\gamma$ with $\beta > 0$ and $\gamma > 1$. For this specific choice of the inductor, Eq.~\eqref{eq_1d_Lagrangian_nl_L} reduces to
\begin{equation}
    \mathcal{L} = \mathcal{L}_1(\phi_1, \dot{\phi}_1) 
    + \frac{(\gamma - 1)G^{\gamma/(\gamma-1)}}{\gamma(\beta \gamma)^{1/(\gamma-1)}} 
    |\dot{\phi}_1|^{\gamma/(\gamma-1)}.
\end{equation}

To compare with Tellegen's results, we consider a linear shunting inductance, $f(\phi_2)=\phi_2^2/2L$, which results in the Lagrangian 
\begin{equation}
    \mathcal{L} = \mathcal{L}_1(\phi_1, \dot{\phi}_1) + L G^2 \dot{\phi}_1^2/2.
\end{equation}
This result agrees with Tellegen's conclusion that the inductively terminated gyrator is equivalent to an effective capacitance $C_\text{eff}=LG^2$.

\subsubsection{Approach to Singularity: Nonlinear Resonator Shunting a Gyrator}\label{sec_nl_L_with_C}
We consider the circuit of the preceding section with a (parasitic) capacitance $C_2$ added, then consider the limit $C_2\rightarrow 0$; we again see that this physical limit differs from the prediction of Dirac-Bergmann that we have just obtained.

\begin{figure}[t!]  
	\centering
	\includegraphics[]{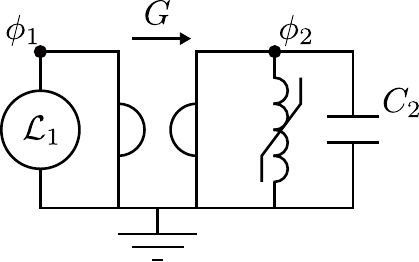}	
	\caption{Gyrator coupling a general network and a nonlinear resonator.}
	\label{fig_Tellegen_nonlin_L_with_Cshunt}
\end{figure}

Thus, we must analyze the circuit consisting of a gyrator coupling a generic electrical network at one port with a nonlinear resonator at the other port; see Fig.~\ref{fig_Tellegen_nonlin_L_with_Cshunt}. The resonator contains a linear capacitance and a nonlinear flux-controlled inductor; the system is described by the total Lagrangian 
\begin{equation}
    \mathcal{L} = 
    \mathcal{L}_1(\phi_1, \dot{\phi}_1)
    - G \dot{\phi}_1 \phi_2
    + \frac{C_2 \dot{\phi}_2^2}{2} - f ( \phi_2 ).
\end{equation}
Defining the conjugate charges $Q_i = \partial \mathcal{L} / \partial \dot{\phi}_i$, and assuming the generic electrical network $\mathcal{L}_1(\phi_1, \dot{\phi}_1)$ to be regular, the total Lagrangian is regular as well, and the corresponding Hamiltonian reads
\begin{equation}\label{eq_nonreciprocal_H_for_BO_approx}
    H 
    = H_1(\phi_1, Q_1+G\phi_2)
    + \frac{Q_2^2}{2C_2} 
    + f(\phi_2).
\end{equation}

We proceed to study the limit of a vanishingly small capacitance at the secondary port, i.e., $C_2 \rightarrow 0$. Moreover, for concreteness, we specify the electrical network at the primary port of the gyrator to be an $LC$-resonator, i.e., $\mathcal{L}_1(\phi_1, \dot{\phi}_1) = C_1 \dot{\phi}_1^2/2 - \phi_1^2/2L_1$, so that
\begin{equation}\label{eq_LC_gyrator_nonlin_LC}
    H 
    = \frac{(Q_1+G\phi_2)^2}{2C_1} 
    + \frac{Q_2^2}{2C_2} 
    + \frac{\phi_1^2}{2L_1}
    + f(\phi_2).
\end{equation}
Identifying $\phi_1$ and $\phi_2$ as slow and fast variables, respectively, the Born-Oppenheimer approximation is applicable, and we treat $\phi_1$ and $Q_1$ as fixed parameters; cf.~Sec.~\ref{sec_BO_approx_introduction} and compare Eq.~\eqref{eq_LC_gyrator_nonlin_LC} with Eq.~\eqref{eq_Hamiltonian_r}. 
In the following, we sequentially analyze each case in which the nonlinear inductor described by $f(\phi_2)$ is categorized as one of the types that we introduced in Sec.~\ref{sec_types_of_inductors}.

Specializing to a generic nonlinear inductor of type~1, it follows from a modification of Theorem~1 that in the limit $C_2 \rightarrow 0$, the system is effectively described by
\begin{equation}
    H_{\text{eff}} 
    = \frac{\phi_1^2}{2L_1},
\end{equation}
i.e., the nonlinear resonator at the secondary port of the gyrator is effectively replaced by a nonlinear inductor approaching an open circuit. Then, in accordance with the Tellegen rule, the $LC$-resonator at the primary port of the gyrator is effectively shunted by a non-linear capacitor that approaches a short.

Similarly, for a generic nonlinear inductor of type~2, it follows from a modified version of Theorem~3 that in the limit $C_2 \rightarrow 0$, the system is effectively described by
\begin{equation}
    H_{\text{eff}} 
    = \frac{Q_1^2}{2C_1} 
    + \frac{\phi_1^2}{2L_1},
\end{equation}
i.e., the nonlinear resonator at the secondary port of the gyrator is effectively replaced by a short circuit. Thus, the $LC$-resonator at the primary port of the gyrator is unaffected by the rest of the circuit in the limit $C_2 \rightarrow 0$.

Finally, for a generic type-L nonlinear inductor satisfying
$\lim_{\phi_2 \to \pm \infty} f(\phi_2)/\phi_2^2=1/{2 \mathfrak{L}}$ with $\mathfrak{L} > 0$, a straightforward modification of Theorem~4 states that in the limit $C_2 \rightarrow 0$, the system is effectively described by
\begin{equation}
    H_{\text{eff}} 
    = \frac{Q_1^2}{2(C_1 + \mathfrak{L} G^2)} 
    + \frac{\phi_1^2}{2 L_1},
\end{equation}
i.e., the nonlinear resonator at the secondary port of the gyrator is effectively replaced by a linear inductance $\mathfrak{L}$. Then, as given by the Tellegen rule, the $LC$-resonator at the primary port of the gyrator is effectively shunted by a linear capacitance $\mathfrak{L}G^2$.

The above results for a nonlinear inductor of type~1, type~2 or type~L, respectively, are all in contradiction to the singular Dirac-Bergmann description of a purely inductively shunted gyrator; see Sec.~\ref{sec_inductive_shunt}. In particular, the dynamics of the singular description (Fig.~\ref{fig_Tellegen_nonlin_L}) strongly depend on the details of the nonlinear inductor [cf.~Eq.~\eqref{eq_1d_Lagrangian_nl_L}], and a classification into one of the three types that we provide does not suffice for a complete description of the system. The disagreement in the dynamics between the singular circuit description and that of the regular one in the limit $C_2 \rightarrow 0$ originates from the fact that the Dirac-Bergmann algorithm reduces to an elimination of variables using the classical equation of motion in Eq.~\eqref{eq_gyrator_nonlin_inductance_Kirchhoff}. Thus, for the singular case with absent shunting capacitance ($C_2 = 0$), the Dirac-Bergmann algorithm completely misses the quantum fluctuations in $\phi_2$, which in the regular case ($C_2 > 0$) become large as $C_2 \rightarrow 0$.

Note that this observation is in full agreement with our analysis of singular reciprocal networks in Sec.~\ref{sec_failure_Dirac}, and it illustrates that one must not change the topology of electrical circuits based on classically obtained laws.

\subsubsection{Gyrator Coupling Two Singular Networks}\label{SubSubSec_Gyrator_GKP}
\begin{figure}[t!]  
	\centering
	\includegraphics[]{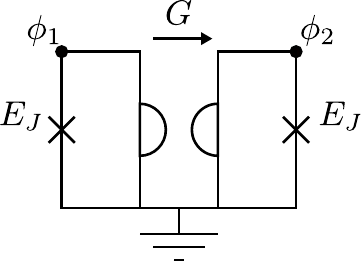}
	\caption{Two Josephson junctions coupled by a gyrator. For simplicity, the terminals of the gyrator are commonly grounded and the Josephson junctions are identical.}
	\label{fig_JJGJJ}
\end{figure}

Similar inconsistencies between the results of the Dirac-Bergmann algorithm and a regularized approach, which removes the singularities by adding tiny capacitances, occur in the quantization of electrical networks that constitute two singular circuits that are coupled by a gyrator. Here, we look only at the case of a gyrator that is terminated by a Josephson junction at each port; see Fig.~\ref{fig_JJGJJ}.

The total Lagrangian of this network,
\begin{equation}
    \mathcal{L} = - G \dot{\phi}_1 \phi_2
    + E_{J,1} \cos(\frac{2\pi}{\Phi_0}\phi_1)
    + E_{J,2} \cos(\frac{2\pi}{\Phi_0}\phi_2)
\end{equation}
is singular, and the application of the Dirac-Bergmann algorithm (Appendix~\ref{sec_general_formalism}) results in the Hamiltonian
\begin{equation}\label{eq_Harper_Hamiltonian}
    H = 
    - E_{J,1} \cos \left( \frac{2\pi}{\Phi_0} \phi_1 \right)
	- E_{J,2} \cos \left( \frac{2\pi}{\Phi_0} \frac{Q_1}{G} \right),
\end{equation}
which depends on one pair of conjugate variables $\phi_1$ and $Q_1$ only and is known as the almost Mathieu operator \cite{BellissardSimon}. The Schrödinger equation associated with the almost Mathieu operator is not a differential equation but a finite-difference equation, which, in the setting of a crystal electron in a magnetic field, is known as the Harper equation. For the special choice $G=\pi\hbar/\Phi_0^2$, the system hardware-encodes a continuous variable quantum error-correcting code as the Hamiltonian in Eq.~\eqref{eq_Harper_Hamiltonian} becomes the GKP Hamiltonian \cite{Rymarz, GKP}.

If, however, internal capacitances of the Josephson junctions are included, the system becomes regular, and it is described by a Hamiltonian that depends on two pairs of conjugate variables. In the limit of small capacitances, a low-energy projection similar to the lowest Landau-Level projection (corresponding to the Born-Oppenheimer approximation) reveals a rescaling of the Josephson energy that is missed by the singular treatment; see Ref.~\cite{Rymarz} for more details.

\subsection{Two Cascaded Gyrators -- Tellegen's Construction of a Transformer}\label{sec_Tellegen_transformer}
\begin{figure}[t!]  
	\centering
	\subfloat{\parbox{0.85\linewidth}{a) \hspace{30pt} \hfill \null \\[-3ex] \includegraphics[]{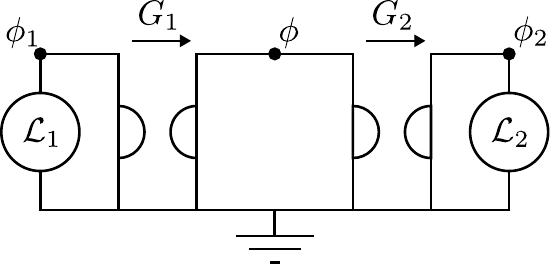}\label{fig_Transfomer_a}}}			\\
	\vspace{15pt}
	\centering
	\subfloat{\parbox{0.55\linewidth}{b) \hspace{30pt} \hfill \null \\[-3ex] \includegraphics[]{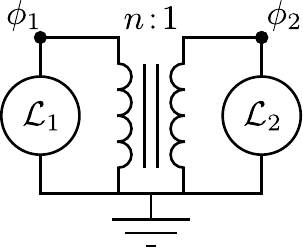}\label{fig_Transfomer_b}}}
	\label{fig_Transfomer}
	\caption{a) Two general electrical networks with Lagrangians $\mathcal{L}_{1,2}$ coupled by two cascaded gyrators. For simplicity, all terminals of the gyrators are commonly grounded. The Lagrangian describing the circuit is singular. b) Tellegen's equivalent circuit of a transformer coupling two general electrical networks with Lagrangians $\mathcal{L}_{1,2}$.}
\end{figure}

In addition to the replacement rule for a gyrator that is terminated by a linear one-port network element, Tellegen also observed \cite{Tellegen} from Kirchhoff's current conservation law that two cascaded gyrators with gyration conductances $G_1$ and $G_2$, respectively, are equivalent to a transformer with turns ratio $n$ (see Fig.~\ref{fig_Transfomer}), 
\begin{equation}
n = G_2/G_1.
\label{turns}
\end{equation}

In the following, based on the Lagrangian as well as the Hamiltonian formalism, we analyze two general electrical networks $\mathcal{L}_1$ and $\mathcal{L}_2$ that are coupled by such a cascade of gyrators. In particular, similar to our analysis above, we compare the singular treatment to the regularized case. Again we find that the two treatments can give significantly different results.

\subsubsection{Dirac-Bergmann Treatment}
Choosing a convenient gauge for the involved gyrators, the circuit in Fig.~\ref{fig_Transfomer_a} is described by the total Lagrangian
\begin{equation}\label{eq_Lagrangian_transformer}
		\mathcal{L} = \mathcal{L}_1(\phi_1, \dot{\phi}_1) + \mathcal{L}_2(\phi_2, \dot{\phi}_2)
		+ (G_2 \dot{\phi}_2 - G_1 \dot{\phi}_1) \phi,
\end{equation}
in which we discard the internal degrees of freedom of the electrical networks described by the Lagrangians $\mathcal{L}_1$ and $\mathcal{L}_2$, which we assume to be regular. The Lagrangian $\mathcal{L}$ in this gauge is singular due to the absence of $\dot{\phi}$. The Dirac-Bergmann algorithm involves the evaluation of the classical Euler-Lagrange equation of motion for the $\phi$-degree of freedom,
\begin{equation}\label{eq_def_transformer}
    \dot{\phi}_1 = \frac{G_2}{G_1} \dot{\phi}_2.
\end{equation}
In fact, recalling that $\dot{\phi}_i = V_i$, this equation defines an ideal transformer with turns ratio $n$ [Eq.~\eqref{turns}].

Although Eq.~\eqref{eq_def_transformer} cannot be solved for $\phi$, we can use it to eliminate the dependence of the total Lagrangian $\mathcal{L}$ on both $\phi$ and $\phi_1$ (or $\phi_2$, equivalently). We insert the integration over time, 
\begin{equation}\label{eq_transformer_with_c}
    \phi_1 = n \phi_2 + c,
\end{equation}
in the total Lagrangian and obtain
\begin{equation}\label{eq_Lagrangian_singular_transformer}
	\mathcal{L} = \mathcal{L}_1(n \phi_2 + c, n\dot{\phi}_2) + \mathcal{L}_2(\phi_2, \dot{\phi}_2).
\end{equation}
Note that $c$ is a yet unspecified constant that is fixed by the initialization of the system and can be interpreted as an effective external magnetic flux through a loop. For electrical networks $\mathcal{L}_1$ and $\mathcal{L}_2$ that contain only linear inductances, the constant $c$ has no effect and can be set to zero. However, if nonlinear inductors are involved, e.g., Josephson junctions, the actual value of $c$ plays a crucial role and can change the dynamics of the system. The physical origin of $c$ can be understood by a regularized treatment of the system.

\subsubsection{Regular Case}
In the following, we modify the previously analyzed electrical network in Fig.~\ref{fig_Transfomer_a}, considering a small but finite linear capacitance $C$ between the node $\phi$ and ground. The Lagrangian in Eq.~\eqref{eq_Lagrangian_transformer} is modified by an additional term $C \dot{\phi}^2/2$ that lifts the singularity of the system. As a result, the Hamiltonian is straightforwardly obtained by an ordinary Legendre transformation and reads
\begin{equation}\label{eq_total_H_regular_transformer}
    H = 
    H_1(\phi_1, Q_1-G_1\phi)
    + H_2(\phi_2, Q_2+G_2\phi)
    + \frac{Q^2}{2C}.
\end{equation}
As in our analysis above, we are interested in the limit of a vanishingly small capacitance $C$. In this regime we can perform a Born-Oppenheimer approximation treating all the variables as constant parameters except from $\phi$ and $Q$. After a shift of the fast variable $\phi$, the potential for this degree of freedom is given by
\begin{equation}\label{eq_Ueff_transformer}
    U_\text{fast}(\phi)
    = H_1(\phi_1, Q_1 + (G_1/G_2) Q_2 - G_1\phi)
    + H_2(\phi_2, G_2\phi).
\end{equation}
Note that the generalized momenta $Q_1$ and $Q_2$ do not appear individually in $U_\text{fast}(\phi)$, but only in the linear combination $Q_1 + (G_1/G_2) Q_2$. Thus, both the Born-Oppenheimer potential and the effective Hamiltonian for the slow degrees of freedom depend on this linear combination $Q_1 + Q_2/n$ only [cf.~Eq.~\eqref{turns}]. As a consequence, the effective Hamiltonian for the slow degrees of freedom commutes with $\phi_1 - n\phi_2$. Therefore, this expression is constant in time, and we reproduce the result in Eq.~\eqref{eq_transformer_with_c}. 

However, it is clear from our analysis above that for nonlinear electrical networks $\mathcal{L}_i$, the Hamiltonian obtained from the Lagrangian in Eq.~\eqref{eq_Lagrangian_singular_transformer} generally differs from the effective Hamiltonian that results from the Born-Oppenheimer approximation applied to Eq.~\eqref{eq_total_H_regular_transformer}, even if the limit $C \rightarrow 0$ is considered. In fact, the regularized approach not only takes account of the additional constant of motion that is missed by Tellegen's replacement rule, but it also captures the quantum fluctuation in $\phi$ that is neglected by the the Dirac-Bergmann algorithm in the singular case.

\section{Conclusion and Outlook}\label{sec_conclusion}
\begin{figure}[t!]  
	\centering	
	\includegraphics[width=0.8\textwidth]{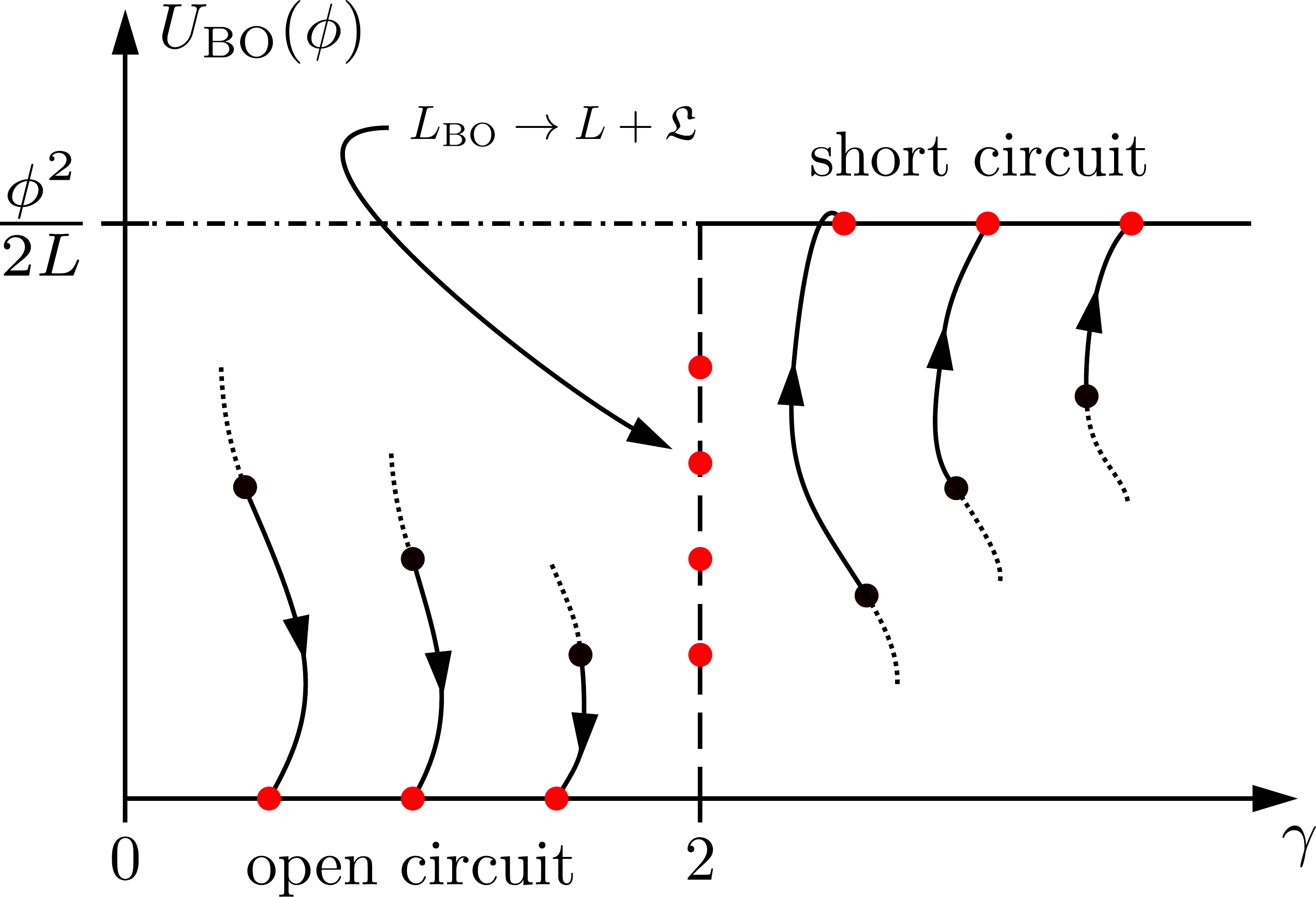}
	\caption{A conceptual flow diagram describing the fixed-point structure of a nearly-singular circuit. Referring to Fig.~\ref{fig_Non_Linear_Inductance}, flows are parameterized by the small capacitor $C'$, with the fixed points (red dots) reached when $C'\rightarrow 0$. Actual computations of the flows are performed by finding the ground state of the fast-variable Schr{\"o}dinger equation in the Born-Oppenheimer treatment. The object that flows is the entire Born-Oppenheimer potential function $U_{BO}(\cdot)$, exemplified here as just one parameter, the value $U_{BO}(\phi)$ at some particular value of the slow variable $\phi$. The flows are shown for different forms of the potential of the nonlinear inductor, schematized here by parameter $\gamma$, as in the potential form $U_{nl}(\phi_c)\sim|\phi_c|^\gamma$; cf. Eq.~\eqref{eq_family_type_2_inductors}. Flows start at some nonuniversal values (black dots); the flows are in reality in a high-dimensional parameter space, and are not necessarily monotone, as schematized by the waviness of the flow lines. We have show that sublinear cases ($0 < \gamma < 2$, type 1) all flow to the universal open-circuit fixed point with $U_{BO}(\phi)=0$; superlinear cases ($\gamma>2$, type 2) all flow to the universal short-circuit fixed point with $U_{BO}(\phi)=\phi^2/2L$. Linear or nearly linear cases (type~L) remain fixed at $\gamma=2$, flowing to a point on the dashed line determined by the starting parameter $\mathfrak{L}$; cf. Theorem 4.}
	\label{fig_RG_flow}
\end{figure}

Taking a final look at the fixed-point structure that our work has uncovered, we offer a schematic ``flow diagram" in Fig.~\ref{fig_RG_flow}. We of course do not use the tools of renormalization group theory here, but flows are well defined in our work, obtained implicitly from the calculation of ground state energies of fast-variable Hamiltonians in the Born-Oppenheimer treatment. While these flows are in a function space (i.e., infinite dimensional), they can usefully be schematized in the two-dimensional space shown.

Indeed, we have striven to render this illustration so that it resembles the well-known renormalization group flows and fixed point structure of the quantum conductance of the 1D interacting Luttinger liquid, as reported by Kane and Fisher \cite{KaneFisher}. At some level, the resemblance is accidental, as the physical problems considered are very different. But there is an intriguing similarity in the phenomenology of the two cases: in Kane and Fisher, the fixed-point cases are those of perfect quantum conductors vs.~insulators, resembling our short- and open-circuit fixed points. Their marginal case is the noninteracting electron case, closely analogous to the ``type-L" case of our work. Our flows have, in some sense, a higher level of complexity in that they are not captured by just the two parameters shown. But it would be intriguing to consider whether the two problems have some deeper level of resemblance.

To give a final perspective on our work: we present a general approach, based on Born-Oppenheimer theory, to derive the Hamiltonian description of a large class of electrical networks, both reciprocal and nonreciprocal, which are nearly singular due to the occurrence of a hierarchy of capacitance values. We also offer a full development of the well-known treatment of ``exactly singular" systems, as applied to our electric circuits, 
based on the Dirac-Bergmann algorithm. 
We compare the quantum dynamics obtained by the two treatments. The discrepancy of the two approaches is absolute: Dirac-Bergmann is a failure. In the language of electric circuit theory, this main finding of our work can be stated as follows:

\textit{The equations of motion obtained from the Lagrangian description of an electrical network must coincide with Kirchhoff's laws of current (voltage) conservation for each node (loop). However, these classical conservation laws must {\bf\em not} be used to eliminate variables or to change the topology of the electrical network.}

In particular, such an elimination of variables on the Lagrangian level is a widespread mistake that is frequently made without justification or knowledge of the consequences, which often results in a wrong description of the system. The classical elimination of variables, motivated by the singularity of the Lagrangian, must be deemed an incorrect procedure as it completely misses quantum-fluctuation effects, which become large as the singular limit is approached. 

As an alternative, we provide the techniques to analyze the regularized system in the singular limit. More specifically, we generalize results from analytic perturbation theory to study large classes of generic nonanalytic perturbations. This allows us to classify network elements with the same effective quantum dynamics as the singular limit is approached, i.e., in the limit of a small intrinsic capacitance, identifying the dominant idea of fixed-point behavior as the correct overall paradigm. 

We finally feature one direction in which our work is incomplete, and offers intriguing new questions for circuit quantization. We have only very slightly touched on what happens if nonlinear capacitors occur in the circuit. While this has not been viewed as very relevant for superconducting qubit circuits, the phenomenon of quantum capacitance could provide a route to such nonlinearities \cite{Graphene_Capacitance_Fang, Graphene_Capacitance_Xia, Graphene_Capacitance_Stampfer}. More relevant may be the phenomenon of quantum phase slip junctions; it is argued \cite{PhysRevA.100.062321} that such junctions act as effective nonlinear capacitors, whose contribution to the kinetic energy has the interesting convex functional form $\,\propto\dot x\arcsin{\dot x}+\sqrt{1-\dot x^2}$, with the dimensionless generalized velocity $\dot x=\dot\phi/V_c$. 

When considering not only parasitic capacitances but parasitic inductances, we note instances of circuits that, even though formally nonsingular, show problems in treating small parameter values going to zero. We expect that a full study of both node (flux) quantization and the dual (loop charge) quantization is necessary to fully understand these problems \cite{Ulrich}. Finally, we point out that if one had strongly nonlinear, nonconvex, characteristics in both inductor and capacitor in a simple resonator, no consistent performance of the Legendre transformation could be done to obtain a Hamiltonian for this circuit; this leaves open the question of how one should quantize this perfectly well defined lossless classical circuit.

To conclude, our work highlights the importance of critically examining the validity of well-tried theorems or simplifications from classical network synthesis, which build on Kirchhoff's conservation laws, prior to applying circuit quantization. We can envision our work to provide the basis for the quantization of unconventional electrical networks yet to be designed that, e.g., involve nonreciprocal elements or more general nonlinear elements going beyond the Josephson junction. Encouraged by the continuous progress in fabricating novel network elements such as on-chip nonreciprocal devices \cite{Rosenthal, Chapman, Lecocq, Barzanjeh, Mahoney}, nonlinear kinetic inductances \cite{Grimsmo_DPA, Manucharyan_nanowire} or nonlinear quantum capacitances \cite{Graphene_Capacitance_Fang, Graphene_Capacitance_Xia, Graphene_Capacitance_Stampfer}, continuing development of the theoretical description of electrical networks containing such elements is highly motivated by the vast new possibilities they offer. Perhaps, someday, superconducting qubits with a nonlinear capacitor \cite{Khorasani_1, Khorasani_2} or intrinsically protected qubits based on the nonreciprocity of the device \cite{Rymarz} might open new, exciting pathways to the realization of a large-scale quantum processor.

\section{Acknowledgments}
We are most grateful to our colleague Fabian Hassler, for many important suggestions and impulses at the early stages of the work reported here, as well as for suggesting the use of Youla's decomposition to effectively represent anti-symmetric matrices.
We thank Daniele Tampieri of TTlabs for insightful observations about convergence criteria for series.
We thank Barry Simon for pointing out Ref.~\cite{Simon_article_long}.
Thanks also to Alex Altland for observing the resemblance of the present work to the renormalization group analysis of Kane and Fisher. We acknowledge support from the Deutsche Forschungsgemeinschaft (DFG, German Research Foundation) under Germany's Excellence Strategy - Cluster of Excellence Matter and Light for Quantum Computing (ML4Q) EXC 2004/1 - 390534769. DDV gratefully acknowledges funding by the German Federal Ministry
of Education and Research within the funding program "Photonic Research
Germany" under contract number 13N14891, and within the funding program "Quantum Technologies - From Basic Research to the Market" (project GeQCoS), contract number
13N15685.

\appendix

\section{Dirac-Bergmann Algorithm for Singular Nonreciprocal Circuits}\label{sec_general_formalism}
In this appendix, which is based on the results presented in Ref.~\cite{Rymarz_Master}, we apply the Dirac-Bergmann algorithm to derive the Hamiltonian description of general electrical networks without ohmic contacts that can be both singular and nonreciprocal. Since the nonreciprocity of any classical linear electrical network can be captured by gyrators as only nonreciprocal circuit elements, their Lagrangian contribution is the first step towards the Hamiltonian formalism.

The Lagrangian of a general electrical network with $m+1$ nodes that is built of linear capacitances, gyrators and any sort of inductances (linear, nonlinear and mutual) can be written in matrix notation as \cite{Rymarz_Master, Duinker, FestschriftChua, Parra4}
\begin{equation}\label{eq_general_network_Lagrangian}
	\mathcal{L} 
	= \frac{1}{2} \dot{\bm{\phi}}^T \bm{C} \dot{\bm{\phi}}
	+ \dot{\bm{\phi}}^T \bm{A} \bm{\phi} 
	- U(\bm{\phi}),
\end{equation}
where due to Kirchhoff's voltage law the electrical network is fully described by at most $m$ independent variables, which we write as $\bm{\phi} = (\phi_1, \ldots, \phi_m)^T$. The Lagrangian contributions of the capacitances, gyrators and inductances are captured by matrices $\bm{C}$, $\bm{A}$ and function $U(\bm{\phi})$, respectively. Since the contribution of the gyrators to the Lagrangian in Eq.~\eqref{eq_general_network_Lagrangian} is akin to that of a vector potential, we will call $\bm{A}$ the vector potential matrix. However, note that the actual analog to the vector potential is given by the product $\bm{A} \bm{\phi}$. As an example, in $m=3$ dimensions, the components of the resulting magnetic field analog read
\begin{equation}
	B_i = \sum_{\mu,\nu=1}^3 \epsilon_{i\mu\nu} A_{\nu\mu},
	\qquad i = 1,2,3.
\end{equation}
Also, the potential $U(\bm{\phi})$ can depend on external magnetic fluxes threading superconducting loops. 

\subsection{Regular Case}
First, we consider the Lagrangian in Eq.~\eqref{eq_general_network_Lagrangian} to be regular, i.e., we assume that the capacitance matrix $\bm{C}$ is invertible. In this case, the Lagrangian is not singular and one does not need to employ the Dirac-Bergmann algorithm (there are no primary constraints). In fact, the corresponding Hamiltonian is obtained by an ordinary Legendre transformation, resulting in
\begin{equation}\label{eq_general_non_singular_Hamiltonian}
	H = \frac{1}{2}
	\big( \bm{Q} - \bm{A} \bm{\phi} \big)^T \bm{C}^{-1} \big( \bm{Q} - \bm{A} \bm{\phi}\big)
	+ U(\bm{\phi}),
\end{equation}
where the canonical conjugate charges are defined as
\begin{equation}\label{eq_conjugate_charges}
	Q_\nu = \frac{\partial \mathcal{L}}{\partial \dot{\phi}_\nu} ,
	\qquad \nu = 1, 2, \ldots m.
\end{equation}
The Hamiltonian in Eq.~\eqref{eq_general_non_singular_Hamiltonian} is quantized by imposing the canonical commutation relation $[\phi_\mu, Q_\nu] = i \hbar \delta_{\mu \nu}$ for $\mu, \nu = 1, 2, \ldots m$. Clearly, the nonreciprocity of the circuit causes a coupling of flux and charge variables.

\subsection{Singular Case}
Next, we allow the Lagrangian in Eq.~\eqref{eq_general_network_Lagrangian} to be singular. In the following, we derive the associated Hamiltonian, which, in the limit of non-singularity, reduces to the Hamiltonian in Eq.~\eqref{eq_general_non_singular_Hamiltonian}. In Appendix~\ref{subsection_basis_construction}, we present a general procedure to construct a basis in which the capacitance matrix $\bm{C}$ is block diagonal and reads
\begin{equation}\label{eq_desired_C}
	\bm{C}
	=
	\left(
		\begin{array}{c|c}
			\bm{C'}							&		\bm{0}_{n \times k}		\\ \hline
			\bm{0}_{k \times n}		&		\bm{0}_{k \times k}
		\end{array}
	\right),
	\qquad
\end{equation}
where $\bm{C'}$ is a symmetric, positive definite $n \times n$ matrix and therefore invertible. Thus, the trivial action of $\bm{C}$ on its $k$-dimensional kernel is separated from the remaining $n$-dimensional subspace, satisfying the rank-nullity theorem $n+k=m$. Simultaneously, in the same basis, the vector potential matrix $\bm{A}$ is block upper triangular and given by
\begin{equation}\label{eq_desired_A}
	\bm{A}
	=
	\left(
		\begin{array}{c|c}
			\bm{A'}					&		\widetilde{\bm{A}}		\\ \hline
			\bm{0}_{k \times n}		&		\bm{A''}
		\end{array}
	\right),
\end{equation}
where $\bm{A'}$ can be chosen to be an anti-symmetric $n \times n$ matrix and $\widetilde{\bm{A}}$ is a not further specified $n \times k $ matrix. The $k \times k$ matrix $\bm{A''}$, which exclusively acts on the kernel of $\bm{C}$, is of the form
\begin{equation}\label{eq_Lambda}
	\bm{A''}
	=
	\left(
		\begin{array}{c|c|c}
			\bm{0}_{l \times l}	&	\bm{d} 					&	\bm{0}_{l \times j}		\\ \hline
			\bm{0}_{l \times l}	&	\bm{0}_{l \times l} &	\bm{0}_{l \times j}		\\ \hline	
			\bm{0}_{j \times l}	&	\bm{0}_{j \times l}	&	\bm{0}_{j \times j}	
		\end{array}
	\right),
\end{equation}
with the diagonal matrix $\bm{d} = \textbf{diag}(\lambda_1, \lambda_2, \ldots, \lambda_l)$ in which all elements are non-vanishing, $\lambda_1, \lambda_2, \ldots, \lambda_l \neq 0$. The integer index $l$ is restricted to $0\leq l \leq k/2$ for even $k$ and $0 \leq l \leq (k-1)/2 $ for odd $k$, respectively, satisfying $2l+j=k$. With our choice of the basis, $l$ is uniquely determined and we can conclude that $j \geq 0$ for even $k$ and $j \geq 1$ for odd $k$. We will elaborate on that later.

The Lagrangian in Eq.~\eqref{eq_general_network_Lagrangian}, expressed in the basis in which the capacitance matrix and the vector potential matrix are of the forms given in Eq.~\eqref{eq_desired_C} and Eq.~\eqref{eq_desired_A}, respectively, is the starting point for the derivation of the corresponding Hamiltonian by means of the Dirac-Bergmann algorithm.

Due to the $k$-dimensional kernel of the capacitance matrix, Eq.~\eqref{eq_conjugate_charges} is not invertible (for $k>0$), and we have to treat the first $n$ and the last $k$ variables differently. For the sake of convenience, we introduce the notation
\begin{equation}
	\bm{\phi} = 
	\begin{pmatrix}
		\bm{\phi'}		\\
		\bm{\phi''}
	\end{pmatrix},		\qquad
	\bm{Q} = 
	\begin{pmatrix}
		\bm{Q'}			\\
		\bm{Q''}
	\end{pmatrix},		\qquad
\end{equation}
in which the single and double prime compactly denote the separation according to $\bm{\phi'}=(\phi_1, \phi_2, \ldots, \phi_n)^T$ and $\bm{\phi''}=(\phi_{n+1}, \phi_{n+2}, \ldots, \phi_{n+k})^T$, respectively, and similarly for $\bm{Q'}$ and $\bm{Q''}$.
Thus, $\dot{\bm{\phi'}}$ collects the variables of the image of $\bm{C}$, whereas $\dot{\bm{\phi''}}$ represents the variables in the kernel of $\bm{C}$; see Eq.~\eqref{eq_desired_C}. An evaluation of Eq.~\eqref{eq_conjugate_charges} for the first $n$ variables yields
\begin{equation}\label{eq_phi_dot_prime}
\begin{split}
	\bm{Q'} &= \bm{C'} \dot{\bm{\phi'}}
	+ \bm{A'} \bm{\phi'} + \widetilde{\bm{A}} \bm{\phi''}		\\
	\Rightarrow \qquad
	\dot{\bm{\phi'}} &= \bm{C'}^{-1}
	( \bm{Q'} - \bm{A'} \bm{\phi'} - \widetilde{\bm{A}} \bm{\phi''} ).
\end{split}
\end{equation}
We deduce that the first $n$ generalized velocities can be expressed as functions of fluxes and conjugate charges only, such that an ordinary Legendre transformation is practicable for those variables.

In contrast to this result, the last $k$ variables evaluate inherently differently. For these variables, Eq.~\eqref{eq_conjugate_charges} evaluates to
\begin{equation}\label{eq_singular_conjugate_charges}
	\bm{Q''} = \bm{A''} \bm{\phi''},
\end{equation}
which does not contain any generalized velocity at all. For this reason, Eq.~\eqref{eq_singular_conjugate_charges} constitutes $k$ primary constraints. In order to proceed, it is convenient to further separate the variables in the style of Eq.~\eqref{eq_Lambda}, i.e., we subdivide $\bm{\phi''}$ into $\bm{\phi''}_a=(\phi_{n+1}, \phi_{n+2}, \ldots, \phi_{n+l})^T$, $\bm{\phi''}_b=(\phi_{n+l+1}, \phi_{n+l+2}, \ldots, \phi_{n+2l})^T$ and $\bm{\phi''}_c=(\phi_{n+2l+1}, \phi_{n+2l+2}, \ldots, \phi_{n+2l+j})^T$, such that we can write $\bm{\phi''} = ( {\bm{\phi''}_a}^T, {\bm{\phi''}_b}^T, {\bm{\phi''}_c}^T )^T$. We proceed similarly with the charge variables and write $\bm{Q''}=({\bm{Q''}_a}^T, {\bm{Q''}_b}^T, {\bm{Q''}_c}^T)^T$. Also, we make the subdivision of $\widetilde{\bm{A}}=(\bm{A}_a, \bm{A}_b, \bm{A}_c)$ in which $\bm{A}_{a,b,c}$ are of the form $n \times l$, $n \times l$ and $n \times j$, respectively. Considering the explicit form of $\bm{A''}$ [see Eq.~\eqref{eq_Lambda}], the subdivision of the doubly primed variables reduces Eq.~\eqref{eq_singular_conjugate_charges} to
\begin{equation}\label{eq_dirac_momenta}
	\bm{Q''}_a = \bm{d} \bm{\phi''}_b, 	\qquad
	\bm{Q''}_b = \bm{0}_{l \times 1},	\qquad
	\bm{Q''}_c = \bm{0}_{j \times 1}.
\end{equation}
This particular form allows us to determine the underlying primary constraints of the system so as to reduce the number of dynamically independent variables.

First, we note that Eq.~\eqref{eq_dirac_momenta} can be used to solve for the flux variables
\begin{equation}\label{eq_new_momenta}
	\bm{\phi''}_b = \bm{d}^{-1} \bm{Q''}_a,
\end{equation}
while the corresponding charges $\bm{Q''}_b$ vanish. This is fundamental for the subsequent derivation of the Hamiltonian and matches the special class of singular Lagrangians discussed in Sec.~\ref{sec_singular_circuits}.

Furthermore, also the conjugate charges $\bm{Q''}_c$ vanish. However, Eq.~\eqref{eq_dirac_momenta} does not provide expressions for the corresponding fluxes $\bm{\phi''}_c$ as functions of all the other fluxes and conjugate charges. In this case, the Dirac-Bergmann algorithm reduces to an evaluation of the classical Euler-Lagrange equations of motion for the $\bm{\phi''}_c$ variables (see Appendix~\ref{app_multivalued} for an exemplification), which simplify to
\begin{equation}\label{eq_phi_c_eom}
	0 =  \frac{\partial U(\bm{\phi})}{\partial \left( \bm{\phi''}_c \right)_i} 
	- \left( \bm{A}_c^T \dot{\bm{\phi'}} \right)_{i}
	,	\qquad i=1,2,\ldots,j.
\end{equation}
These equations are then used to determine the $\bm{\phi''}_c$ variables as functions of all the other variables, i.e., to find constraints of the form
\begin{equation}\label{eq_phi_c_constraints}
	\bm{\phi''}_c \equiv \bm{\phi''}_c(\bm{\phi'}, \bm{\phi''}_a, \bm{\phi''}_b,  \bm{Q'}).
\end{equation}

Given these constraints, it is clear that the fluxes $\bm{\phi''}_c$ are no independent variables, and one can reduce the number of dynamical degrees of freedom by $j$. Note that, however, the relation in Eq.~\eqref{eq_phi_c_constraints} is not necessarily single-valued and might result in a branched Hamiltonian; see Sec.~\ref{sec_branched_Hamiltonian} for an example.

Furthermore, it is also possible that one cannot derive at all constrains in the form of Eq.~\eqref{eq_phi_c_constraints} from Eq.~\eqref{eq_phi_c_eom}, e.g., if components of $\bm{\phi''}_c$ do not enter the potential $U(\bm{\phi})$. In that case, one has to look for constraints of different form or additional constraints that remove the dependence on $\bm{\phi''}_c$ in the subsequent formalism; see Sec.~\ref{sec_Tellegen_transformer} for an example. We note that constraints of different form than Eq.~\eqref{eq_phi_c_constraints} can even lead to a further reduction of the number of degrees of freedom.

In the following, we assume that we can solve for constraints in form of Eq.~\eqref{eq_phi_c_constraints}.

As explained in Sec.~\ref{sec_singular_circuits}, following the Dirac-Bergmann algorithm, a well-defined Hamiltonian formalism corresponding to the Lagrangian in Eq.~\eqref{eq_general_network_Lagrangian} is established by introducing the Hamiltonian $H = \dot{\bm{\phi}}^T \bm{Q} - \mathcal{L}$ in which we have to insert both the expression for $\dot{\bm{\phi'}}$ [see Eq.~\eqref{eq_phi_dot_prime}] and the constraints for $\bm{\phi'}_c$ [see Eq.~\eqref{eq_phi_c_constraints}] and substitute $\bm{\phi''}_b$ by $\bm{d}^{-1} \bm{Q''}_a$ [see Eq.~\eqref{eq_new_momenta}]. This eventually results in the final Hamiltonian
\begin{widetext}
\begin{equation}\label{eq_general_final_Hamiltonian}
\begin{split}
	H = \frac{1}{2}
	&\big( \bm{Q'} - \bm{A'} \bm{\phi'} - \bm{A}_{a} \bm{\phi''}_a - \bm{A}_{b} \bm{d}^{-1} \bm{Q''}_a - \bm{A}_{c} \bm{\phi''}_c \big)^T
	\bm{C'}^{-1} 
	\big( \bm{Q'} - \bm{A'} \bm{\phi'} - \bm{A}_{a} \bm{\phi''}_a - \bm{A}_{b} \bm{d}^{-1} \bm{Q''}_a - \bm{A}_{c} \bm{\phi''}_c \big) \\
	&+ U \Big( \big( \bm{\phi'}^T, {\bm{\phi''}_a}^T, {\bm{Q''}_a}^T \bm{d}^{-1}, {\bm{\phi''}_c}^T \big)^T \Big).
\end{split}
\end{equation}
\end{widetext}
Conveniently and by construction, the Dirac brackets, which need to be introduced in the process of the Dirac-Bergmann algorithm, do not differ from the conventional Poisson brackets. They are given by $\{ \phi_\mu, Q_\nu \} = \delta_{\mu \nu}$ for the fluxes and charges appearing in the final Hamiltonian in Eq.~\eqref{eq_general_final_Hamiltonian}, i.e., for the indices $\mu, \nu=1,2,\ldots,n+l$. Recall that the effective number of degrees of freedom is reduced to $m-j-l = n+l$.

By considering the final Hamiltonian in Eq.~\eqref{eq_general_final_Hamiltonian}, we observe that charge and flux variables are not easily separable anymore. In particular, for singular circuits, charge variables belonging to $\bm{Q}_a''$ can also appear in the potential energy term of the Hamiltonian [see second line of Eq.~\eqref{eq_general_final_Hamiltonian}], originating from the inductances in the electrical network. In principle, this can yield to physically as well as mathematically interesting properties of the system; see Sec.~\ref{SubSubSec_Gyrator_GKP} for instance.

Finally, to quantize the theory obtained by the Dirac-Bergmann algorithm, the Dirac brackets are used to impose the canonical commutation relations $[ \phi_\mu, Q_\nu ] = i \hbar \delta_{\mu \nu}$.
Although this step might be formally legitimate, in the main text we demonstrate that the quantum theory thereby obtained does not give rise to a correct description of physical electrical networks. In fact, the singularity of electrical networks is lifted by taking into account the small but finite intrinsic capacitances or stray capacitances that make the capacitance matrix in Eq.~\eqref{eq_desired_C} invertible. As a result, the otherwise constrained variables become dynamical, which in the quantum mechanical case give rise to large quantum fluctuations that are absent in the singular description according to the Dirac-Bergmann algorithm.

\subsection{Construction of an Appropriate Basis for the Dirac-Bergmann Algorithm}\label{subsection_basis_construction}
As already stated before, the Lagrangian of a general electrical network that is built out of linear capacitances, any sort of inductances (linear, nonlinear and mutual) and gyrators can be written as in Eq.~\eqref{eq_general_network_Lagrangian}. Since every single capacitor's contribution to the Lagrangian is a positive quadratic form, the capacitance matrix $\bm{C}$ is positive semi-definite ($\bm{C} \geq 0$) and can be chosen to be symmetric ($\bm{C}=\bm{C}^T$) in any initial basis. Similarly, every single gyrator's contribution to the Lagrangian can be written in symmetric gauge such that the resulting vector potential matrix $\bm{A}$ is anti-symmetric ($\bm{A}=-\bm{A}^T$). The potential $U(\bm{\phi})$ takes all the inductances of the network into account.

In the following, we show how to construct the basis for the Lagrangian in Eq.~\eqref{eq_general_network_Lagrangian} such that $\bm{C}$ and $\bm{A}$ are of the form claimed in Eqs.~\eqref{eq_desired_C}~and~\eqref{eq_desired_A}, respectively. We begin with the basis transformation that separates the invertible part of the capacitance matrix from its kernel.

\subsubsection{Transformation of the Capacitance Matrix}
Since the capacitance matrix is symmetric, the spectral theorem states that its eigenvectors can be constructed to form a complete orthonormal basis. Therefore, there exists an orthogonal matrix $\bm{B}$ such that
\begin{equation}\label{eq_C_kernel}
	\bm{B}^T \bm{C} \bm{B} =
	\left(
	\begin{array}{c|c}
	\bm{C'}							&		\bm{0}_{n \times k}		\\
	\hline
	\bm{0}_{k \times n}		&		\bm{0}_{k \times k}
	\end{array}
	\right),
\end{equation}
with a symmetric and positive definite $n \times n$ matrix $\bm{C'}$. The dimensions of the block matrices in Eq.~\eqref{eq_C_kernel} satisfy the rank-nullity theorem $m=n+k$ with $n=\text{rank}(\bm{C})$ and $k=\dim[\ker(\bm{C})]$, i.e., $k$ is the total number of vanishing eigenvalues of $\bm{C}$. Note that the matrix $\bm{B}$ can be constructed by taking its first $n$ columns to be normalized, linearly independent linear combinations of the $n$ eigenvectors of $\bm{C}$ with eigenvalues greater than zero, while its last $k$ columns are normalized, linearly independent linear combinations of the $k$ eigenvectors of $\bm{C}$ with vanishing eigenvalue. Probably the most convenient choice of $\bm{B}$ would be to take the bare $n$ eigenvectors of $\bm{C}$ with eigenvalues greater than zero to be the first $n$ columns of $\bm{B}$, such that $\bm{C'}$ becomes diagonal with strictly positive diagonal-elements only. However, we will proceed with the general case in which $\bm{C'}$ is not necessarily diagonal. Nevertheless, $\bm{C'}$ is positive definite and therefore invertible.

We use $\bm{B}$ to define the variable transformation $\widehat{\bm{\phi}} = \bm{B}^T \bm{\phi}$ as well as
\begin{equation}\label{eq_general_Legendre_first_trafo}
	\widehat{\bm{C}} = \bm{B}^T \bm{C} \bm{B}
	,	\qquad
	\widehat{\bm{A}} = \bm{B}^T \bm{A} \bm{B}
	,	\qquad
	\widehat{U} ( \widehat{\bm{\phi}} ) = U(\bm{B} \widehat{\bm{\phi}}),
\end{equation}
such that $\widehat{\bm{C}}$ is in the desired form of Eq.~\eqref{eq_C_kernel} and $\widehat{\bm{A}}$ remains anti-symmetric. In the following, we will omit the hats for reasons of clarity. By doing so, the transformed Lagrangian remains in the form of Eq.~\eqref{eq_general_network_Lagrangian}. \\

We proceed with a basis transformation that prepares the vector potential matrix such that it can be brought into the form of Eq.~\eqref{eq_desired_A} afterwards.

\subsubsection{Transformation of the Vector Potential Matrix}
Since the vector potential matrix is anti-symmetric, it can be expressed in block-matrix form as
\begin{equation}
	\bm{A} = 
	\left(
	\begin{array}{c|c}
	\bm{A'}				&		\bm{A}_{12}		\\
	\hline
	-\bm{A}_{12}^T		&		\bm{\Lambda}
	\end{array}
	\right),
\end{equation}
in which $\bm{A'}$ and $\bm{\Lambda}$ are anti-symmetric $n \times n$ and $k \times k$ matrices, respectively, while $\bm{A}_{12}$ is an unspecified $n \times k$ matrix. The dimensions of these block-matrices are chosen to match the subdivision of the capacitance matrix; see Eq.~\eqref{eq_C_kernel}. Because of the anti-symmetry of $\bm{\Lambda}$, there exists an orthogonal matrix $\bm{D''}$ that transforms $\bm{\Lambda}$ into a form akin to the Youla normal form \cite{Youla}; in particular,
\begin{equation}\label{eq_Youla}
	\bm{D''}^T \bm{\Lambda} \bm{D''}
	= \frac{1}{2} \left( \bm{A''} - \bm{A''}^T \right),
\end{equation}
with $\bm{A''}$ defined in Eq.~\eqref{eq_Lambda}. For that matter, $\bm{D''}$ can be constructed by taking its columns to be the orthonormal eigenvectors of ${\bm{\Lambda}}^2$. Note that Eq.~\eqref{eq_Youla} differs from the conventional Youla normal form by a permutation of the columns and rows, which, however, preserves the orthogonality of the transformation matrix $\bm{D''}$. We choose this particular form for reasons of convenience with regard to the derivation of the Hamiltonian by means of the Dirac-Bergmann algorithm.

We remark that the nonzero, purely imaginary eigenvalues of $\bm{\Lambda}$ appear pairwise and are given by $\pm i \lambda_1/2, \pm i \lambda_2/2, \ldots, \pm i \lambda_l/2$; cf. Eq.~\eqref{eq_Lambda}. Thus, we can conclude that $k=2l+j$, where $j$ denotes the total number of vanishing eigenvalues of $\bm{\Lambda}$. Especially, for odd $k$, it follows directly that $j \geq 1$.

We use $\bm{D''}$ to define the orthogonal matrix
\begin{equation}
	\bm{D} = 
	\left(
	\begin{array}{c|c}
	\bm{1}_{n \times n}		&		\bm{0}_{n \times k}		\\
	\hline
	\bm{0}_{k \times n}		&		\bm{D''}
	\end{array}
	\right),
\end{equation}
which, in turn, defines the variable transformation $\widehat{\bm{\phi}} = \bm{D}^T \bm{\phi}$ together with
\begin{equation}\label{eq_general_Legendre_second_trafo}
\widehat{\bm{C}} = \bm{D}^T \bm{C} \bm{D}
	,	\qquad
	\widehat{\bm{A}} = \bm{D}^T \bm{A} \bm{D}
	,	\qquad
	\widehat{U} ( \widehat{\bm{\phi}} ) = U(\bm{D} \widehat{\bm{\phi}}).
\end{equation}
Importantly, this transformation leaves the capacitance matrix unaffected, i.e., $\widehat{\bm{C}} = \bm{C}$. Furthermore, defining $\widetilde{\bm{A}} = 2 \bm{A}_{12} \bm{D''}$, we evaluate the transformed vector potential matrix to read
\begin{equation}\label{eq_vector_pot_mat_almost_finished}
	\widehat{\bm{A}}
	=
	\left(
	\begin{array}{c|c}
	\bm{A'}											&	\frac{1}{2} \widetilde{\bm{A}}			\\
	\hline
	-\frac{1}{2} \widetilde{\bm{A}}^T	&	\frac{1}{2} \left( \bm{A''} - \bm{A''}^T \right)
	\end{array}
	\right).
\end{equation}
As previously, we omit the hats in further calculations such that the transformed Lagrangian stays in form of Eq.~\eqref{eq_general_network_Lagrangian}. \\

In the last step, we perform a gauge transformation that finally brings the vector potential matrix into the desired form of Eq.~\eqref{eq_desired_A}.

\subsubsection{Gauge Transformation}
Similar to the case of a single gyrator, we can exploit the gauge freedom of the vector potential matrix and add a total time derivative to the Lagrangian without affecting the classical equations of motion \cite{Goldstein}. In particular, we consider the Lagrangian
\begin{equation}
	\widehat{\mathcal{L}}
	= \mathcal{L} + \frac{d}{dt} \big( \bm{\phi}^T \bm{F} \bm{\phi} \big)	
	= \mathcal{L} + \dot{\bm{\phi}}^T ( \bm{F} + \bm{F}^T) \bm{\phi},
\end{equation}
with a time-independent $m \times m$ matrix $\bm{F}$. Its explicit appearance in the Lagrangian $\widehat{\mathcal{L}}$ can be eliminated by a gauge transformation of the vector potential matrix,
\begin{equation}\label{eq_gauge_transformation}
	\widehat{\bm{A}} = \bm{A} + \left( \bm{F} + \bm{F}^T \right).
\end{equation}
With the specific choice of
\begin{equation}
	\bm{F}
	=
	\left(
	\begin{array}{c|c}
	\bm{\chi'}					&	\frac{1}{2} \widetilde{\bm{A}}				\\
	\hline
	\bm{0}_{k\times n}	&	\frac{1}{2} \bm{A''}
	\end{array}
	\right),
\end{equation}
in which $\bm{\chi'}$ is an arbitrary, time-independent $n \times n$ matrix, the gauge transformation in Eq.~\eqref{eq_gauge_transformation} transforms the vector potential matrix in Eq.~\eqref{eq_vector_pot_mat_almost_finished} into
\begin{equation}\label{eq_Landau_like_gauge}
	\widehat{\bm{A}}
	=
	\left(
	\begin{array}{c|c}
	\widehat{\bm{A'}}		&	\widetilde{\bm{A}}				\\
	\hline
	\bm{0}_{k\times n}	&	\bm{A''}
	\end{array}
	\right)
	,	\qquad
	\widehat{\bm{A'}} = \bm{A'} + \bm{\chi'} + \bm{\chi'}^T,
\end{equation}
which is the desired form of Eq.~\eqref{eq_desired_A}. Furthermore, we observe that the symmetric part of the upper left block matrix $\widehat{\bm{A'}}$ is only defined up to a gauge, which can be chosen arbitrarily.

Finally, by omitting the hats, the Lagrangian is still in the form of Eq.~\eqref{eq_general_network_Lagrangian} and the capacitance matrix as well as the vector potential matrix are in the form supposed for the presented application of the Dirac-Bergmann algorithm.

\section{Avoiding the Branched Hamiltonian}\label{app_multivalued}
In this appendix, we further analyze the singular Lagrangian
\begin{equation}\label{eq_initial_singular_L}
	\mathcal{L} = \frac{m\dot{x}^2}{2} - \frac{(x-y)^2}{2} + \beta\cos(y), 
\end{equation}
which is a relabeled version of the Lagrangian in Eq.~\eqref{eq_Lagrangian_multivalued_potential} describing the circuit in Fig.~\ref{fig_multivalued_a}. As elaborated in Sec.~\ref{sec_branched_Hamiltonian}, after eliminating the variable $y$ in the Lagrangian, the corresponding one-dimensional Hamiltonian $H_s$ [see Eq.~\eqref{eq_Hamiltonian_s}] is branched if $\beta>1$.

As opposed to this result, here, we demonstrate that such a branched Hamiltonian can be avoided at the price of choosing different variables that are non-canonical. To this end, we apply the Dirac-Bergmann algorithm. The conjugate momenta $p_x = \partial \mathcal{L} / \partial \dot{x}$, $p_y = \partial \mathcal{L} / \partial \dot{y}$ give rise to one primary constraint,
\begin{equation}\label{eq_prim_constrain}
	G_1 = p_y \simeq 0,
\end{equation}
in which the weak equality sign $\simeq$ reminds us that one must not use the equation before the Poisson brackets are evaluated. Accounting for this primary constraint via the Lagrange multiplier $\mu$, the primary Hamiltonian of the system reads
\begin{equation}\label{eq_H_primary}
	H_P = p_x \dot{x} + p_y \dot{y} - \mathcal{L} + \mu G_1,
\end{equation}
and it governs the time evolution of the system. In particular, requiring that the time evolution of the primary constraint 
\begin{equation}
	\dot{G}_1 \simeq \{ G_1, H_P \},
\end{equation}
vanishes, results in a consistency condition that leads to the secondary constraint
\begin{equation}\label{eq_sec_constrain}
	G_2 = x - y - \beta \sin(y) \simeq 0,
\end{equation}
which is essentially the Euler-Lagrange equation of motion for the $y$-degree of freedom. Note that $\dot{G}_2$ does not give rise to further constraints.

In the following, both the primary constraint and the secondary constraint will be used to define the Dirac brackets -- a redefinition of the Poisson brackets. To this end, we introduce the matrix 
\begin{equation}\label{eq_DB_matrix_M}
	\bm{M}
	=
	\begin{pmatrix}
	0		&		\{ G_1, G_2 \}	\\
	\{ G_2, G_1 \}	&		0		
	\end{pmatrix},
\end{equation}
which collects the mutual Poisson brackets of the constrains. In particular, we find
\begin{equation}
	\{ G_1, G_2 \} = 1 + \beta \cos(y),
\end{equation}
which, in general, does not vanish. Thus, both constraints $G_1$ and $G_2$ are second class constraints. Next, Dirac's version of the Poisson brackets of two general functions $A$ and $B$ are defined as
\begin{equation}
	\{ A, B \}_D \coloneqq \{ A, B \} - \sum_{i,j=1}^2 \{ A, G_i \} (\bm{M}^{-1})_{ij} \{ G_j, B \}.
\end{equation}
The only non-vanishing Dirac brackets of position and momentum variables are
\begin{equation}\label{eq_singuler_Dirac_bracket}
	\{ x, p_x \}_D = 1,	\qquad
	\{ y, p_x \}_D = \frac{1}{1+\beta\cos(y)}.
\end{equation}

Having introduced the Dirac brackets, the weak equality signs in Eqs.~\eqref{eq_prim_constrain}~and~\eqref{eq_sec_constrain} can be replaced by strong equality signs, keeping in mind that one must not work with the usual Poisson brackets but Dirac's version. As a consequence, $x$ and $p_y$ can be eliminated in the primary Hamiltonian in Eq.~\eqref{eq_H_primary}, thus resulting in the final Hamiltonian
\begin{equation}\label{eq_final_H_without_branches}
	H = \frac{p_x^2}{2m} + \frac{\beta^2 \sin^2(y)}{2} - \beta \cos(y),
\end{equation}
which is not branched but a function of non-symplectic coordinates.

The classical Hamiltonian equations of motion for $y$ and $p_x$ are two coupled differential equations of first order and evaluate to
\begin{subequations}\label{eq_cl_eq_motion}
\begin{align}
	\dot{y} &= \{ y, H \}_D = \frac{p_x}{m[1+\beta\cos(y)]},		\\
	\dot{p}_x &= \{ p_x, H \}_D = -\beta \sin(y).
\end{align}
\end{subequations}
As a consistency check, they can be combined to obtain the second order differential equation
\begin{equation}\label{eq_second_order_eom}
	m\ddot{y} = \frac{\beta\sin(y)}{1+\beta\cos(y)} \left( m \dot{y}^2 -1 \right),
\end{equation}
which coincides with the Euler-Lagrange equation of motion derived from the Lagrangian in Eq.~\eqref{eq_initial_singular_L} after eliminating the $x$-degree of freedom.

At this point, for $\beta \geq 1$, we note that the matrix $\bm{M}$ in Eq.~\eqref{eq_DB_matrix_M} is not invertible if
\begin{equation}\label{eq_condition_singularity_M}
    1+\beta\cos(y)=0,    
\end{equation}
which results in a singularity of the Dirac brackets in Eq.~\eqref{eq_singuler_Dirac_bracket}. However, the values of $y$ that satisfy the condition in Eq.~\eqref{eq_condition_singularity_M} correspond to critical values of the final Hamiltonian in Eq.~\eqref{eq_final_H_without_branches} \footnote{The critical values of the final Hamiltonian in Eq.~\eqref{eq_final_H_without_branches} coincide with those of the secondary constraint $x(y)=y+\beta\sin(y)$ [cf. Eq.~\eqref{eq_sec_constrain}]. Therefore, these values correspond to the branching points of the multi-valued potential.} where the symplectic structure of phase space vanishes \cite{Zhao}. In fact, as the agreement of the classical Hamiltonian and Lagrangian equations of motion shows, such a singularity does not affect the validity of the Dirac-Bergmann algorithm.

Finally, we remark that one can also construct a single-valued Hamiltonian with a set of symplectic coordinates by introducing the new momentum
\begin{equation}
	p = p_x [1 + \beta \cos(y)]
\end{equation}
such that $\{ y, p \}_D = 1$. Given these variables, however, the system is described by the Hamiltonian
\begin{equation}\label{eq_Hamiltonian_with_pos_dep_mass}
	H' = \frac{p^2}{2m(y)} + \frac{\beta^2 \sin^2(y)}{2} - \beta \cos(y),
\end{equation}
which involves the position-dependent mass
\begin{equation}
	m(y) = m [1 + \beta \cos(y)]^2
\end{equation}
that vanishes at $1+\beta\cos(y)=0$.

Similar to our approach in the main text in Sec.~\ref{sec_branched_Hamiltonian}, here, we do not attempt to provide a quantized theory for the system that is described by the Lagrangian in Eq.~\eqref{eq_initial_singular_L}. Despite the aforementioned singularities for $\beta \geq 1$, it turns out that the quantization of $H$ in Eq.~\eqref{eq_final_H_without_branches} would require a systematic non-canonical quantization process, while $H'$ in Eq.~\eqref{eq_Hamiltonian_with_pos_dep_mass} would need to be brought into a Hermitian form prior quantization. As we have shown in the main text, for the description of electrical networks both these procedures can be circumvented by removing the singularity of the initial Lagrangian.

\section{Evaluation of the Radius of Convergence}\label{app_lower_bound_radius}
The proof of Theorem~1 in Sec.~\ref{subsec_type_1} requires the Rayleigh-Schrödinger series in Eq.~\eqref{eq_RS_series_type_1} to converge. Here, we derive a lower bound of its radius of convergence $\lambda_n(\phi, \alpha)$. In the following, we consider all parameters and variables to be dimensionless.

By definition, any analytic family of type (A) in the sense of Kato can be written as \cite{Kato}
\begin{equation}\label{eq_Kato_family_A}
    T(\lambda) = T + \lambda T^{(1)} + \lambda^2 T^{(2)} + \ldots
\end{equation}
with $T$ being an closable operator with domain $D$ and $T^{(k)}$ being operators with domains containing $D$. Furthermore, for any analytic family of type (A), there exist constants $a, b, c > 0$ such that (see p.~378, Remark 2.8 in Ref. \cite{Kato})
\begin{equation}\label{eq_upper_T_bound}
    \Vert T^{(k)} u \Vert
    \leq
    c^{k-1}
    (a \Vert u \Vert + b \Vert T u \Vert),
    \quad u \in D, \, k \in \mathbb{N}.
\end{equation}
A comparison of $\lambda^{2q} H_\text{aux}$ in Eq.~\eqref{eq_for_Kato} with Eq.~\eqref{eq_Kato_family_A} identifies $T=H_0$, $T^{({k})} = 0$ for $k \notin \{q, 2q, 2q-p\}$, and
\begin{equation}\label{eq_T(k)}
    T^{(q)} = -\frac{y \phi}{\sqrt{L}}, \quad
    T^{(2q)} = \frac{\phi^2}{2L}, \quad
    T^{(2q-p)} = \alpha^\gamma U_{nl} \left( \tfrac{\sqrt{L}y}{\alpha} \right).
\end{equation}
In the following, we show that each operator in Eq.~\eqref{eq_T(k)} is $T$-{\em bounded}. As $T^{(2q)} \propto \mathbb{1}$, it follows trivially that
\begin{equation}
    \Vert T^{(2q)} u \Vert
    \leq 
    \frac{\phi^2}{2L} \Vert u \Vert.
\end{equation}
For constants $A,B>0$ satisfying $4AB \geq 1$, the inequality $|y| \leq A + B y^2$ holds, and therefore (cf. Sec.~II.1. and Sec.~II.9. in Ref.~\cite{Simon_article_long})
\begin{equation}
    \Vert T^{(q)} u \Vert
    \leq 
    \frac{\phi (A+2B)}{\sqrt{L}} \Vert u \Vert
    + \frac{2 \phi B}{\sqrt{L}} \Vert T u \Vert.
\end{equation}
Recall that $U_{nl}(\phi_c)$ describes a nonlinear inductor of type~1. Lemma~1 guarantees the existence of constants $\beta, M > 0$ such that $|U_{nl}(\phi_c)| \leq \beta |\phi_c|^\gamma + M$. Furthermore, for $\gamma \in (0,2)$, there exist constants $A',B' > 0$ such that $|y|^\gamma \leq B'y^2 + A'$. It follows that
\begin{equation}
\begin{split}
    \Vert T^{(2q-p)} u \Vert
    \leq
    &\left[ \alpha^\gamma M + L^{\gamma/2} \beta (A' + 2 B') \right]
    \Vert u \Vert       \\
    &+
    2 L^{\gamma/2} \beta B' \Vert T u \Vert.
\end{split}
\end{equation}
Thus, the inequality in Eq.~\eqref{eq_upper_T_bound} is satisfied with the choice
\begin{align}
    a &= \max \left\{
    \frac{\phi^2}{2L},
    \frac{\phi (A+2B)}{\sqrt{L}},
    \alpha^\gamma M + L^{\gamma/2} \beta (A' + 2 B')
    \right\},                            \\
    b &= \max \left\{
    \frac{2 \phi B}{\sqrt{L}},
    2 L^{\gamma/2} \beta B'
    \right\},                            \\
    c &= 1.
\end{align}
In the following, we set $A = \phi/2\sqrt{L}$ and $B = \sqrt{L}/2\phi$. Furthermore, for any value of $\phi$, we can chose $\beta$ such that $a = \alpha^\gamma M + L^{\gamma/2} \beta (A' + 2 B')$ and $b = 2 L^{\gamma/2} \beta B'$. Note that $a \equiv a(\alpha)$ increases monotonically as $\alpha$ increases.

Following p.~379, Remark 2.9 in Ref.~\cite{Kato}, the Rayleight-Schrödinger series Eq.~\eqref{eq_RS_series_type_1} is convergent at least for 
\begin{equation}
    \lambda < \min_{\zeta \in \Gamma} 
    (a \Vert R(\zeta) \Vert + b \Vert T R(\zeta) \Vert + c)^{-1},
\end{equation}
where $\Gamma$ is a closed curve in the complex plane separating the $n^{\rm{th}}$ eigenenergy of $T$ from the rest of its spectrum and $R(\zeta) = (T-\zeta)^{-1}$ is the resolvent of $T$. 

Recall that $T=H_0$ is the unperturbed harmonic-oscillator Hamiltonian with the spectrum $n+1/2$, $n \in \mathbb{N}_0$. W choose $\Gamma$ to be a circle with radius $1/2$ centered at $n+1/2$. We find $\Vert R(\zeta) \Vert = 2$ and the (unsharp) inequality $\Vert T R(\zeta) \Vert \leq 3+2n$.
Thus, the radius of convergence of the Rayleigh-Schrödinger series is lower-bounded by 
\begin{equation}
    r_n(\phi, \alpha) = 
    \frac{1}{2a(\alpha) + (3n+2)b +1 },
\end{equation}
which is always positive and increases as $\alpha$ decreases.

\bibliography{bib_file.bib}

\begin{thebibliography}{105}%
\makeatletter
\providecommand \@ifxundefined [1]{%
 \@ifx{#1\undefined}
}%
\providecommand \@ifnum [1]{%
 \ifnum #1\expandafter \@firstoftwo
 \else \expandafter \@secondoftwo
 \fi
}%
\providecommand \@ifx [1]{%
 \ifx #1\expandafter \@firstoftwo
 \else \expandafter \@secondoftwo
 \fi
}%
\providecommand \natexlab [1]{#1}%
\providecommand \enquote  [1]{``#1''}%
\providecommand \bibnamefont  [1]{#1}%
\providecommand \bibfnamefont [1]{#1}%
\providecommand \citenamefont [1]{#1}%
\providecommand \href@noop [0]{\@secondoftwo}%
\providecommand \href [0]{\begingroup \@sanitize@url \@href}%
\providecommand \@href[1]{\@@startlink{#1}\@@href}%
\providecommand \@@href[1]{\endgroup#1\@@endlink}%
\providecommand \@sanitize@url [0]{\catcode `\\12\catcode `\$12\catcode
  `\&12\catcode `\#12\catcode `\^12\catcode `\_12\catcode `\%12\relax}%
\providecommand \@@startlink[1]{}%
\providecommand \@@endlink[0]{}%
\providecommand \url  [0]{\begingroup\@sanitize@url \@url }%
\providecommand \@url [1]{\endgroup\@href {#1}{\urlprefix }}%
\providecommand \urlprefix  [0]{URL }%
\providecommand \Eprint [0]{\href }%
\providecommand \doibase [0]{http://dx.doi.org/}%
\providecommand \selectlanguage [0]{\@gobble}%
\providecommand \bibinfo  [0]{\@secondoftwo}%
\providecommand \bibfield  [0]{\@secondoftwo}%
\providecommand \translation [1]{[#1]}%
\providecommand \BibitemOpen [0]{}%
\providecommand \bibitemStop [0]{}%
\providecommand \bibitemNoStop [0]{.\EOS\space}%
\providecommand \EOS [0]{\spacefactor3000\relax}%
\providecommand \BibitemShut  [1]{\csname bibitem#1\endcsname}%
\let\auto@bib@innerbib\@empty
\bibitem [{\citenamefont {Arute}\ \emph {et~al.}(2019)\citenamefont {Arute},
  \citenamefont {Arya}, \citenamefont {Babbush} \emph {et~al.}}]{Google}%
  \BibitemOpen
  \bibfield  {author} {\bibinfo {author} {\bibfnamefont {F.}~\bibnamefont
  {Arute}}, \bibinfo {author} {\bibfnamefont {K.}~\bibnamefont {Arya}},
  \bibinfo {author} {\bibfnamefont {R.}~\bibnamefont {Babbush}},  \emph
  {et~al.},\ }\bibfield  {title} {\enquote {\bibinfo {title} {{Quantum
  supremacy using a programmable superconducting processor}},}\ }\href
  {\doibase 10.1038/s41586-019-1666-5} {\bibfield  {journal} {\bibinfo
  {journal} {Nature}\ }\textbf {\bibinfo {volume} {574}},\ \bibinfo {pages}
  {505--510} (\bibinfo {year} {2019})}\BibitemShut {NoStop}%
\bibitem [{\citenamefont {Wu}\ \emph {et~al.}(2021)\citenamefont {Wu} \emph
  {et~al.}}]{Zuchongzhi2_0}%
  \BibitemOpen
  \bibfield  {author} {\bibinfo {author} {\bibfnamefont {Y.}~\bibnamefont {Wu}}
  \emph {et~al.},\ }\bibfield  {title} {\enquote {\bibinfo {title} {{Strong
  Quantum Computational Advantage Using a Superconducting Quantum
  Processor}},}\ }\href {\doibase 10.1103/PhysRevLett.127.180501} {\bibfield
  {journal} {\bibinfo  {journal} {Phys. Rev. Lett.}\ }\textbf {\bibinfo
  {volume} {127}},\ \bibinfo {pages} {180501} (\bibinfo {year}
  {2021})}\BibitemShut {NoStop}%
\bibitem [{\citenamefont {Zhu}\ \emph {et~al.}(2021)\citenamefont {Zhu},
  \citenamefont {Cao}, \citenamefont {Chen} \emph {et~al.}}]{Zuchongzhi2_1}%
  \BibitemOpen
  \bibfield  {author} {\bibinfo {author} {\bibfnamefont {Q.}~\bibnamefont
  {Zhu}}, \bibinfo {author} {\bibfnamefont {S.}~\bibnamefont {Cao}}, \bibinfo
  {author} {\bibfnamefont {F.}~\bibnamefont {Chen}},  \emph {et~al.},\
  }\href@noop {} {\enquote {\bibinfo {title} {{Quantum Computational Advantage
  via 60-Qubit 24-Cycle Random Circuit Sampling}},}\ } (\bibinfo {year}
  {2021}),\ \Eprint {http://arxiv.org/abs/2109.03494} {arXiv:2109.03494}
  \BibitemShut {NoStop}%
\bibitem [{\citenamefont {Feynman}(1982)}]{Feynman}%
  \BibitemOpen
  \bibfield  {author} {\bibinfo {author} {\bibfnamefont {R.~P.}\ \bibnamefont
  {Feynman}},\ }\bibfield  {title} {\enquote {\bibinfo {title} {{Simulating
  Physics with Computers}},}\ }\href@noop {} {\bibfield  {journal} {\bibinfo
  {journal} {International journal of theoretical physics}\ }\textbf {\bibinfo
  {volume} {21}},\ \bibinfo {pages} {467--488} (\bibinfo {year}
  {1982})}\BibitemShut {NoStop}%
\bibitem [{\citenamefont {Nielsen}\ and\ \citenamefont
  {Chuang}(2010)}]{Nielsen}%
  \BibitemOpen
  \bibfield  {author} {\bibinfo {author} {\bibfnamefont {M.~A.}\ \bibnamefont
  {Nielsen}}\ and\ \bibinfo {author} {\bibfnamefont {I.~L.}\ \bibnamefont
  {Chuang}},\ }\href {\doibase 10.1017/CBO9780511976667} {\emph {\bibinfo
  {title} {{Quantum Computation and Quantum Information: 10th Anniversary
  Edition}}}}\ (\bibinfo  {publisher} {Cambridge University Press},\ \bibinfo
  {year} {2010})\BibitemShut {NoStop}%
\bibitem [{\citenamefont {Shnirman}\ \emph {et~al.}(1997)\citenamefont
  {Shnirman}, \citenamefont {Sch\"on},\ and\ \citenamefont
  {Hermon}}]{ShnirmanCPB}%
  \BibitemOpen
  \bibfield  {author} {\bibinfo {author} {\bibfnamefont {A.}~\bibnamefont
  {Shnirman}}, \bibinfo {author} {\bibfnamefont {G.}~\bibnamefont {Sch\"on}}, \
  and\ \bibinfo {author} {\bibfnamefont {Z.}~\bibnamefont {Hermon}},\
  }\bibfield  {title} {\enquote {\bibinfo {title} {{Quantum Manipulations of
  Small Josephson Junctions}},}\ }\href {\doibase 10.1103/PhysRevLett.79.2371}
  {\bibfield  {journal} {\bibinfo  {journal} {Phys. Rev. Lett.}\ }\textbf
  {\bibinfo {volume} {79}},\ \bibinfo {pages} {2371--2374} (\bibinfo {year}
  {1997})}\BibitemShut {NoStop}%
\bibitem [{\citenamefont {Bouchiat}\ \emph {et~al.}(1998)\citenamefont
  {Bouchiat}, \citenamefont {Vion}, \citenamefont {Joyez}, \citenamefont
  {Esteve},\ and\ \citenamefont {Devoret}}]{DevoretCPB}%
  \BibitemOpen
  \bibfield  {author} {\bibinfo {author} {\bibfnamefont {V.}~\bibnamefont
  {Bouchiat}}, \bibinfo {author} {\bibfnamefont {D.}~\bibnamefont {Vion}},
  \bibinfo {author} {\bibfnamefont {P.}~\bibnamefont {Joyez}}, \bibinfo
  {author} {\bibfnamefont {D.}~\bibnamefont {Esteve}}, \ and\ \bibinfo {author}
  {\bibfnamefont {M.~H.}\ \bibnamefont {Devoret}},\ }\bibfield  {title}
  {\enquote {\bibinfo {title} {{Quantum Coherence with a Single Cooper
  Pair}},}\ }\href {\doibase 10.1238/physica.topical.076a00165} {\bibfield
  {journal} {\bibinfo  {journal} {Physica Scripta}\ }\textbf {\bibinfo {volume}
  {T76}},\ \bibinfo {pages} {165} (\bibinfo {year} {1998})}\BibitemShut
  {NoStop}%
\bibitem [{\citenamefont {Nakamura}\ \emph {et~al.}(1999)\citenamefont
  {Nakamura}, \citenamefont {Pashkin},\ and\ \citenamefont
  {Tsai}}]{NakamuraCPB}%
  \BibitemOpen
  \bibfield  {author} {\bibinfo {author} {\bibfnamefont {Y.}~\bibnamefont
  {Nakamura}}, \bibinfo {author} {\bibfnamefont {Y.~A.}\ \bibnamefont
  {Pashkin}}, \ and\ \bibinfo {author} {\bibfnamefont {J.~S.}\ \bibnamefont
  {Tsai}},\ }\bibfield  {title} {\enquote {\bibinfo {title} {{Coherent control
  of macroscopic quantum states in a single-Cooper-pair box}},}\ }\href
  {\doibase 10.1038/19718} {\bibfield  {journal} {\bibinfo  {journal} {Nature}\
  }\textbf {\bibinfo {volume} {398}},\ \bibinfo {pages} {786--788} (\bibinfo
  {year} {1999})}\BibitemShut {NoStop}%
\bibitem [{\citenamefont {Preskill}(2018)}]{PreskillNISQ}%
  \BibitemOpen
  \bibfield  {author} {\bibinfo {author} {\bibfnamefont {J.}~\bibnamefont
  {Preskill}},\ }\bibfield  {title} {\enquote {\bibinfo {title} {{Quantum
  {C}omputing in the {NISQ} era and beyond}},}\ }\href {\doibase
  10.22331/q-2018-08-06-79} {\bibfield  {journal} {\bibinfo  {journal}
  {{Quantum}}\ }\textbf {\bibinfo {volume} {2}},\ \bibinfo {pages} {79}
  (\bibinfo {year} {2018})}\BibitemShut {NoStop}%
\bibitem [{\citenamefont {Manucharyan}\ \emph {et~al.}(2009)\citenamefont
  {Manucharyan}, \citenamefont {Koch}, \citenamefont {Glazman},\ and\
  \citenamefont {Devoret}}]{Manucharyan}%
  \BibitemOpen
  \bibfield  {author} {\bibinfo {author} {\bibfnamefont {V.~E.}\ \bibnamefont
  {Manucharyan}}, \bibinfo {author} {\bibfnamefont {J.}~\bibnamefont {Koch}},
  \bibinfo {author} {\bibfnamefont {L.~I.}\ \bibnamefont {Glazman}}, \ and\
  \bibinfo {author} {\bibfnamefont {M.~H.}\ \bibnamefont {Devoret}},\
  }\bibfield  {title} {\enquote {\bibinfo {title} {{Fluxonium: Single
  Cooper-Pair Circuit Free of Charge Offsets}},}\ }\href {\doibase
  10.1126/science.1175552} {\bibfield  {journal} {\bibinfo  {journal}
  {Science}\ }\textbf {\bibinfo {volume} {326}},\ \bibinfo {pages} {113--116}
  (\bibinfo {year} {2009})}\BibitemShut {NoStop}%
\bibitem [{\citenamefont {Somoroff}\ \emph {et~al.}(2021)\citenamefont
  {Somoroff}, \citenamefont {Ficheux}, \citenamefont {Mencia}, \citenamefont
  {Xiong}, \citenamefont {Kuzmin},\ and\ \citenamefont
  {Manucharyan}}]{Somoroff}%
  \BibitemOpen
  \bibfield  {author} {\bibinfo {author} {\bibfnamefont {A.}~\bibnamefont
  {Somoroff}}, \bibinfo {author} {\bibfnamefont {Q.}~\bibnamefont {Ficheux}},
  \bibinfo {author} {\bibfnamefont {R.~A.}\ \bibnamefont {Mencia}}, \bibinfo
  {author} {\bibfnamefont {H.}~\bibnamefont {Xiong}}, \bibinfo {author}
  {\bibfnamefont {R.~V.}\ \bibnamefont {Kuzmin}}, \ and\ \bibinfo {author}
  {\bibfnamefont {V.~E.}\ \bibnamefont {Manucharyan}},\ }\href@noop {}
  {\enquote {\bibinfo {title} {{Millisecond coherence in a superconducting
  qubit}},}\ } (\bibinfo {year} {2021}),\ \Eprint
  {http://arxiv.org/abs/2103.08578} {arXiv:2103.08578} \BibitemShut {NoStop}%
\bibitem [{\citenamefont {Devoret}\ \emph {et~al.}(1995)\citenamefont {Devoret}
  \emph {et~al.}}]{Devoret1}%
  \BibitemOpen
  \bibfield  {author} {\bibinfo {author} {\bibfnamefont {M.}~\bibnamefont
  {Devoret}} \emph {et~al.},\ }\bibfield  {title} {\enquote {\bibinfo {title}
  {{Quantum Fluctuations in Electrical Circuits}},}\ }\href@noop {} {\bibfield
  {journal} {\bibinfo  {journal} {Les Houches, Session LXIII}\ }\textbf
  {\bibinfo {volume} {7}} (\bibinfo {year} {1995})}\BibitemShut {NoStop}%
\bibitem [{\citenamefont {Vool}\ and\ \citenamefont
  {Devoret}(2017)}]{Devoret2}%
  \BibitemOpen
  \bibfield  {author} {\bibinfo {author} {\bibfnamefont {U.}~\bibnamefont
  {Vool}}\ and\ \bibinfo {author} {\bibfnamefont {M.}~\bibnamefont {Devoret}},\
  }\bibfield  {title} {\enquote {\bibinfo {title} {{Introduction to quantum
  electromagnetic circuits}},}\ }\href {\doibase 10.1002/cta.2359} {\bibfield
  {journal} {\bibinfo  {journal} {International Journal of Circuit Theory and
  Applications}\ }\textbf {\bibinfo {volume} {45}},\ \bibinfo {pages}
  {897–934} (\bibinfo {year} {2017})}\BibitemShut {NoStop}%
\bibitem [{\citenamefont {Girvin}(2014)}]{Girvin_cQED}%
  \BibitemOpen
  \bibfield  {author} {\bibinfo {author} {\bibfnamefont {S.~M.}\ \bibnamefont
  {Girvin}},\ }\bibfield  {title} {\enquote {\bibinfo {title} {{Circuit {QED}:
  superconducting qubits coupled to microwave photons}},}\ }in\ \href {\doibase
  10.1093/acprof:oso/9780199681181.003.0003} {\emph {\bibinfo {booktitle}
  {{Quantum Machines: Measurement and Control of Engineered Quantum
  Systems}}}}\ (\bibinfo  {publisher} {Oxford University Press},\ \bibinfo
  {year} {2014})\ pp.\ \bibinfo {pages} {113--256}\BibitemShut {NoStop}%
\bibitem [{\citenamefont {Blais}\ \emph {et~al.}(2021)\citenamefont {Blais},
  \citenamefont {Grimsmo}, \citenamefont {Girvin},\ and\ \citenamefont
  {Wallraff}}]{cQED_Review}%
  \BibitemOpen
  \bibfield  {author} {\bibinfo {author} {\bibfnamefont {A.}~\bibnamefont
  {Blais}}, \bibinfo {author} {\bibfnamefont {A.~L.}\ \bibnamefont {Grimsmo}},
  \bibinfo {author} {\bibfnamefont {S.~M.}\ \bibnamefont {Girvin}}, \ and\
  \bibinfo {author} {\bibfnamefont {A.}~\bibnamefont {Wallraff}},\ }\bibfield
  {title} {\enquote {\bibinfo {title} {{Circuit quantum electrodynamics}},}\
  }\href {\doibase 10.1103/RevModPhys.93.025005} {\bibfield  {journal}
  {\bibinfo  {journal} {Rev. Mod. Phys.}\ }\textbf {\bibinfo {volume} {93}},\
  \bibinfo {pages} {025005} (\bibinfo {year} {2021})}\BibitemShut {NoStop}%
\bibitem [{\citenamefont {Rymarz}\ \emph {et~al.}(2021)\citenamefont {Rymarz},
  \citenamefont {Bosco}, \citenamefont {Ciani},\ and\ \citenamefont
  {DiVincenzo}}]{Rymarz}%
  \BibitemOpen
  \bibfield  {author} {\bibinfo {author} {\bibfnamefont {M.}~\bibnamefont
  {Rymarz}}, \bibinfo {author} {\bibfnamefont {S.}~\bibnamefont {Bosco}},
  \bibinfo {author} {\bibfnamefont {A.}~\bibnamefont {Ciani}}, \ and\ \bibinfo
  {author} {\bibfnamefont {D.~P.}\ \bibnamefont {DiVincenzo}},\ }\bibfield
  {title} {\enquote {\bibinfo {title} {{Hardware-Encoding Grid States in a
  Nonreciprocal Superconducting Circuit}},}\ }\href {\doibase
  10.1103/PhysRevX.11.011032} {\bibfield  {journal} {\bibinfo  {journal} {Phys.
  Rev. X}\ }\textbf {\bibinfo {volume} {11}},\ \bibinfo {pages} {011032}
  (\bibinfo {year} {2021})}\BibitemShut {NoStop}%
\bibitem [{\citenamefont {Roth}\ \emph {et~al.}(2021)\citenamefont {Roth},
  \citenamefont {Ma},\ and\ \citenamefont {Chew}}]{Roth_IEEE}%
  \BibitemOpen
  \bibfield  {author} {\bibinfo {author} {\bibfnamefont {T.~E.}\ \bibnamefont
  {Roth}}, \bibinfo {author} {\bibfnamefont {R.}~\bibnamefont {Ma}}, \ and\
  \bibinfo {author} {\bibfnamefont {W.~C.}\ \bibnamefont {Chew}},\ }\href@noop
  {} {\enquote {\bibinfo {title} {{An Introduction to the Transmon Qubit for
  Electromagnetic Engineers}},}\ } (\bibinfo {year} {2021}),\ \Eprint
  {http://arxiv.org/abs/2106.11352} {arXiv:2106.11352} \BibitemShut {NoStop}%
\bibitem [{\citenamefont {Brooks}\ \emph {et~al.}(2013)\citenamefont {Brooks},
  \citenamefont {Kitaev},\ and\ \citenamefont {Preskill}}]{Brooks}%
  \BibitemOpen
  \bibfield  {author} {\bibinfo {author} {\bibfnamefont {P.}~\bibnamefont
  {Brooks}}, \bibinfo {author} {\bibfnamefont {A.}~\bibnamefont {Kitaev}}, \
  and\ \bibinfo {author} {\bibfnamefont {J.}~\bibnamefont {Preskill}},\
  }\bibfield  {title} {\enquote {\bibinfo {title} {{Protected gates for
  superconducting qubits}},}\ }\href {\doibase 10.1103/PhysRevA.87.052306}
  {\bibfield  {journal} {\bibinfo  {journal} {Phys. Rev. A}\ }\textbf {\bibinfo
  {volume} {87}},\ \bibinfo {pages} {052306} (\bibinfo {year}
  {2013})}\BibitemShut {NoStop}%
\bibitem [{\citenamefont {Viola}\ and\ \citenamefont
  {Catelani}(2015)}]{ViolaCatelani}%
  \BibitemOpen
  \bibfield  {author} {\bibinfo {author} {\bibfnamefont {G.}~\bibnamefont
  {Viola}}\ and\ \bibinfo {author} {\bibfnamefont {G.}~\bibnamefont
  {Catelani}},\ }\bibfield  {title} {\enquote {\bibinfo {title} {{Collective
  modes in the fluxonium qubit}},}\ }\href {\doibase
  10.1103/PhysRevB.92.224511} {\bibfield  {journal} {\bibinfo  {journal} {Phys.
  Rev. B}\ }\textbf {\bibinfo {volume} {92}},\ \bibinfo {pages} {224511}
  (\bibinfo {year} {2015})}\BibitemShut {NoStop}%
\bibitem [{\citenamefont {Frattini}\ \emph {et~al.}(2017)\citenamefont
  {Frattini}, \citenamefont {Vool}, \citenamefont {Shankar}, \citenamefont
  {Narla}, \citenamefont {Sliwa},\ and\ \citenamefont {Devoret}}]{SNAIL1}%
  \BibitemOpen
  \bibfield  {author} {\bibinfo {author} {\bibfnamefont {N.~E.}\ \bibnamefont
  {Frattini}}, \bibinfo {author} {\bibfnamefont {U.}~\bibnamefont {Vool}},
  \bibinfo {author} {\bibfnamefont {S.}~\bibnamefont {Shankar}}, \bibinfo
  {author} {\bibfnamefont {A.}~\bibnamefont {Narla}}, \bibinfo {author}
  {\bibfnamefont {K.~M.}\ \bibnamefont {Sliwa}}, \ and\ \bibinfo {author}
  {\bibfnamefont {M.~H.}\ \bibnamefont {Devoret}},\ }\bibfield  {title}
  {\enquote {\bibinfo {title} {{3-wave mixing Josephson dipole element}},}\
  }\href {\doibase 10.1063/1.4984142} {\bibfield  {journal} {\bibinfo
  {journal} {Applied Physics Letters}\ }\textbf {\bibinfo {volume} {110}},\
  \bibinfo {pages} {222603} (\bibinfo {year} {2017})}\BibitemShut {NoStop}%
\bibitem [{\citenamefont {Frattini}\ \emph {et~al.}(2018)\citenamefont
  {Frattini}, \citenamefont {Sivak}, \citenamefont {Lingenfelter},
  \citenamefont {Shankar},\ and\ \citenamefont {Devoret}}]{SNAIL2}%
  \BibitemOpen
  \bibfield  {author} {\bibinfo {author} {\bibfnamefont {N.~E.}\ \bibnamefont
  {Frattini}}, \bibinfo {author} {\bibfnamefont {V.~V.}\ \bibnamefont {Sivak}},
  \bibinfo {author} {\bibfnamefont {A.}~\bibnamefont {Lingenfelter}}, \bibinfo
  {author} {\bibfnamefont {S.}~\bibnamefont {Shankar}}, \ and\ \bibinfo
  {author} {\bibfnamefont {M.~H.}\ \bibnamefont {Devoret}},\ }\bibfield
  {title} {\enquote {\bibinfo {title} {{Optimizing the Nonlinearity and
  Dissipation of a SNAIL Parametric Amplifier for Dynamic Range}},}\ }\href
  {\doibase 10.1103/physrevapplied.10.054020} {\bibfield  {journal} {\bibinfo
  {journal} {Physical Review Applied}\ }\textbf {\bibinfo {volume} {10}}
  (\bibinfo {year} {2018}),\ 10.1103/physrevapplied.10.054020}\BibitemShut
  {NoStop}%
\bibitem [{\citenamefont {Sivak}\ \emph {et~al.}(2019)\citenamefont {Sivak},
  \citenamefont {Frattini}, \citenamefont {Joshi}, \citenamefont
  {Lingenfelter}, \citenamefont {Shankar},\ and\ \citenamefont
  {Devoret}}]{SNAIL3}%
  \BibitemOpen
  \bibfield  {author} {\bibinfo {author} {\bibfnamefont {V.~V.}\ \bibnamefont
  {Sivak}}, \bibinfo {author} {\bibfnamefont {N.~E.}\ \bibnamefont {Frattini}},
  \bibinfo {author} {\bibfnamefont {V.~R.}\ \bibnamefont {Joshi}}, \bibinfo
  {author} {\bibfnamefont {A.}~\bibnamefont {Lingenfelter}}, \bibinfo {author}
  {\bibfnamefont {S.}~\bibnamefont {Shankar}}, \ and\ \bibinfo {author}
  {\bibfnamefont {M.~H.}\ \bibnamefont {Devoret}},\ }\bibfield  {title}
  {\enquote {\bibinfo {title} {{Kerr-Free Three-Wave Mixing in Superconducting
  Quantum Circuits}},}\ }\href {\doibase 10.1103/physrevapplied.11.054060}
  {\bibfield  {journal} {\bibinfo  {journal} {Physical Review Applied}\
  }\textbf {\bibinfo {volume} {11}} (\bibinfo {year} {2019}),\
  10.1103/physrevapplied.11.054060}\BibitemShut {NoStop}%
\bibitem [{\citenamefont {Chitta}\ \emph {et~al.}(2022)\citenamefont {Chitta},
  \citenamefont {Zhao}, \citenamefont {Huang}, \citenamefont {Mondragon-Shem},\
  and\ \citenamefont {Koch}}]{Koch_Chitta}%
  \BibitemOpen
  \bibfield  {author} {\bibinfo {author} {\bibfnamefont {S.~P.}\ \bibnamefont
  {Chitta}}, \bibinfo {author} {\bibfnamefont {T.}~\bibnamefont {Zhao}},
  \bibinfo {author} {\bibfnamefont {Z.}~\bibnamefont {Huang}}, \bibinfo
  {author} {\bibfnamefont {I.}~\bibnamefont {Mondragon-Shem}}, \ and\ \bibinfo
  {author} {\bibfnamefont {J.}~\bibnamefont {Koch}},\ }\href@noop {} {\enquote
  {\bibinfo {title} {{Computer-aided quantization and numerical analysis of
  superconducting circuits}},}\ } (\bibinfo {year} {2022}),\ \Eprint
  {http://arxiv.org/abs/2206.08320} {arXiv:2206.08320} \BibitemShut {NoStop}%
\bibitem [{\citenamefont {Bergmann}(1949)}]{Bergmann1}%
  \BibitemOpen
  \bibfield  {author} {\bibinfo {author} {\bibfnamefont {P.~G.}\ \bibnamefont
  {Bergmann}},\ }\bibfield  {title} {\enquote {\bibinfo {title} {{Non-Linear
  Field Theories}},}\ }\href {\doibase 10.1103/PhysRev.75.680} {\bibfield
  {journal} {\bibinfo  {journal} {Phys. Rev.}\ }\textbf {\bibinfo {volume}
  {75}},\ \bibinfo {pages} {680--685} (\bibinfo {year} {1949})}\BibitemShut
  {NoStop}%
\bibitem [{\citenamefont {Bergmann}\ and\ \citenamefont
  {Brunings}(1949)}]{Bergmann2}%
  \BibitemOpen
  \bibfield  {author} {\bibinfo {author} {\bibfnamefont {P.~G.}\ \bibnamefont
  {Bergmann}}\ and\ \bibinfo {author} {\bibfnamefont {J.~H.~M.}\ \bibnamefont
  {Brunings}},\ }\bibfield  {title} {\enquote {\bibinfo {title} {{Non-Linear
  Field Theories II. Canonical Equations and Quantization}},}\ }\href {\doibase
  10.1103/RevModPhys.21.480} {\bibfield  {journal} {\bibinfo  {journal} {Rev.
  Mod. Phys.}\ }\textbf {\bibinfo {volume} {21}},\ \bibinfo {pages} {480--487}
  (\bibinfo {year} {1949})}\BibitemShut {NoStop}%
\bibitem [{\citenamefont {Dirac}(1950)}]{Dirac1}%
  \BibitemOpen
  \bibfield  {author} {\bibinfo {author} {\bibfnamefont {P.~A.~M.}\
  \bibnamefont {Dirac}},\ }\bibfield  {title} {\enquote {\bibinfo {title}
  {{Generalized Hamiltonian Dynamics}},}\ }\href {\doibase
  10.4153/CJM-1950-012-1} {\bibfield  {journal} {\bibinfo  {journal} {Canadian
  Journal of Mathematics}\ }\textbf {\bibinfo {volume} {2}},\ \bibinfo {pages}
  {129–148} (\bibinfo {year} {1950})}\BibitemShut {NoStop}%
\bibitem [{\citenamefont {Anderson}\ and\ \citenamefont
  {Bergmann}(1951)}]{Bergmann3}%
  \BibitemOpen
  \bibfield  {author} {\bibinfo {author} {\bibfnamefont {J.~L.}\ \bibnamefont
  {Anderson}}\ and\ \bibinfo {author} {\bibfnamefont {P.~G.}\ \bibnamefont
  {Bergmann}},\ }\bibfield  {title} {\enquote {\bibinfo {title} {{Constraints
  in Covariant Field Theories}},}\ }\href {\doibase 10.1103/PhysRev.83.1018}
  {\bibfield  {journal} {\bibinfo  {journal} {Phys. Rev.}\ }\textbf {\bibinfo
  {volume} {83}},\ \bibinfo {pages} {1018--1025} (\bibinfo {year}
  {1951})}\BibitemShut {NoStop}%
\bibitem [{\citenamefont {Dirac}(1958)}]{Dirac2}%
  \BibitemOpen
  \bibfield  {author} {\bibinfo {author} {\bibfnamefont {P.~A.~M.}\
  \bibnamefont {Dirac}},\ }\bibfield  {title} {\enquote {\bibinfo {title}
  {{Generalized Hamiltonian dynamics}},}\ }\href {\doibase
  10.1098/rspa.1958.0141} {\bibfield  {journal} {\bibinfo  {journal}
  {Proceedings of the Royal Society of London. Series A. Mathematical and
  Physical Sciences}\ }\textbf {\bibinfo {volume} {246}},\ \bibinfo {pages}
  {326--332} (\bibinfo {year} {1958})}\BibitemShut {NoStop}%
\bibitem [{\citenamefont {Dirac}(2001)}]{Dirac3}%
  \BibitemOpen
  \bibfield  {author} {\bibinfo {author} {\bibfnamefont {P.~A.~M.}\
  \bibnamefont {Dirac}},\ }\href@noop {} {\emph {\bibinfo {title} {{Lectures on
  Quantum Mechanics}}}}\ (\bibinfo  {publisher} {Dover Publications},\ \bibinfo
  {year} {2001})\BibitemShut {NoStop}%
\bibitem [{\citenamefont {Henneaux}\ and\ \citenamefont
  {Teitelboim}(1992)}]{HenneauxBook}%
  \BibitemOpen
  \bibfield  {author} {\bibinfo {author} {\bibfnamefont {M.}~\bibnamefont
  {Henneaux}}\ and\ \bibinfo {author} {\bibfnamefont {C.}~\bibnamefont
  {Teitelboim}},\ }\href {\doibase 10.1515/9780691213866} {\emph {\bibinfo
  {title} {{Quantization of Gauge Systems}}}}\ (\bibinfo  {publisher}
  {Princeton University Press},\ \bibinfo {year} {1992})\BibitemShut {NoStop}%
\bibitem [{\citenamefont {Rothe}\ and\ \citenamefont
  {Rothe}(2010)}]{RotheBook}%
  \BibitemOpen
  \bibfield  {author} {\bibinfo {author} {\bibfnamefont {H.~J.}\ \bibnamefont
  {Rothe}}\ and\ \bibinfo {author} {\bibfnamefont {K.~D.}\ \bibnamefont
  {Rothe}},\ }\href {\doibase 10.1142/7689} {\emph {\bibinfo {title}
  {{Classical and Quantum Dynamics of Constrained Hamiltonian Systems}}}}\
  (\bibinfo  {publisher} {World Scientific},\ \bibinfo {year}
  {2010})\BibitemShut {NoStop}%
\bibitem [{\citenamefont {Brown}(2022)}]{Brown}%
  \BibitemOpen
  \bibfield  {author} {\bibinfo {author} {\bibfnamefont {J.~D.}\ \bibnamefont
  {Brown}},\ }\bibfield  {title} {\enquote {\bibinfo {title} {{Singular
  Lagrangians, Constrained Hamiltonian Systems and Gauge Invariance: An Example
  of the Dirac{\textendash}Bergmann Algorithm}},}\ }\href {\doibase
  10.3390/universe8030171} {\bibfield  {journal} {\bibinfo  {journal}
  {Universe}\ }\textbf {\bibinfo {volume} {8}} (\bibinfo {year} {2022}),\
  10.3390/universe8030171}\BibitemShut {NoStop}%
\bibitem [{\citenamefont {Henneaux}\ \emph {et~al.}(1987)\citenamefont
  {Henneaux}, \citenamefont {Teitelboim},\ and\ \citenamefont
  {Zanelli}}]{Henneaux}%
  \BibitemOpen
  \bibfield  {author} {\bibinfo {author} {\bibfnamefont {M.}~\bibnamefont
  {Henneaux}}, \bibinfo {author} {\bibfnamefont {C.}~\bibnamefont
  {Teitelboim}}, \ and\ \bibinfo {author} {\bibfnamefont {J.}~\bibnamefont
  {Zanelli}},\ }\bibfield  {title} {\enquote {\bibinfo {title} {{Quantum
  mechanics for multivalued Hamiltonians}},}\ }\href {\doibase
  10.1103/PhysRevA.36.4417} {\bibfield  {journal} {\bibinfo  {journal} {Phys.
  Rev. A}\ }\textbf {\bibinfo {volume} {36}},\ \bibinfo {pages} {4417--4420}
  (\bibinfo {year} {1987})}\BibitemShut {NoStop}%
\bibitem [{\citenamefont {Shapere}\ and\ \citenamefont
  {Wilczek}(2012{\natexlab{a}})}]{Wilczek_branched}%
  \BibitemOpen
  \bibfield  {author} {\bibinfo {author} {\bibfnamefont {A.}~\bibnamefont
  {Shapere}}\ and\ \bibinfo {author} {\bibfnamefont {F.}~\bibnamefont
  {Wilczek}},\ }\bibfield  {title} {\enquote {\bibinfo {title} {{Branched
  Quantization}},}\ }\href {\doibase 10.1103/PhysRevLett.109.200402} {\bibfield
   {journal} {\bibinfo  {journal} {Phys. Rev. Lett.}\ }\textbf {\bibinfo
  {volume} {109}},\ \bibinfo {pages} {200402} (\bibinfo {year}
  {2012}{\natexlab{a}})}\BibitemShut {NoStop}%
\bibitem [{\citenamefont {Peikari}(1974)}]{Peikari}%
  \BibitemOpen
  \bibfield  {author} {\bibinfo {author} {\bibfnamefont {B.}~\bibnamefont
  {Peikari}},\ }\href@noop {} {\emph {\bibinfo {title} {{Fundamentals of
  Network Analysis and Synthesis}}}}\ (\bibinfo  {publisher} {Prentice-Hall
  Englewood Cliffs, New Jersey},\ \bibinfo {year} {1974})\BibitemShut {NoStop}%
\bibitem [{\citenamefont {Chua}\ \emph {et~al.}(1987)\citenamefont {Chua},
  \citenamefont {Desoer},\ and\ \citenamefont {Kuh}}]{DesoerKuh1}%
  \BibitemOpen
  \bibfield  {author} {\bibinfo {author} {\bibfnamefont {L.~O.}\ \bibnamefont
  {Chua}}, \bibinfo {author} {\bibfnamefont {C.~A.}\ \bibnamefont {Desoer}}, \
  and\ \bibinfo {author} {\bibfnamefont {E.~S.}\ \bibnamefont {Kuh}},\
  }\href@noop {} {\emph {\bibinfo {title} {{Linear and Nonlinear Circuits}}}}\
  (\bibinfo  {publisher} {McGraw-Hill College},\ \bibinfo {year}
  {1987})\BibitemShut {NoStop}%
\bibitem [{\citenamefont {Desoer}\ and\ \citenamefont
  {Kuh}(1969)}]{DesoerKuh2}%
  \BibitemOpen
  \bibfield  {author} {\bibinfo {author} {\bibfnamefont {C.~A.}\ \bibnamefont
  {Desoer}}\ and\ \bibinfo {author} {\bibfnamefont {E.~S.}\ \bibnamefont
  {Kuh}},\ }\href@noop {} {\emph {\bibinfo {title} {{Basic Circuit Theory}}}}\
  (\bibinfo  {publisher} {McGraw-Hill International Book Company},\ \bibinfo
  {year} {1969})\BibitemShut {NoStop}%
\bibitem [{Note1()}]{Note1}%
  \BibitemOpen
  \bibinfo {note} {When a flux-controlled inductor has a characteristic
  $f(\cdot )$ that is periodic in $\phi $, it may be appropriate (and important
  for quantization) to consider $\phi $ to be a compact rather than extended
  variable \cite {PhysRevLett.103.217004}. This subtle issue does not affect
  any of the conclusions of this paper, and we will not consider the
  compactness question any further here}\BibitemShut {NoStop}%
\bibitem [{\citenamefont {Kato}(1995)}]{Kato}%
  \BibitemOpen
  \bibfield  {author} {\bibinfo {author} {\bibfnamefont {T.}~\bibnamefont
  {Kato}},\ }\href@noop {} {\emph {\bibinfo {title} {{Perturbation Theory for
  Linear Operators}}}}\ (\bibinfo  {publisher} {Springer},\ \bibinfo {year}
  {1995})\BibitemShut {NoStop}%
\bibitem [{\citenamefont {Reed}\ and\ \citenamefont {Simon}(1978)}]{ReedSimon}%
  \BibitemOpen
  \bibfield  {author} {\bibinfo {author} {\bibfnamefont {M.}~\bibnamefont
  {Reed}}\ and\ \bibinfo {author} {\bibfnamefont {B.}~\bibnamefont {Simon}},\
  }\href@noop {} {\emph {\bibinfo {title} {Methods of Modern Mathematical
  Physics. IV Analysis of Operators}}}\ (\bibinfo  {publisher} {Academic
  Press},\ \bibinfo {year} {1978})\BibitemShut {NoStop}%
\bibitem [{Note2()}]{Note2}%
  \BibitemOpen
  \bibinfo {note} {There are exceptions to this rule. For example, as we see in
  Sec.~\ref {sec_Tellegen_transformer}, ideal transformers do not enter the
  Lagrangian of a system as an additional term but enforce a constraint on the
  variables.}\BibitemShut {Stop}%
\bibitem [{Note3()}]{Note3}%
  \BibitemOpen
  \bibinfo {note} {Typically, the derivation of a Hamiltonian describing a
  superconducting circuit involves the inversion of a capacitance matrix, which
  conventional textbooks usually assume to be accomplishable.}\BibitemShut
  {Stop}%
\bibitem [{Note4()}]{Note4}%
  \BibitemOpen
  \bibinfo {note} {We note that an alternatively proposed procedure -- the
  Faddeev-Jackiw method \cite {RotheBook, Faddeev_Jackiw}, which is equivalent
  to the Dirac-Bergmann algorithm \cite {Garcia, Non_Equivalence} -- promises a
  simplified approach as no classification of constraints is required. In this
  paper, however, we work with the Dirac-Bergmann algorithm as its application
  turns out to be straightforward for our systems.}\BibitemShut {Stop}%
\bibitem [{Note5()}]{Note5}%
  \BibitemOpen
  \bibinfo {note} {Note that different approaches of circuit quantization might
  also involve loop charges or mixtures of both sets of variables; see
  Ref.~\cite {Ulrich}. Again, we stress that the singularity of a system might
  depend on the chosen set of variables describing the system.}\BibitemShut
  {Stop}%
\bibitem [{\citenamefont {Rymarz}(2018)}]{Rymarz_Master}%
  \BibitemOpen
  \bibfield  {author} {\bibinfo {author} {\bibfnamefont {M.}~\bibnamefont
  {Rymarz}},\ }\emph {\bibinfo {title} {The Quantum Electrodynamics of Singular
  and Nonreciprocal Superconducting Circuits}},\ \href
  {https://www.quantuminfo.physik.rwth-aachen.de/global/show_document.asp?id=aaaaaaaaabgovci}
  {\bibinfo {type} {Master's thesis}},\ \bibinfo  {school} {RWTH Aachen
  University} (\bibinfo {year} {2018})\BibitemShut {NoStop}%
\bibitem [{\citenamefont {Parra-Rodriguez}\ \emph {et~al.}(2019)\citenamefont
  {Parra-Rodriguez}, \citenamefont {Egusquiza}, \citenamefont {DiVincenzo},\
  and\ \citenamefont {Solano}}]{Parra1}%
  \BibitemOpen
  \bibfield  {author} {\bibinfo {author} {\bibfnamefont {A.}~\bibnamefont
  {Parra-Rodriguez}}, \bibinfo {author} {\bibfnamefont {I.~L.}\ \bibnamefont
  {Egusquiza}}, \bibinfo {author} {\bibfnamefont {D.~P.}\ \bibnamefont
  {DiVincenzo}}, \ and\ \bibinfo {author} {\bibfnamefont {E.}~\bibnamefont
  {Solano}},\ }\bibfield  {title} {\enquote {\bibinfo {title} {{Canonical
  circuit quantization with linear nonreciprocal devices}},}\ }\href {\doibase
  10.1103/PhysRevB.99.014514} {\bibfield  {journal} {\bibinfo  {journal} {Phys.
  Rev. B}\ }\textbf {\bibinfo {volume} {99}},\ \bibinfo {pages} {014514}
  (\bibinfo {year} {2019})}\BibitemShut {NoStop}%
\bibitem [{\citenamefont {Parra-Rodriguez}\ and\ \citenamefont
  {Egusquiza}(2022{\natexlab{a}})}]{Parra2}%
  \BibitemOpen
  \bibfield  {author} {\bibinfo {author} {\bibfnamefont {A.}~\bibnamefont
  {Parra-Rodriguez}}\ and\ \bibinfo {author} {\bibfnamefont {I.~L.}\
  \bibnamefont {Egusquiza}},\ }\bibfield  {title} {\enquote {\bibinfo {title}
  {{Canonical quantisation of telegrapher's equations coupled by ideal
  nonreciprocal elements}},}\ }\href {\doibase 10.22331/q-2022-04-04-681}
  {\bibfield  {journal} {\bibinfo  {journal} {{Quantum}}\ }\textbf {\bibinfo
  {volume} {6}},\ \bibinfo {pages} {681} (\bibinfo {year}
  {2022}{\natexlab{a}})}\BibitemShut {NoStop}%
\bibitem [{\citenamefont {Parra-Rodriguez}\ and\ \citenamefont
  {Egusquiza}(2022{\natexlab{b}})}]{Parrax}%
  \BibitemOpen
  \bibfield  {author} {\bibinfo {author} {\bibfnamefont {A.}~\bibnamefont
  {Parra-Rodriguez}}\ and\ \bibinfo {author} {\bibfnamefont {I.~L.}\
  \bibnamefont {Egusquiza}},\ }\bibfield  {title} {\enquote {\bibinfo {title}
  {Quantum fluctuations in electrical multiport linear systems},}\ }\href
  {\doibase 10.1103/PhysRevB.106.054504} {\bibfield  {journal} {\bibinfo
  {journal} {Phys. Rev. B}\ }\textbf {\bibinfo {volume} {106}},\ \bibinfo
  {pages} {054504} (\bibinfo {year} {2022}{\natexlab{b}})}\BibitemShut
  {NoStop}%
\bibitem [{\citenamefont {Egusquiza}\ and\ \citenamefont
  {Parra-Rodriguez}(2022)}]{Parra4}%
  \BibitemOpen
  \bibfield  {author} {\bibinfo {author} {\bibfnamefont {I.~L.}\ \bibnamefont
  {Egusquiza}}\ and\ \bibinfo {author} {\bibfnamefont {A.}~\bibnamefont
  {Parra-Rodriguez}},\ }\bibfield  {title} {\enquote {\bibinfo {title}
  {{Algebraic canonical quantization of lumped superconducting networks}},}\
  }\href {\doibase 10.1103/PhysRevB.106.024510} {\bibfield  {journal} {\bibinfo
   {journal} {Phys. Rev. B}\ }\textbf {\bibinfo {volume} {106}},\ \bibinfo
  {pages} {024510} (\bibinfo {year} {2022})}\BibitemShut {NoStop}%
\bibitem [{\citenamefont {Clarke}\ and\ \citenamefont
  {Braginski}(2006)}]{Squid_handbook}%
  \BibitemOpen
  \bibfield  {author} {\bibinfo {author} {\bibfnamefont {J.}~\bibnamefont
  {Clarke}}\ and\ \bibinfo {author} {\bibfnamefont {A.~I.}\ \bibnamefont
  {Braginski}},\ }\href@noop {} {\emph {\bibinfo {title} {{The SQUID Handbook:
  Applications of SQUIDs and SQUID systems}}}}\ (\bibinfo  {publisher} {John
  Wiley \& Sons},\ \bibinfo {year} {2006})\BibitemShut {NoStop}%
\bibitem [{\citenamefont {Minev}\ \emph {et~al.}(2020)\citenamefont {Minev},
  \citenamefont {Leghtas}, \citenamefont {Mundhada}, \citenamefont
  {Christakis}, \citenamefont {Pop},\ and\ \citenamefont {Devoret}}]{Minev}%
  \BibitemOpen
  \bibfield  {author} {\bibinfo {author} {\bibfnamefont {Z.~K.}\ \bibnamefont
  {Minev}}, \bibinfo {author} {\bibfnamefont {Z.}~\bibnamefont {Leghtas}},
  \bibinfo {author} {\bibfnamefont {S.~O.}\ \bibnamefont {Mundhada}}, \bibinfo
  {author} {\bibfnamefont {L.}~\bibnamefont {Christakis}}, \bibinfo {author}
  {\bibfnamefont {I.~M.}\ \bibnamefont {Pop}}, \ and\ \bibinfo {author}
  {\bibfnamefont {M.~H.}\ \bibnamefont {Devoret}},\ }\href@noop {} {\enquote
  {\bibinfo {title} {{Energy-participation quantization of Josephson
  circuits}},}\ } (\bibinfo {year} {2020}),\ \Eprint
  {http://arxiv.org/abs/2010.00620v3} {arXiv:2010.00620v3} \BibitemShut
  {NoStop}%
\bibitem [{\citenamefont {Richer}\ \emph {et~al.}(2017)\citenamefont {Richer},
  \citenamefont {Maleeva}, \citenamefont {Skacel}, \citenamefont {Pop},\ and\
  \citenamefont {DiVincenzo}}]{Richer2}%
  \BibitemOpen
  \bibfield  {author} {\bibinfo {author} {\bibfnamefont {S.}~\bibnamefont
  {Richer}}, \bibinfo {author} {\bibfnamefont {N.}~\bibnamefont {Maleeva}},
  \bibinfo {author} {\bibfnamefont {S.~T.}\ \bibnamefont {Skacel}}, \bibinfo
  {author} {\bibfnamefont {I.~M.}\ \bibnamefont {Pop}}, \ and\ \bibinfo
  {author} {\bibfnamefont {D.}~\bibnamefont {DiVincenzo}},\ }\bibfield  {title}
  {\enquote {\bibinfo {title} {{Inductively shunted transmon qubit with tunable
  transverse and longitudinal coupling}},}\ }\href {\doibase
  10.1103/PhysRevB.96.174520} {\bibfield  {journal} {\bibinfo  {journal} {Phys.
  Rev. B}\ }\textbf {\bibinfo {volume} {96}},\ \bibinfo {pages} {174520}
  (\bibinfo {year} {2017})}\BibitemShut {NoStop}%
\bibitem [{\citenamefont {Teitelboim}\ and\ \citenamefont
  {Zanelli}(1987)}]{Teitelboim}%
  \BibitemOpen
  \bibfield  {author} {\bibinfo {author} {\bibfnamefont {C.}~\bibnamefont
  {Teitelboim}}\ and\ \bibinfo {author} {\bibfnamefont {J.}~\bibnamefont
  {Zanelli}},\ }\bibfield  {title} {\enquote {\bibinfo {title} {{Dimensionally
  continued topological gravitation theory in Hamiltonian form}},}\ }\href
  {\doibase 10.1088/0264-9381/4/4/010} {\bibfield  {journal} {\bibinfo
  {journal} {Classical and Quantum Gravity}\ }\textbf {\bibinfo {volume} {4}},\
  \bibinfo {pages} {L125} (\bibinfo {year} {1987})}\BibitemShut {NoStop}%
\bibitem [{\citenamefont {Shapere}\ and\ \citenamefont
  {Wilczek}(2012{\natexlab{b}})}]{Wilczek_classical}%
  \BibitemOpen
  \bibfield  {author} {\bibinfo {author} {\bibfnamefont {A.}~\bibnamefont
  {Shapere}}\ and\ \bibinfo {author} {\bibfnamefont {F.}~\bibnamefont
  {Wilczek}},\ }\bibfield  {title} {\enquote {\bibinfo {title} {{Classical Time
  Crystals}},}\ }\href {\doibase 10.1103/PhysRevLett.109.160402} {\bibfield
  {journal} {\bibinfo  {journal} {Phys. Rev. Lett.}\ }\textbf {\bibinfo
  {volume} {109}},\ \bibinfo {pages} {160402} (\bibinfo {year}
  {2012}{\natexlab{b}})}\BibitemShut {NoStop}%
\bibitem [{Note6()}]{Note6}%
  \BibitemOpen
  \bibinfo {note} {The multi-valuedness of a branched Hamiltonian in the
  momentum might originate from a non-convex velocity dependence of the
  corresponding Lagrangian.}\BibitemShut {Stop}%
\bibitem [{\citenamefont {Zhao}\ \emph {et~al.}(2013)\citenamefont {Zhao},
  \citenamefont {Yu},\ and\ \citenamefont {Xu}}]{Zhao}%
  \BibitemOpen
  \bibfield  {author} {\bibinfo {author} {\bibfnamefont {L.}~\bibnamefont
  {Zhao}}, \bibinfo {author} {\bibfnamefont {P.}~\bibnamefont {Yu}}, \ and\
  \bibinfo {author} {\bibfnamefont {W.}~\bibnamefont {Xu}},\ }\bibfield
  {title} {\enquote {\bibinfo {title} {{Hamiltonian description of singular
  Lagrangian systems with spontaneously broken time translation symmetry}},}\
  }\href {\doibase 10.1142/s0217732313500028} {\bibfield  {journal} {\bibinfo
  {journal} {Modern Physics Letters A}\ }\textbf {\bibinfo {volume} {28}},\
  \bibinfo {pages} {1350002} (\bibinfo {year} {2013})}\BibitemShut {NoStop}%
\bibitem [{\citenamefont {Wilczek}(2012)}]{Wilczek_quantum}%
  \BibitemOpen
  \bibfield  {author} {\bibinfo {author} {\bibfnamefont {F.}~\bibnamefont
  {Wilczek}},\ }\bibfield  {title} {\enquote {\bibinfo {title} {{Quantum Time
  Crystals}},}\ }\href {\doibase 10.1103/PhysRevLett.109.160401} {\bibfield
  {journal} {\bibinfo  {journal} {Phys. Rev. Lett.}\ }\textbf {\bibinfo
  {volume} {109}},\ \bibinfo {pages} {160401} (\bibinfo {year}
  {2012})}\BibitemShut {NoStop}%
\bibitem [{\citenamefont {Bruno}(2013)}]{Bruno}%
  \BibitemOpen
  \bibfield  {author} {\bibinfo {author} {\bibfnamefont {P.}~\bibnamefont
  {Bruno}},\ }\bibfield  {title} {\enquote {\bibinfo {title} {{Impossibility of
  Spontaneously Rotating Time Crystals: A No-Go Theorem}},}\ }\href {\doibase
  10.1103/PhysRevLett.111.070402} {\bibfield  {journal} {\bibinfo  {journal}
  {Phys. Rev. Lett.}\ }\textbf {\bibinfo {volume} {111}},\ \bibinfo {pages}
  {070402} (\bibinfo {year} {2013})}\BibitemShut {NoStop}%
\bibitem [{\citenamefont {Watanabe}\ and\ \citenamefont
  {Oshikawa}(2015)}]{Watanabe}%
  \BibitemOpen
  \bibfield  {author} {\bibinfo {author} {\bibfnamefont {H.}~\bibnamefont
  {Watanabe}}\ and\ \bibinfo {author} {\bibfnamefont {M.}~\bibnamefont
  {Oshikawa}},\ }\bibfield  {title} {\enquote {\bibinfo {title} {{Absence of
  Quantum Time Crystals}},}\ }\href {\doibase 10.1103/PhysRevLett.114.251603}
  {\bibfield  {journal} {\bibinfo  {journal} {Phys. Rev. Lett.}\ }\textbf
  {\bibinfo {volume} {114}},\ \bibinfo {pages} {251603} (\bibinfo {year}
  {2015})}\BibitemShut {NoStop}%
\bibitem [{Note7()}]{Note7}%
  \BibitemOpen
  \bibinfo {note} {Working with flux variables as generalized positions, the
  parasitic capacitance of an inductive element adds a term to the Lagrangian
  that is quadratic in the generalized velocity. This is basically equivalent
  to adding a mass to an otherwise massless particle, which, of course, is not
  a correct step in all physical theories.}\BibitemShut {Stop}%
\bibitem [{\citenamefont {Burkard}\ \emph {et~al.}(2004)\citenamefont
  {Burkard}, \citenamefont {Koch},\ and\ \citenamefont {DiVincenzo}}]{BKD}%
  \BibitemOpen
  \bibfield  {author} {\bibinfo {author} {\bibfnamefont {G.}~\bibnamefont
  {Burkard}}, \bibinfo {author} {\bibfnamefont {R.~H.}\ \bibnamefont {Koch}}, \
  and\ \bibinfo {author} {\bibfnamefont {D.~P.}\ \bibnamefont {DiVincenzo}},\
  }\bibfield  {title} {\enquote {\bibinfo {title} {{Multilevel quantum
  description of decoherence in superconducting qubits}},}\ }\href {\doibase
  10.1103/PhysRevB.69.064503} {\bibfield  {journal} {\bibinfo  {journal} {Phys.
  Rev. B}\ }\textbf {\bibinfo {volume} {69}},\ \bibinfo {pages} {064503}
  (\bibinfo {year} {2004})}\BibitemShut {NoStop}%
\bibitem [{\citenamefont {{Born}}\ and\ \citenamefont
  {{Oppenheimer}}(1927)}]{Born_Oppenheimer}%
  \BibitemOpen
  \bibfield  {author} {\bibinfo {author} {\bibfnamefont {M.}~\bibnamefont
  {{Born}}}\ and\ \bibinfo {author} {\bibfnamefont {R.}~\bibnamefont
  {{Oppenheimer}}},\ }\bibfield  {title} {\enquote {\bibinfo {title} {{Zur
  Quantentheorie der Molekeln}},}\ }\href {\doibase 10.1002/andp.19273892002}
  {\bibfield  {journal} {\bibinfo  {journal} {Annalen der Physik}\ }\textbf
  {\bibinfo {volume} {389}},\ \bibinfo {pages} {457--484} (\bibinfo {year}
  {1927})}\BibitemShut {NoStop}%
\bibitem [{\citenamefont {DiVincenzo}\ \emph {et~al.}(2006)\citenamefont
  {DiVincenzo}, \citenamefont {Brito},\ and\ \citenamefont
  {Koch}}]{PhysRevB.74.014514}%
  \BibitemOpen
  \bibfield  {author} {\bibinfo {author} {\bibfnamefont {D.~P.}\ \bibnamefont
  {DiVincenzo}}, \bibinfo {author} {\bibfnamefont {F.}~\bibnamefont {Brito}}, \
  and\ \bibinfo {author} {\bibfnamefont {R.~H.}\ \bibnamefont {Koch}},\
  }\bibfield  {title} {\enquote {\bibinfo {title} {{Decoherence rates in
  complex Josephson qubit circuits}},}\ }\href {\doibase
  10.1103/PhysRevB.74.014514} {\bibfield  {journal} {\bibinfo  {journal} {Phys.
  Rev. B}\ }\textbf {\bibinfo {volume} {74}},\ \bibinfo {pages} {014514}
  (\bibinfo {year} {2006})}\BibitemShut {NoStop}%
\bibitem [{Note8()}]{Note8}%
  \BibitemOpen
  \bibinfo {note} {In the molecular physics literature, $H_\protect \text
  {fast}$ is called the electronic Hamiltonian as it describes the motion of
  the fast (light) electrons for fixed positions of the slow (heavy)
  nuclei.}\BibitemShut {Stop}%
\bibitem [{Note9()}]{Note9}%
  \BibitemOpen
  \bibinfo {note} {These definitions are structured to exclude from
  consideration potential functions that deviate very slowly from a linear
  inductance, e.g., $U_{nl}(\phi _c) \sim \phi _c^2/\log {(\phi _c^2+4)}$. We
  suspect that this potential behaves like a type-1 case, but the proofs of our
  theorems do not apply. Note that such functions are also excluded by the
  second part of the definition of type-L inductors.}\BibitemShut {Stop}%
\bibitem [{Note10()}]{Note10}%
  \BibitemOpen
  \bibinfo {note} {Sublinear and superlinear refer to the form of the classical
  two-terminal current-versus-flux characteristic. When this relation is
  linear, the coefficient of proportionality is an inverse inductance, referred
  to as a reluctance in the electrical literature.}\BibitemShut {Stop}%
\bibitem [{\citenamefont {{Morse}}\ and\ \citenamefont
  {{Feshbach}}(1953)}]{1953MF}%
  \BibitemOpen
  \bibfield  {author} {\bibinfo {author} {\bibfnamefont {P.~M.}\ \bibnamefont
  {{Morse}}}\ and\ \bibinfo {author} {\bibfnamefont {H.}~\bibnamefont
  {{Feshbach}}},\ }\href@noop {} {\emph {\bibinfo {title} {{Methods of
  Theoretical Physics}}}}\ (\bibinfo  {publisher} {International Series in Pure
  and Applied Physics},\ \bibinfo {year} {1953})\BibitemShut {NoStop}%
\bibitem [{Note11()}]{Note11}%
  \BibitemOpen
  \bibinfo {note} {Assume $H = H_0 + \lambda ^{2q}A + \lambda ^{q}B + \lambda
  ^{2q-p}C$ to be an analytic family in the sense of Kato, where the operators
  $H_0, A, B,C$ do not depend on $\lambda $. Then, within some finite radius of
  convergence, the Rayleigh-Schr\IeC {\"o}dinger series is analytic in $\lambda
  $, and the only powers of $\lambda $ are given by $k=i(2q)+j(q)+l(2q-p)$ with
  $i,j,l \in \protect \mathbb {N}_0$. Thus, with $q<2q-p<2q$, there are only
  four possible values of $k$ such that $k \leq 2q$.}\BibitemShut {Stop}%
\bibitem [{\citenamefont {Golubov}\ \emph {et~al.}(2004)\citenamefont
  {Golubov}, \citenamefont {Kupriyanov},\ and\ \citenamefont
  {Il'ichev}}]{RMP_Golubov_Kupriyanov_Ilichev}%
  \BibitemOpen
  \bibfield  {author} {\bibinfo {author} {\bibfnamefont {A.~A.}\ \bibnamefont
  {Golubov}}, \bibinfo {author} {\bibfnamefont {M.~Yu.}\ \bibnamefont
  {Kupriyanov}}, \ and\ \bibinfo {author} {\bibfnamefont {E.}~\bibnamefont
  {Il'ichev}},\ }\bibfield  {title} {\enquote {\bibinfo {title} {{The
  current-phase relation in Josephson junctions}},}\ }\href {\doibase
  10.1103/RevModPhys.76.411} {\bibfield  {journal} {\bibinfo  {journal} {Rev.
  Mod. Phys.}\ }\textbf {\bibinfo {volume} {76}},\ \bibinfo {pages} {411--469}
  (\bibinfo {year} {2004})}\BibitemShut {NoStop}%
\bibitem [{\citenamefont {Simon}\ and\ \citenamefont
  {Dicke}(1970)}]{Simon_article_long}%
  \BibitemOpen
  \bibfield  {author} {\bibinfo {author} {\bibfnamefont {B.}~\bibnamefont
  {Simon}}\ and\ \bibinfo {author} {\bibfnamefont {A.}~\bibnamefont {Dicke}},\
  }\bibfield  {title} {\enquote {\bibinfo {title} {{Coupling constant
  analyticity for the anharmonic oscillator}},}\ }\href {\doibase
  https://doi.org/10.1016/0003-4916(70)90240-X} {\bibfield  {journal} {\bibinfo
   {journal} {Annals of Physics}\ }\textbf {\bibinfo {volume} {58}},\ \bibinfo
  {pages} {76--136} (\bibinfo {year} {1970})}\BibitemShut {NoStop}%
\bibitem [{\citenamefont {Simon}(1982)}]{Simon_article_short}%
  \BibitemOpen
  \bibfield  {author} {\bibinfo {author} {\bibfnamefont {B.}~\bibnamefont
  {Simon}},\ }\bibfield  {title} {\enquote {\bibinfo {title} {{Large orders and
  summability of eigenvalue perturbation theory: A mathematical overview}},}\
  }\href {\doibase https://doi.org/10.1002/qua.560210103} {\bibfield  {journal}
  {\bibinfo  {journal} {International Journal of Quantum Chemistry}\ }\textbf
  {\bibinfo {volume} {21}},\ \bibinfo {pages} {3--25} (\bibinfo {year}
  {1982})}\BibitemShut {NoStop}%
\bibitem [{\citenamefont {Noguchi}\ \emph {et~al.}(2020)\citenamefont
  {Noguchi}, \citenamefont {Osada}, \citenamefont {Masuda}, \citenamefont
  {Kono}, \citenamefont {Heya}, \citenamefont {Wolski}, \citenamefont
  {Takahashi}, \citenamefont {Sugiyama}, \citenamefont {Lachance-Quirion},\
  and\ \citenamefont {Nakamura}}]{NakamuraSNAIL}%
  \BibitemOpen
  \bibfield  {author} {\bibinfo {author} {\bibfnamefont {A.}~\bibnamefont
  {Noguchi}}, \bibinfo {author} {\bibfnamefont {A.}~\bibnamefont {Osada}},
  \bibinfo {author} {\bibfnamefont {S.}~\bibnamefont {Masuda}}, \bibinfo
  {author} {\bibfnamefont {S.}~\bibnamefont {Kono}}, \bibinfo {author}
  {\bibfnamefont {K.}~\bibnamefont {Heya}}, \bibinfo {author} {\bibfnamefont
  {S.~P.}\ \bibnamefont {Wolski}}, \bibinfo {author} {\bibfnamefont
  {H.}~\bibnamefont {Takahashi}}, \bibinfo {author} {\bibfnamefont
  {T.}~\bibnamefont {Sugiyama}}, \bibinfo {author} {\bibfnamefont
  {D.}~\bibnamefont {Lachance-Quirion}}, \ and\ \bibinfo {author}
  {\bibfnamefont {Y.}~\bibnamefont {Nakamura}},\ }\bibfield  {title} {\enquote
  {\bibinfo {title} {{Fast parametric two-qubit gates with suppressed residual
  interaction using the second-order nonlinearity of a cubic transmon}},}\
  }\href {\doibase 10.1103/PhysRevA.102.062408} {\bibfield  {journal} {\bibinfo
   {journal} {Phys. Rev. A}\ }\textbf {\bibinfo {volume} {102}},\ \bibinfo
  {pages} {062408} (\bibinfo {year} {2020})}\BibitemShut {NoStop}%
\bibitem [{Note12()}]{Note12}%
  \BibitemOpen
  \bibinfo {note} {Although SNAILs are mostly operated with $N=3$ large
  Josephson junctions, we focus on the case $N=2$ for the sake of
  simplicity.}\BibitemShut {Stop}%
\bibitem [{\citenamefont {Koch}\ \emph {et~al.}(2007)\citenamefont {Koch},
  \citenamefont {Yu}, \citenamefont {Gambetta}, \citenamefont {Houck},
  \citenamefont {Schuster}, \citenamefont {Majer}, \citenamefont {Blais},
  \citenamefont {Devoret}, \citenamefont {Girvin},\ and\ \citenamefont
  {Schoelkopf}}]{Koch}%
  \BibitemOpen
  \bibfield  {author} {\bibinfo {author} {\bibfnamefont {J.}~\bibnamefont
  {Koch}}, \bibinfo {author} {\bibfnamefont {T.~M.}\ \bibnamefont {Yu}},
  \bibinfo {author} {\bibfnamefont {J.}~\bibnamefont {Gambetta}}, \bibinfo
  {author} {\bibfnamefont {A.~A.}\ \bibnamefont {Houck}}, \bibinfo {author}
  {\bibfnamefont {D.~I.}\ \bibnamefont {Schuster}}, \bibinfo {author}
  {\bibfnamefont {J.}~\bibnamefont {Majer}}, \bibinfo {author} {\bibfnamefont
  {A.}~\bibnamefont {Blais}}, \bibinfo {author} {\bibfnamefont {M.~H.}\
  \bibnamefont {Devoret}}, \bibinfo {author} {\bibfnamefont {S.~M.}\
  \bibnamefont {Girvin}}, \ and\ \bibinfo {author} {\bibfnamefont {R.~J.}\
  \bibnamefont {Schoelkopf}},\ }\bibfield  {title} {\enquote {\bibinfo {title}
  {{Charge-insensitive qubit design derived from the Cooper pair box}},}\
  }\href {\doibase 10.1103/PhysRevA.76.042319} {\bibfield  {journal} {\bibinfo
  {journal} {Phys. Rev. A}\ }\textbf {\bibinfo {volume} {76}},\ \bibinfo
  {pages} {042319} (\bibinfo {year} {2007})}\BibitemShut {NoStop}%
\bibitem [{Note13()}]{Note13}%
  \BibitemOpen
  \bibinfo {note} {A similar renormalization of the Josephson energy also
  occurs in the analysis of nonreciprocal circuits; see Sec.~\ref
  {SubSubSec_Gyrator_GKP}}\BibitemShut {NoStop}%
\bibitem [{\citenamefont {Di~Paolo}\ \emph {et~al.}(2021)\citenamefont
  {Di~Paolo}, \citenamefont {Baker}, \citenamefont {Foley}, \citenamefont
  {S{\'e}n{\'e}chal},\ and\ \citenamefont {Blais}}]{DiPaoloBlais}%
  \BibitemOpen
  \bibfield  {author} {\bibinfo {author} {\bibfnamefont {A.}~\bibnamefont
  {Di~Paolo}}, \bibinfo {author} {\bibfnamefont {T.~E.}\ \bibnamefont {Baker}},
  \bibinfo {author} {\bibfnamefont {A.}~\bibnamefont {Foley}}, \bibinfo
  {author} {\bibfnamefont {D.}~\bibnamefont {S{\'e}n{\'e}chal}}, \ and\
  \bibinfo {author} {\bibfnamefont {A.}~\bibnamefont {Blais}},\ }\bibfield
  {title} {\enquote {\bibinfo {title} {{Efficient modeling of superconducting
  quantum circuits with tensor networks}},}\ }\href {\doibase
  10.1038/s41534-020-00352-4} {\bibfield  {journal} {\bibinfo  {journal} {npj
  Quantum Information}\ }\textbf {\bibinfo {volume} {7}},\ \bibinfo {pages}
  {11} (\bibinfo {year} {2021})}\BibitemShut {NoStop}%
\bibitem [{Note14()}]{Note14}%
  \BibitemOpen
  \bibinfo {note} {Note that the boundary conditions in Eq.~\protect \textup
  {\hbox {\mathsurround \z@ \protect \normalfont (\ignorespaces \ref
  {eq_SNAIL_BC_old}\unskip \@@italiccorr )}} imply a preceding unitary gauge
  transformation of the initial wave function as the offset charges $\nu _1,
  \nu _2$ do not enter the Hamiltonian in Eq.~\protect \textup {\hbox
  {\mathsurround \z@ \protect \normalfont (\ignorespaces \ref
  {eq_Hamiltonian_SNAIL}\unskip \@@italiccorr )}}.}\BibitemShut {Stop}%
\bibitem [{\citenamefont {Tellegen}(1948)}]{Tellegen}%
  \BibitemOpen
  \bibfield  {author} {\bibinfo {author} {\bibfnamefont {B.}~\bibnamefont
  {Tellegen}},\ }\bibfield  {title} {\enquote {\bibinfo {title} {{The gyrator,
  a new electric network element}},}\ }\href@noop {} {\bibfield  {journal}
  {\bibinfo  {journal} {Philips Res. Rep}\ }\textbf {\bibinfo {volume} {3}},\
  \bibinfo {pages} {81--101} (\bibinfo {year} {1948})}\BibitemShut {NoStop}%
\bibitem [{\citenamefont {{Duinker, S.}}(1959)}]{Duinker}%
  \BibitemOpen
  \bibfield  {author} {\bibinfo {author} {\bibnamefont {{Duinker, S.}}},\
  }\bibfield  {title} {\enquote {\bibinfo {title} {{Traditors, a new class of
  non-energic non-linear network elements}},}\ }\href@noop {} {\bibfield
  {journal} {\bibinfo  {journal} {Philips Res. Rep.}\ }\textbf {\bibinfo
  {volume} {14}},\ \bibinfo {pages} {29--51} (\bibinfo {year}
  {1959})}\BibitemShut {NoStop}%
\bibitem [{\citenamefont {{A. Adamatzky and G. Chen}}(2013)}]{FestschriftChua}%
  \BibitemOpen
  \bibfield  {author} {\bibinfo {author} {\bibnamefont {{A. Adamatzky and G.
  Chen}}},\ }\href@noop {} {\emph {\bibinfo {title} {{Chaos, CNN, Memristors
  and Beyond: A Festschrift for Leon Chua}}}}\ (\bibinfo  {publisher} {World
  Scientific},\ \bibinfo {year} {2013})\BibitemShut {NoStop}%
\bibitem [{Note15()}]{Note15}%
  \BibitemOpen
  \bibinfo {note} {We have confirmed that the Dirac-Bergmann result here is
  gauge invariant.}\BibitemShut {Stop}%
\bibitem [{\citenamefont {Bellissard}\ and\ \citenamefont
  {Simon}(1982)}]{BellissardSimon}%
  \BibitemOpen
  \bibfield  {author} {\bibinfo {author} {\bibfnamefont {J.}~\bibnamefont
  {Bellissard}}\ and\ \bibinfo {author} {\bibfnamefont {B.}~\bibnamefont
  {Simon}},\ }\bibfield  {title} {\enquote {\bibinfo {title} {{Cantor spectrum
  for the almost Mathieu equation}},}\ }\href {\doibase
  https://doi.org/10.1016/0022-1236(82)90094-5} {\bibfield  {journal} {\bibinfo
   {journal} {Journal of Functional Analysis}\ }\textbf {\bibinfo {volume}
  {48}},\ \bibinfo {pages} {408--419} (\bibinfo {year} {1982})}\BibitemShut
  {NoStop}%
\bibitem [{\citenamefont {Gottesman}\ \emph {et~al.}(2001)\citenamefont
  {Gottesman}, \citenamefont {Kitaev},\ and\ \citenamefont {Preskill}}]{GKP}%
  \BibitemOpen
  \bibfield  {author} {\bibinfo {author} {\bibfnamefont {D.}~\bibnamefont
  {Gottesman}}, \bibinfo {author} {\bibfnamefont {A.}~\bibnamefont {Kitaev}}, \
  and\ \bibinfo {author} {\bibfnamefont {J.}~\bibnamefont {Preskill}},\
  }\bibfield  {title} {\enquote {\bibinfo {title} {{Encoding a qubit in an
  oscillator}},}\ }\href {\doibase 10.1103/PhysRevA.64.012310} {\bibfield
  {journal} {\bibinfo  {journal} {Phys. Rev. A}\ }\textbf {\bibinfo {volume}
  {64}},\ \bibinfo {pages} {012310} (\bibinfo {year} {2001})}\BibitemShut
  {NoStop}%
\bibitem [{\citenamefont {Kane}\ and\ \citenamefont
  {Fisher}(1992)}]{KaneFisher}%
  \BibitemOpen
  \bibfield  {author} {\bibinfo {author} {\bibfnamefont {C.~L.}\ \bibnamefont
  {Kane}}\ and\ \bibinfo {author} {\bibfnamefont {M.~P.~A.}\ \bibnamefont
  {Fisher}},\ }\bibfield  {title} {\enquote {\bibinfo {title} {{Transport in a
  one-channel Luttinger liquid}},}\ }\href {\doibase
  10.1103/PhysRevLett.68.1220} {\bibfield  {journal} {\bibinfo  {journal}
  {Phys. Rev. Lett.}\ }\textbf {\bibinfo {volume} {68}},\ \bibinfo {pages}
  {1220--1223} (\bibinfo {year} {1992})}\BibitemShut {NoStop}%
\bibitem [{\citenamefont {Fang}\ \emph {et~al.}(2007)\citenamefont {Fang},
  \citenamefont {Konar}, \citenamefont {Xing},\ and\ \citenamefont
  {Jena}}]{Graphene_Capacitance_Fang}%
  \BibitemOpen
  \bibfield  {author} {\bibinfo {author} {\bibfnamefont {T.}~\bibnamefont
  {Fang}}, \bibinfo {author} {\bibfnamefont {A.}~\bibnamefont {Konar}},
  \bibinfo {author} {\bibfnamefont {H.}~\bibnamefont {Xing}}, \ and\ \bibinfo
  {author} {\bibfnamefont {D.}~\bibnamefont {Jena}},\ }\bibfield  {title}
  {\enquote {\bibinfo {title} {{Carrier statistics and quantum capacitance of
  graphene sheets and ribbons}},}\ }\href {\doibase 10.1063/1.2776887}
  {\bibfield  {journal} {\bibinfo  {journal} {Applied Physics Letters}\
  }\textbf {\bibinfo {volume} {91}},\ \bibinfo {pages} {092109} (\bibinfo
  {year} {2007})}\BibitemShut {NoStop}%
\bibitem [{\citenamefont {Xia}\ \emph {et~al.}(2009)\citenamefont {Xia},
  \citenamefont {Chen}, \citenamefont {Li},\ and\ \citenamefont
  {Tao}}]{Graphene_Capacitance_Xia}%
  \BibitemOpen
  \bibfield  {author} {\bibinfo {author} {\bibfnamefont {J.}~\bibnamefont
  {Xia}}, \bibinfo {author} {\bibfnamefont {F.}~\bibnamefont {Chen}}, \bibinfo
  {author} {\bibfnamefont {J.}~\bibnamefont {Li}}, \ and\ \bibinfo {author}
  {\bibfnamefont {N.}~\bibnamefont {Tao}},\ }\bibfield  {title} {\enquote
  {\bibinfo {title} {{Measurement of the quantum capacitance of graphene}},}\
  }\href {\doibase 10.1038/nnano.2009.177} {\bibfield  {journal} {\bibinfo
  {journal} {Nature Nanotechnology}\ }\textbf {\bibinfo {volume} {4}},\
  \bibinfo {pages} {505--509} (\bibinfo {year} {2009})}\BibitemShut {NoStop}%
\bibitem [{\citenamefont {Dröscher}\ \emph {et~al.}(2010)\citenamefont
  {Dröscher}, \citenamefont {Roulleau}, \citenamefont {Molitor}, \citenamefont
  {Studerus}, \citenamefont {Stampfer}, \citenamefont {Ensslin},\ and\
  \citenamefont {Ihn}}]{Graphene_Capacitance_Stampfer}%
  \BibitemOpen
  \bibfield  {author} {\bibinfo {author} {\bibfnamefont {S.}~\bibnamefont
  {Dröscher}}, \bibinfo {author} {\bibfnamefont {P.}~\bibnamefont {Roulleau}},
  \bibinfo {author} {\bibfnamefont {F.}~\bibnamefont {Molitor}}, \bibinfo
  {author} {\bibfnamefont {P.}~\bibnamefont {Studerus}}, \bibinfo {author}
  {\bibfnamefont {C.}~\bibnamefont {Stampfer}}, \bibinfo {author}
  {\bibfnamefont {K.}~\bibnamefont {Ensslin}}, \ and\ \bibinfo {author}
  {\bibfnamefont {T.}~\bibnamefont {Ihn}},\ }\bibfield  {title} {\enquote
  {\bibinfo {title} {{Quantum capacitance and density of states of
  graphene}},}\ }\href {\doibase 10.1063/1.3391670} {\bibfield  {journal}
  {\bibinfo  {journal} {Applied Physics Letters}\ }\textbf {\bibinfo {volume}
  {96}},\ \bibinfo {pages} {152104} (\bibinfo {year} {2010})}\BibitemShut
  {NoStop}%
\bibitem [{\citenamefont {Le}\ \emph {et~al.}(2019)\citenamefont {Le},
  \citenamefont {Grimsmo}, \citenamefont {M\"uller},\ and\ \citenamefont
  {Stace}}]{PhysRevA.100.062321}%
  \BibitemOpen
  \bibfield  {author} {\bibinfo {author} {\bibfnamefont {D.~T.}\ \bibnamefont
  {Le}}, \bibinfo {author} {\bibfnamefont {A.}~\bibnamefont {Grimsmo}},
  \bibinfo {author} {\bibfnamefont {C.}~\bibnamefont {M\"uller}}, \ and\
  \bibinfo {author} {\bibfnamefont {T.~M.}\ \bibnamefont {Stace}},\ }\bibfield
  {title} {\enquote {\bibinfo {title} {Doubly nonlinear superconducting
  qubit},}\ }\href {\doibase 10.1103/PhysRevA.100.062321} {\bibfield  {journal}
  {\bibinfo  {journal} {Phys. Rev. A}\ }\textbf {\bibinfo {volume} {100}},\
  \bibinfo {pages} {062321} (\bibinfo {year} {2019})}\BibitemShut {NoStop}%
\bibitem [{\citenamefont {Ulrich}\ and\ \citenamefont
  {Hassler}(2016)}]{Ulrich}%
  \BibitemOpen
  \bibfield  {author} {\bibinfo {author} {\bibfnamefont {J.}~\bibnamefont
  {Ulrich}}\ and\ \bibinfo {author} {\bibfnamefont {F.}~\bibnamefont
  {Hassler}},\ }\bibfield  {title} {\enquote {\bibinfo {title} {{Dual approach
  to circuit quantization using loop charges}},}\ }\href {\doibase
  10.1103/PhysRevB.94.094505} {\bibfield  {journal} {\bibinfo  {journal} {Phys.
  Rev. B}\ }\textbf {\bibinfo {volume} {94}},\ \bibinfo {pages} {094505}
  (\bibinfo {year} {2016})}\BibitemShut {NoStop}%
\bibitem [{\citenamefont {Rosenthal}\ \emph {et~al.}(2017)\citenamefont
  {Rosenthal}, \citenamefont {Chapman}, \citenamefont {Higginbotham},
  \citenamefont {Kerckhoff},\ and\ \citenamefont {Lehnert}}]{Rosenthal}%
  \BibitemOpen
  \bibfield  {author} {\bibinfo {author} {\bibfnamefont {E.~I.}\ \bibnamefont
  {Rosenthal}}, \bibinfo {author} {\bibfnamefont {B.~J.}\ \bibnamefont
  {Chapman}}, \bibinfo {author} {\bibfnamefont {A.~P.}\ \bibnamefont
  {Higginbotham}}, \bibinfo {author} {\bibfnamefont {J.}~\bibnamefont
  {Kerckhoff}}, \ and\ \bibinfo {author} {\bibfnamefont {K.~W.}\ \bibnamefont
  {Lehnert}},\ }\bibfield  {title} {\enquote {\bibinfo {title} {Breaking
  lorentz reciprocity with frequency conversion and delay},}\ }\href {\doibase
  10.1103/PhysRevLett.119.147703} {\bibfield  {journal} {\bibinfo  {journal}
  {Phys. Rev. Lett.}\ }\textbf {\bibinfo {volume} {119}},\ \bibinfo {pages}
  {147703} (\bibinfo {year} {2017})}\BibitemShut {NoStop}%
\bibitem [{\citenamefont {Chapman}\ \emph {et~al.}(2017)\citenamefont
  {Chapman}, \citenamefont {Rosenthal}, \citenamefont {Kerckhoff},
  \citenamefont {Moores}, \citenamefont {Vale}, \citenamefont {Mates},
  \citenamefont {Hilton}, \citenamefont {Lalumi\`ere}, \citenamefont {Blais},\
  and\ \citenamefont {Lehnert}}]{Chapman}%
  \BibitemOpen
  \bibfield  {author} {\bibinfo {author} {\bibfnamefont {B.~J.}\ \bibnamefont
  {Chapman}}, \bibinfo {author} {\bibfnamefont {E.~I.}\ \bibnamefont
  {Rosenthal}}, \bibinfo {author} {\bibfnamefont {J.}~\bibnamefont
  {Kerckhoff}}, \bibinfo {author} {\bibfnamefont {B.~A.}\ \bibnamefont
  {Moores}}, \bibinfo {author} {\bibfnamefont {L.~R.}\ \bibnamefont {Vale}},
  \bibinfo {author} {\bibfnamefont {J.~A.~B.}\ \bibnamefont {Mates}}, \bibinfo
  {author} {\bibfnamefont {G.~C.}\ \bibnamefont {Hilton}}, \bibinfo {author}
  {\bibfnamefont {K.}~\bibnamefont {Lalumi\`ere}}, \bibinfo {author}
  {\bibfnamefont {A.}~\bibnamefont {Blais}}, \ and\ \bibinfo {author}
  {\bibfnamefont {K.~W.}\ \bibnamefont {Lehnert}},\ }\bibfield  {title}
  {\enquote {\bibinfo {title} {{Widely Tunable On-Chip Microwave Circulator for
  Superconducting Quantum Circuits}},}\ }\href {\doibase
  10.1103/PhysRevX.7.041043} {\bibfield  {journal} {\bibinfo  {journal} {Phys.
  Rev. X}\ }\textbf {\bibinfo {volume} {7}},\ \bibinfo {pages} {041043}
  (\bibinfo {year} {2017})}\BibitemShut {NoStop}%
\bibitem [{\citenamefont {Lecocq}\ \emph {et~al.}(2017)\citenamefont {Lecocq},
  \citenamefont {Ranzani}, \citenamefont {Peterson}, \citenamefont {Cicak},
  \citenamefont {Simmonds}, \citenamefont {Teufel},\ and\ \citenamefont
  {Aumentado}}]{Lecocq}%
  \BibitemOpen
  \bibfield  {author} {\bibinfo {author} {\bibfnamefont {F.}~\bibnamefont
  {Lecocq}}, \bibinfo {author} {\bibfnamefont {L.}~\bibnamefont {Ranzani}},
  \bibinfo {author} {\bibfnamefont {G.~A.}\ \bibnamefont {Peterson}}, \bibinfo
  {author} {\bibfnamefont {K.}~\bibnamefont {Cicak}}, \bibinfo {author}
  {\bibfnamefont {R.~W.}\ \bibnamefont {Simmonds}}, \bibinfo {author}
  {\bibfnamefont {J.~D.}\ \bibnamefont {Teufel}}, \ and\ \bibinfo {author}
  {\bibfnamefont {J.}~\bibnamefont {Aumentado}},\ }\bibfield  {title} {\enquote
  {\bibinfo {title} {{Nonreciprocal Microwave Signal Processing with a
  Field-Programmable Josephson Amplifier}},}\ }\href {\doibase
  10.1103/PhysRevApplied.7.024028} {\bibfield  {journal} {\bibinfo  {journal}
  {Phys. Rev. Applied}\ }\textbf {\bibinfo {volume} {7}},\ \bibinfo {pages}
  {024028} (\bibinfo {year} {2017})}\BibitemShut {NoStop}%
\bibitem [{\citenamefont {Barzanjeh}\ \emph {et~al.}(2017)\citenamefont
  {Barzanjeh}, \citenamefont {Wulf}, \citenamefont {Peruzzo}, \citenamefont
  {Kalaee}, \citenamefont {Dieterle}, \citenamefont {Painter},\ and\
  \citenamefont {Fink}}]{Barzanjeh}%
  \BibitemOpen
  \bibfield  {author} {\bibinfo {author} {\bibfnamefont {S.}~\bibnamefont
  {Barzanjeh}}, \bibinfo {author} {\bibfnamefont {M.}~\bibnamefont {Wulf}},
  \bibinfo {author} {\bibfnamefont {M.}~\bibnamefont {Peruzzo}}, \bibinfo
  {author} {\bibfnamefont {M.}~\bibnamefont {Kalaee}}, \bibinfo {author}
  {\bibfnamefont {P.~B.}\ \bibnamefont {Dieterle}}, \bibinfo {author}
  {\bibfnamefont {O.}~\bibnamefont {Painter}}, \ and\ \bibinfo {author}
  {\bibfnamefont {J.~M.}\ \bibnamefont {Fink}},\ }\bibfield  {title} {\enquote
  {\bibinfo {title} {{Mechanical on-chip microwave circulator}},}\ }\href
  {\doibase 10.1038/s41467-017-01304-x} {\bibfield  {journal} {\bibinfo
  {journal} {Nature Communications}\ }\textbf {\bibinfo {volume} {8}},\
  \bibinfo {pages} {953} (\bibinfo {year} {2017})}\BibitemShut {NoStop}%
\bibitem [{\citenamefont {Mahoney}\ \emph {et~al.}(2017)\citenamefont
  {Mahoney}, \citenamefont {Colless}, \citenamefont {Pauka}, \citenamefont
  {Hornibrook}, \citenamefont {Watson}, \citenamefont {Gardner}, \citenamefont
  {Manfra}, \citenamefont {Doherty},\ and\ \citenamefont {Reilly}}]{Mahoney}%
  \BibitemOpen
  \bibfield  {author} {\bibinfo {author} {\bibfnamefont {A.~C.}\ \bibnamefont
  {Mahoney}}, \bibinfo {author} {\bibfnamefont {J.~I.}\ \bibnamefont
  {Colless}}, \bibinfo {author} {\bibfnamefont {S.~J.}\ \bibnamefont {Pauka}},
  \bibinfo {author} {\bibfnamefont {J.~M.}\ \bibnamefont {Hornibrook}},
  \bibinfo {author} {\bibfnamefont {J.~D.}\ \bibnamefont {Watson}}, \bibinfo
  {author} {\bibfnamefont {G.~C.}\ \bibnamefont {Gardner}}, \bibinfo {author}
  {\bibfnamefont {M.~J.}\ \bibnamefont {Manfra}}, \bibinfo {author}
  {\bibfnamefont {A.~C.}\ \bibnamefont {Doherty}}, \ and\ \bibinfo {author}
  {\bibfnamefont {D.~J.}\ \bibnamefont {Reilly}},\ }\bibfield  {title}
  {\enquote {\bibinfo {title} {{On-Chip Microwave Quantum Hall Circulator}},}\
  }\href {\doibase 10.1103/PhysRevX.7.011007} {\bibfield  {journal} {\bibinfo
  {journal} {Phys. Rev. X}\ }\textbf {\bibinfo {volume} {7}},\ \bibinfo {pages}
  {011007} (\bibinfo {year} {2017})}\BibitemShut {NoStop}%
\bibitem [{\citenamefont {Parker}\ \emph {et~al.}(2022)\citenamefont {Parker},
  \citenamefont {Savytskyi}, \citenamefont {Vine}, \citenamefont {Laucht},
  \citenamefont {Duty}, \citenamefont {Morello}, \citenamefont {Grimsmo},\ and\
  \citenamefont {Pla}}]{Grimsmo_DPA}%
  \BibitemOpen
  \bibfield  {author} {\bibinfo {author} {\bibfnamefont {D.~J.}\ \bibnamefont
  {Parker}}, \bibinfo {author} {\bibfnamefont {M.}~\bibnamefont {Savytskyi}},
  \bibinfo {author} {\bibfnamefont {W.}~\bibnamefont {Vine}}, \bibinfo {author}
  {\bibfnamefont {A.}~\bibnamefont {Laucht}}, \bibinfo {author} {\bibfnamefont
  {T.}~\bibnamefont {Duty}}, \bibinfo {author} {\bibfnamefont {A.}~\bibnamefont
  {Morello}}, \bibinfo {author} {\bibfnamefont {A.~L.}\ \bibnamefont
  {Grimsmo}}, \ and\ \bibinfo {author} {\bibfnamefont {J.~J.}\ \bibnamefont
  {Pla}},\ }\bibfield  {title} {\enquote {\bibinfo {title} {{Degenerate
  Parametric Amplification via Three-Wave Mixing Using Kinetic Inductance}},}\
  }\href {\doibase 10.1103/PhysRevApplied.17.034064} {\bibfield  {journal}
  {\bibinfo  {journal} {Phys. Rev. Applied}\ }\textbf {\bibinfo {volume}
  {17}},\ \bibinfo {pages} {034064} (\bibinfo {year} {2022})}\BibitemShut
  {NoStop}%
\bibitem [{\citenamefont {Ku}\ \emph {et~al.}(2010)\citenamefont {Ku},
  \citenamefont {Manucharyan},\ and\ \citenamefont
  {Bezryadin}}]{Manucharyan_nanowire}%
  \BibitemOpen
  \bibfield  {author} {\bibinfo {author} {\bibfnamefont {J.}~\bibnamefont
  {Ku}}, \bibinfo {author} {\bibfnamefont {V.}~\bibnamefont {Manucharyan}}, \
  and\ \bibinfo {author} {\bibfnamefont {A.}~\bibnamefont {Bezryadin}},\
  }\bibfield  {title} {\enquote {\bibinfo {title} {{Superconducting nanowires
  as nonlinear inductive elements for qubits}},}\ }\href {\doibase
  10.1103/PhysRevB.82.134518} {\bibfield  {journal} {\bibinfo  {journal} {Phys.
  Rev. B}\ }\textbf {\bibinfo {volume} {82}},\ \bibinfo {pages} {134518}
  (\bibinfo {year} {2010})}\BibitemShut {NoStop}%
\bibitem [{\citenamefont {Khorasani}\ and\ \citenamefont
  {Koottandavida}(2017)}]{Khorasani_1}%
  \BibitemOpen
  \bibfield  {author} {\bibinfo {author} {\bibfnamefont {S.}~\bibnamefont
  {Khorasani}}\ and\ \bibinfo {author} {\bibfnamefont {A.}~\bibnamefont
  {Koottandavida}},\ }\bibfield  {title} {\enquote {\bibinfo {title}
  {{Nonlinear graphene quantum capacitors for electro-optics}},}\ }\href
  {http://dx.doi.org/10.1038/s41699-017-0011-9} {\bibfield  {journal} {\bibinfo
   {journal} {npj 2D Materials and Applications}\ }\textbf {\bibinfo {volume}
  {1}} (\bibinfo {year} {2017})}\BibitemShut {NoStop}%
\bibitem [{\citenamefont {Khorasani}(2018)}]{Khorasani_2}%
  \BibitemOpen
  \bibfield  {author} {\bibinfo {author} {\bibfnamefont {S.}~\bibnamefont
  {Khorasani}},\ }\bibfield  {title} {\enquote {\bibinfo {title} {{CUBIT:
  Capacitive qUantum BIT}},}\ }\href {\doibase 10.3390/c4030039} {\bibfield
  {journal} {\bibinfo  {journal} {Journal of Carbon Research}\ }\textbf
  {\bibinfo {volume} {4}},\ \bibinfo {pages} {39} (\bibinfo {year}
  {2018})}\BibitemShut {NoStop}%
\bibitem [{\citenamefont {Youla}(1961)}]{Youla}%
  \BibitemOpen
  \bibfield  {author} {\bibinfo {author} {\bibfnamefont {D.~C.}\ \bibnamefont
  {Youla}},\ }\bibfield  {title} {\enquote {\bibinfo {title} {{A Normal form
  for a Matrix under the Unitary Congruence Group}},}\ }\href {\doibase
  10.4153/CJM-1961-059-8} {\bibfield  {journal} {\bibinfo  {journal} {Canadian
  Journal of Mathematics}\ }\textbf {\bibinfo {volume} {13}},\ \bibinfo {pages}
  {694–704} (\bibinfo {year} {1961})}\BibitemShut {NoStop}%
\bibitem [{\citenamefont {{H. Goldstein, C. Poole and J.
  Safko}}(2001)}]{Goldstein}%
  \BibitemOpen
  \bibfield  {author} {\bibinfo {author} {\bibnamefont {{H. Goldstein, C. Poole
  and J. Safko}}},\ }\href@noop {} {\emph {\bibinfo {title} {{Classical
  Mechanics (3rd Edition)}}}}\ (\bibinfo  {publisher} {Pearson},\ \bibinfo
  {year} {2001})\BibitemShut {NoStop}%
\bibitem [{Note16()}]{Note16}%
  \BibitemOpen
  \bibinfo {note} {The critical values of the final Hamiltonian in Eq.~\protect
  \textup {\hbox {\mathsurround \z@ \protect \normalfont (\ignorespaces \ref
  {eq_final_H_without_branches}\unskip \@@italiccorr )}} coincide with those of
  the secondary constraint $x(y)=y+\beta \sin (y)$ [cf. Eq.~\protect \textup
  {\hbox {\mathsurround \z@ \protect \normalfont (\ignorespaces \ref
  {eq_sec_constrain}\unskip \@@italiccorr )}}]. Therefore, these values
  correspond to the branching points of the multi-valued
  potential.}\BibitemShut {Stop}%
\bibitem [{\citenamefont {Koch}\ \emph {et~al.}(2009)\citenamefont {Koch},
  \citenamefont {Manucharyan}, \citenamefont {Devoret},\ and\ \citenamefont
  {Glazman}}]{PhysRevLett.103.217004}%
  \BibitemOpen
  \bibfield  {author} {\bibinfo {author} {\bibfnamefont {J.}~\bibnamefont
  {Koch}}, \bibinfo {author} {\bibfnamefont {V.}~\bibnamefont {Manucharyan}},
  \bibinfo {author} {\bibfnamefont {M.~H.}\ \bibnamefont {Devoret}}, \ and\
  \bibinfo {author} {\bibfnamefont {L.~I.}\ \bibnamefont {Glazman}},\
  }\bibfield  {title} {\enquote {\bibinfo {title} {{Charging Effects in the
  Inductively Shunted Josephson Junction}},}\ }\href {\doibase
  10.1103/PhysRevLett.103.217004} {\bibfield  {journal} {\bibinfo  {journal}
  {Phys. Rev. Lett.}\ }\textbf {\bibinfo {volume} {103}},\ \bibinfo {pages}
  {217004} (\bibinfo {year} {2009})}\BibitemShut {NoStop}%
\bibitem [{\citenamefont {Faddeev}\ and\ \citenamefont
  {Jackiw}(1988)}]{Faddeev_Jackiw}%
  \BibitemOpen
  \bibfield  {author} {\bibinfo {author} {\bibfnamefont {L.}~\bibnamefont
  {Faddeev}}\ and\ \bibinfo {author} {\bibfnamefont {R.}~\bibnamefont
  {Jackiw}},\ }\bibfield  {title} {\enquote {\bibinfo {title} {{Hamiltonian
  reduction of unconstrained and constrained systems}},}\ }\href {\doibase
  10.1103/PhysRevLett.60.1692} {\bibfield  {journal} {\bibinfo  {journal}
  {Phys. Rev. Lett.}\ }\textbf {\bibinfo {volume} {60}},\ \bibinfo {pages}
  {1692--1694} (\bibinfo {year} {1988})}\BibitemShut {NoStop}%
\bibitem [{\citenamefont {García}\ and\ \citenamefont {Pons}(1997)}]{Garcia}%
  \BibitemOpen
  \bibfield  {author} {\bibinfo {author} {\bibfnamefont {J.~A.}\ \bibnamefont
  {García}}\ and\ \bibinfo {author} {\bibfnamefont {J.~M.}\ \bibnamefont
  {Pons}},\ }\bibfield  {title} {\enquote {\bibinfo {title} {{Equivalence of
  Faddeev–Jackiw and Dirac Approaches for Gauge Theories}},}\ }\href
  {\doibase 10.1142/s0217751x97000505} {\bibfield  {journal} {\bibinfo
  {journal} {International Journal of Modern Physics A}\ }\textbf {\bibinfo
  {volume} {12}},\ \bibinfo {pages} {451–464} (\bibinfo {year}
  {1997})}\BibitemShut {NoStop}%
\bibitem [{\citenamefont {Liao}\ and\ \citenamefont
  {Huang}(2007)}]{Non_Equivalence}%
  \BibitemOpen
  \bibfield  {author} {\bibinfo {author} {\bibfnamefont {L.}~\bibnamefont
  {Liao}}\ and\ \bibinfo {author} {\bibfnamefont {Y.~C.}\ \bibnamefont
  {Huang}},\ }\bibfield  {title} {\enquote {\bibinfo {title} {{Non-equivalence
  of Faddeev–Jackiw method and Dirac–Bergmann algorithm and the
  modification of Faddeev–Jackiw method for keeping the equivalence}},}\
  }\href {\doibase https://doi.org/10.1016/j.aop.2006.11.013} {\bibfield
  {journal} {\bibinfo  {journal} {Annals of Physics}\ }\textbf {\bibinfo
  {volume} {322}},\ \bibinfo {pages} {2469--2484} (\bibinfo {year}
  {2007})}\BibitemShut {NoStop}%
\end{thebibliography}%

\end{document}